\documentclass[]{article}
\usepackage[margin=2.5cm]{geometry}
\usepackage{comment}
\usepackage{amsmath,amssymb,extarrows,mathtools,graphicx,subfigure,setspace,bbold}
\usepackage{cite}
\usepackage{slashed}
\usepackage{color}
\usepackage{tikz}
\usepackage{multirow}
\usepackage{hhline}
\makeatletter
\def\vhrulefill#1{\leavevmode\leaders\hrule\@height#1\hfill \kern\z@}
\makeatother
\makeatother

\newcommand{\be}{\begin{equation}}
\newcommand{\bea}{\begin{eqnarray}}
\newcommand{\eea}{\end{eqnarray}}
\newcommand{\ba}{\begin{array}}
\newcommand{\ea}{\end{array}}
\newcommand{\ee}{\end{equation}}
\newcommand{\bes}{\begin{equation*}}
\newcommand{\beas}{\begin{eqnarray*}}
\newcommand{\eeas}{\end{eqnarray*}}
\newcommand{\bas}{\begin{array*}}
\newcommand{\eas}{\end{array*}}
\newcommand{\ees}{\end{equation*}}

\numberwithin{equation}{section}
\begin{document}
\onehalfspacing
\vfill
\begin{titlepage}
\vspace{10mm}
\begin{flushright}
 IPM/P-2014/026 \\
%FPAUO-12/10\\
\end{flushright}
%\vbox{
 %   \halign{#\hfil         
 %          IPM/P-2014/050 
  %         
 %                     }  end of \halign
  %    }  % end of \vbox
\vspace*{20mm}
\begin{center}
{\Large {\bf On the Time Evolution of Holographic $n$-partite Information}\\
}

\vspace*{15mm}
\vspace*{1mm}
{Mohsen Alishahiha${}^a$,   M. Reza Mohammadi Mozaffar$^a$ and Mohammad Reza Tanhayi ${}^{a,b}$}

 \vspace*{1cm}

{\it ${}^a$ School of Physics, Institute for Research in Fundamental Sciences (IPM)\\
P.O. Box 19395-5531, Tehran, Iran \\  
${}^b$ Department of Physics, Faculty of Basic Science, Islamic Azad University Central Tehran Branch (IAUCTB), P.O. Box  14676-86831, Tehran, Iran\\
%${}^c $ School of Particles and Accelerators\\Institute for Research in Fundamental Sciences (IPM)\\
%P.O. Box 19395-5531, Tehran, Iran \\ 
}
 \vspace*{0.5cm}
{E-mails: {\tt alishah@ipm.ir,  m$_{-}$mohammadi@ipm.ir, m-tanhayi@iauctb.ac.ir}}%

\vspace*{1cm}
%%\maketitle
\end{center}

\begin{abstract}
We study various scaling behaviors of $n$-partite information during a process of 
thermalization for $n$ disjoint system consisting of $n$ parallel strips whose
widths  are much larger than the separation between them. By making use of the holographic description
for entanglement entropy we explore holographic description of the $n$-partite information by which 
we show that it has a definite sign: it is positive for even $n$ and negative for odd $n$. 
This might thought of as an intrinsic property of a field theory which has gravity dual.

\end{abstract}

\end{titlepage}

%%%%%%%%%%%%%%%%%%%
%%%%%%%%%%%%%%%%%%%%%%%%%%%%%%%%%%%%%%%%%%%%%%%%%%%%%%%%%%%%%%%%%%%%%%%%%%%%%%
%%%%%%%%%%%%%%%%%%%%%%%%%%%%%%%%%%%%%%%%%%%%%%%%%%%%%%%%%%%%%%%%%%%%%%%%%%%%%%
\section{Introduction}

Entanglement entropy for a spatial region $A$ in a quantum field theory is defined 
by the von Neumann entropy of the corresponding reduced density matrix, 
$S_A=-{\rm Tr}(\rho_A \log\rho_A)$. Here $\rho_A$ is the reduced density matrix given by
$\rho_A={\rm Tr}_{\bar{A}}\,\rho$ with $\bar{A}$ being the complement of $A$ and $\rho$ is the
total density matrix describing the state of the corresponding quantum field theory. In general,
for a local field theory, the 
entanglement entropy is UV divergent and the coefficient of the most divergent term  for spatial dimensions bigger than one is proportional 
to the area
of the entangling region\cite{Srednicki:1993im}, while for the spatial dimension equal to one the divergent 
term is logarithmic (see for example \cite{{Holzhey:1994we},{Calabrese:2004eu}}
for two dimensional CFT).

Entanglement entropy for two disjoint regions has been studied in\cite{{Caraglio:2008pk},{Furukawa:2008uk},{Calabrese:2009ez},{Calabrese:2010he}}. We note, however, that  
for two disjoint regions $A$ and $B$, it is more natural to compute the amount of correlations
(both classical and quantum) between these two regions which is given by the mutual information. It is actually a quantity which measures the amount of information that $A$ and $B$ can share.
In terms of the entanglement entropy it is given by
\be\label{IAB}
I(A,B)=S(A)+S(B)-S({A\cup B}),
\ee
where $S(A), S(B)$ and $S({A\cup B})$ respectively are the entanglement entropies of $A$, $B$ and their
union with the rest of the system. Although the entanglement entropy is UV divergent, the mutual information is finite. Moreover by making use of the subadditivity  property of the entanglement entropy, it is evident  that the mutual information is always non-negative and it is zero for two uncorrelated systems. 

More generally one may want to compute entanglement entropy for a subsystem consisting of 
$n$ disjoint regions 
$A_i,\;i=1,\cdots, n$. Following the notion of mutual information for a system of two disjoint 
regions, it is natural to define a quantity, $n$-partite information,  which could measure the amount of 
information or correlations (both classical and quantum) between them. Intuitively, one would expect 
that for $n$ un-correlated systems the $n$-partite information must be zero. Moreover,  for $n$ separated systems 
it should be finite.  Actually for a given $n$ disjoint regions, there is no a unique way to define
  $n$-partite information and indeed, it can be defined in different ways;  each of them has its own advantage. In particular 
in terms of the entanglement entropy one may define an $n$-partite information 
as follows\cite{Hayden:2011ag}
\be\label{JAB}
I^{[n]}(A_{\{i\}})=\sum_{i=1}^nS(A_i)-\sum_{i<j}^n S(A_i\cup A_j)+\sum_{i<j<k}^n S(A_i\cup A_j\cup A_k)
-\cdots\cdots -(-1)^n S(A_1\cup A_2\cup\cdots\cup A_n),
\ee
where $S(A_i\cup A_j\cdots )$ is the entanglement entropy of the region $A_i\cup A_j\dots$ with the rest of the system. Note that with this definition the
1-partite information and 2-partite information are, indeed, entanglement entropy and mutual information, respectively. It 
is clear that, for this definition, $n$-partite information for $n\geq 2$ is finite. It is worth noting  that 
the $n$-partite information \eqref{JAB} may be
re-expressed in terms of $(n-1)$-partite information as follows
\be\label{IJ}
I^{[n]}(A_{\{i\}})=I^{[n-1]}(A_{\{1,\cdots,n-2\}},A_{n-1})+
I^{[n-1]}(A_{\{1,\cdots,n-2\}},A_n)-I^{[n-1]}(A_{\{1,\cdots,n-2\}},A_{n-1}\cup A_n).
\ee
Therefore the  $n$-partite information $I^{[n]}$ may be thought of a quantity which measures the degree of 
extensivity of the $(n-1)$-partite information. Moreover, in terms of the mutual information, the $n$-partite information \eqref{JAB} may be recast into the following form
\bea
I^{[n]}(A_{\{i\}})&=&\sum_{i=2}^nI^{[2]}(A_1,A_i)-\sum_{i=2<j}^n
I^{[2]}(A_1,A_i\cup A_j)+\sum_{i=2<j<k}^nI^{[2]}(A_1,A_i\cup A_j\cup A_k)-\cdots\cr &&\cr
&+&(-1)^nI^{[2]}(A_1,A_2\cup A_2\cdots\cup A_n).
\eea
It is worth mentioning that although the mutual information is always non-negative, the $n$-partite information
$I^{[n]}$ could have either  signs.

In the literature of information theory for a subsystem consisting of $n$ disjoint regions, one may
define another quantity which, indeed, is a direct generalization of
mutual information  (known as multi-partite entanglement) defined as  (see for example \cite{Horodecki:2009zz})
\be\label{IA}
J^{[n]}(A_{\{i\}})=\sum_{i}^nS(A_i)-S(A_1\cup A_2\cup\cdots \cup A_n).
\ee
In terms of the mutual information it may be recast into the following form
\be
J^{[n]}(A_{\{i\}})=I^{[2]}(A_1, A_2)+I^{[2]}(A_1\cup A_2,A_3)+\cdots +I^{[2]}(A_1\cup A_2\cdots\cup A_{n-1},
A_n).
\ee
Note that this quantity is finite for a system with $n$ disjoint regions and is zero for $n$ un-correlated regions. In this
paper we will mainly study the $n$-partite information defined in equation \eqref{JAB}.

Entanglement entropy (or other measures defined above) may provide a useful quantity in studying the process of  thermalization 
under which an excited state could thermalize to an equilibrated state. 
Indeed, being out of equilibrium the thermodynamical quantities such as temperature, entropy, 
pressure, etc. are not well defined during  the process of thermalization.  Evolution of a system after a global quantum quench  \cite{CC} is an example of the thermalization. In a field theory, global quantum quench is a sudden change in the system which might be caused by turning on/off a parameter in the
Hamiltonian of the system  in an interval $\delta t\rightarrow 0$. This change takes the system to an excited state for a new Hamiltonian with non-zero energy density that could eventually thermalize to an equilibrium state.

Although the entanglement entropy could be useful to probe the thermalization process, in general,
 for a generic quantum system it is difficult to compute it. We note, however, that
for those  strongly coupled systems which have gravitational duals \cite{Maldacena:1997}, in order to compute the entanglement entropy one may employ its holographic description \cite{{RT:2006PRL},{RT:2006}}. Of course  when the system is time-dependent, one should use the covariant proposal of entanglement entropy\cite{Hubeny:2007xt}. For completeness and further use, we have reviewed the holographic computations of the entanglement entropy in an appendix.

In the context of AdS/CFT correspondence, the thermalization process after a global quantum quench may be described by the process of a black hole (brane) formation due to a gravitational collapse of a thin shell of matter. The corresponding metric of a collapsing shell of neutral matter in $d+1$ dimensions is given by the AdS-Vaidya metric
\be\label{Vaidya}
 ds^2=\frac{1}{\rho^2}\bigg(-f(\rho,v)dv^2-2 d\rho dv+d\vec{x}^2\bigg),\;\;\;\;\;\;\;\;\;
f(\rho,v)=1-m\;\theta(v)\;\rho^{d},
\ee
 where $\rho$ is the radial coordinate, $x_i$'s $(i=1,\cdots,d-1)$ are the spatial boundary coordinates and $v$ is the null coordinate, note that the AdS radius is set to be one. Here, $\theta(v)$ is the step function, and hence, one is dealing with an AdS geometry for $v<0$ while
for $v>0$ the geometry is an AdS-Schwarzschild black brane whose horizon is located at $\rho_H=m^{-1/d}$.

The above background having  a theta function on it  ($m(v)=m \;\theta(v)$) could provide a gravitational description for a sudden change in a strongly coupled field theory  which might be caused by  turning on a  source for an operator in an  interval $\delta t\rightarrow 0$.
This change can excite the system to an excited state with non-zero energy density that could
eventually thermalize to an equilibrium thermal state. Since we are considering a sudden change in the
theory,  it is then natural to think of the  process  as a thermalization after {\it a global quantum quench}.

Therefore in order to study  entanglement entropy during the process of thermalization after 
a global quantum quench  one needs to compute  the holographic entanglement entropy in the above time-dependent background using its covariant proposal\cite{Hubeny:2007xt}.
 Indeed, using the above metric, time dependent behaviors of entanglement entropy  has been studied in several works including  \cite{{AbajoArrastia:2010yt},{Albash:2010mv},{Balasubramanian:2010ce},{Aparicio:2011zy},{Galante:2012pv},{Caceres:2012em},{Baron:2012fv},{Fischler:2012ca},{Fischler:2012uv},{Caputa:2013eka},{Fischler:2013fba},{Pedraza:2014moa},{Mukherjee:2014gia}}\footnote{
Time dependent entanglement entropy for field theories whose gravitational duals 
are provided by hyperscaling violating geometries has been studied in\cite{{Alishahiha:2014cwa},{Fonda:2014ula}}.}  where it was shown that the entanglement entropy grows linearly with time
then it saturates to its equilibrium value (see also \cite{Keranen:2014zoa} ).

The above consideration may be compared with the results of  \cite{CC}  where the behavior 
of the entanglement entropy after a  global quantum quench for a two dimension CFT was studied.
Although the quench considered in this case is  different from that studied holographically,
there is an agreement between the results of these two different setups. Of course this agreement 
might be 
understood from the fact that in both cases one is considering  the evolution of an excited state
in a CFT. Keeping this  distinction in our mind, in what follows we will refer to  our  setup as a gravitational description for a thermalization process after a global quantum  quench.

Using the gauge/gravity correspondence the mutual information has also been studied
in\cite{{Headrick:2010zt},{Hubeny:2007re},{Tonni:2010pv}} where it was shown that the mutual
information exhibits a phase transition from a positive value to zero as one increases the 
distance between two regions. Time dependent behavior of mutual information in 
a global quantum quench has been studied in \cite{{Balasubramanian:2011at},{Allais:2011ys},{Callan:2012ip},{Li:2013sia}} where it was numerically  shown that the mutual information in a time 
dependent background is always non-negative if the solution satisfies null energy condition.

 Tripartite information $I^{[3]}$ during a global quantum quench for a strongly coupled field theory
 has been also studied in \cite{{Balasubramanian:2011at},{Allais:2011ys}} using its 
holographic description. It was shown, numerically, that the tripartite information for 
a strongly coupled field theory which has gravity description is always non-positive. Actually,
it was observed in \cite{Hayden:2011ag} that the holographic mutual information is monogamous. 
Therefore one may consider the monogamy condition of mutual information for a strongly 
coupled field theory as a necessary condition for a theory to have a gravity description.

  The main aim of this article is to study different scaling behaviors of the  $n$-partite information
in the thermalization process of a strongly coupled field theory undergoing a global
quantum  quench using the holographic description.  To do so, we will consider $n$ disjoint parallel strips with the same width $\ell$ separated by distance $h$. Motivated by the study of mutual information,
we will consider the case where $\ell\gg h$. In this case and under certain assumptions the expression
for $n$-partite information will be simplified significantly so that we could study its
scaling behaviors analytically. Indeed  taking into account that $n$-partite information may be expressed  in 
terms of the entanglement entropy of different entangling regions,  one may  utilize the procedure of  \cite{{Liu:2013iza},{Liu:2013qca}} 
to compute  the corresponding entanglement entropy and thereby
% in a global quantum quench 
to study the evolution 
of $n$-partite information during a global quantum quench. By making use of this procedure we will
show that the holographic $n$-partite information has definite sign, which following
\cite{Hayden:2011ag} might be thought of as a necessary condition for a theory to have a gravity description.

The paper is organized as follows: in the next section, we will review the computations of holographic
mutual information in static backgrounds. In order to explore the procedure we will first  study the scaling behaviors of mutual information 
in the thermalization process after a global quantum quench in a strongly coupled field theory, in section three. The time evolution of the $n$-partite information will be discussed in section 4 and 5.  In section 6 we will present  numerical results to examine our analytical results. Finally section 7 is devoted to  conclusions and discussions. We have enclosed the paper with some further details of calculations in an appendix.

%%%%%%%%%%%%%%%%%%%%%%%%%%%%%%%%%%%%%%%%%%%%%%%%%%%%%%%%%%%%%%%%%%%
%%%%%%%%%%%%%%%%%%%%%%%%%%%%%%%%%%%%%%%%%%%%%%%%%%%%%%%%%%%%%%%%%%%

\section{Mutual information for static backgrounds}

In this section we review the computation of holographic  mutual information for two
parallel strips in static backgrounds. The backgrounds we will be  considering is either  an AdS geometry or an AdS black brane which could provide  gravitational descriptions for a conformal field theory at the ground state or a thermal state, respectively.  To fix our notation, let us consider two parallel infinite strips with the equal width $\ell$ 
separated by a distance $h$ in a $d$-dimensional
field theory as depicted in Fig.1.  
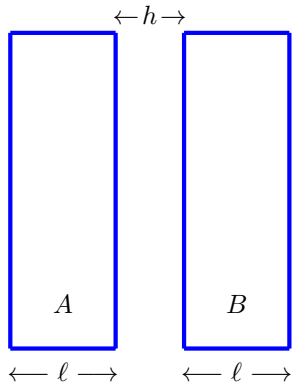
\begin{figure}[h]
\label{fig:1}
\begin{center}
\begin{tikzpicture}[scale=.7]
\draw[ultra thick,blue] (-0.5,0) -- (-0.5,6);
\draw[ultra thick,blue] (-0.5,0) -- (1.5,0);
\draw[ultra thick,blue] (1.5,0) -- (1.5,6);
\draw[ultra thick,blue] (-0.5,6) -- (1.5,6);
\draw[ultra thick,blue] (2.8,0) -- (2.8,6);
\draw[ultra thick,blue] (2.8,0) -- (4.8,0);
\draw[ultra thick,blue] (4.8,0) -- (4.8,6);
\draw[ultra thick,blue] (2.8,6) -- (4.8,6);
\draw[] (0.5,0.5) node[above] {$A$}; 
\draw[] (3.8,0.5) node[above] {$B$}; 
\draw[] (0.5,-0.1) node[below] {$\longleftarrow \ell  \longrightarrow$}; 
\draw[] (3.8,-0.1) node[below] {$\longleftarrow \ell  \longrightarrow$}; 
\draw[] (2.15,6) node[above] {$\leftarrow \! h \! \rightarrow$}; 
\end{tikzpicture}
\caption{Two disjoint entangling regions $A$ and $B$ for computing mutual information.}
\end{center}
\end{figure}

It was argued in \cite{Headrick:2010zt} that the holographic mutual information undergoes a first 
order phase transition as one increases the distance between two strips. Indeed, there is a critical value of $\frac{h}{\ell}$  above which  the mutual information vanishes. This peculiar behavior has to do with 
the definition of entanglement entropy of the union $A\cup B$. Holographically this phase transition
may be understood from the fact that for a given two strips with the widths $\ell$ and distance $h$, there could be 
two minimal hypersurfaces associated with the entanglement entropy $S(A\cup B)$ and thus
the corresponding entanglement entropy behaves differently. More precisely one gets
\bea\label{SAUB}
S({A\cup B})=\Bigg\{ \begin{array}{rcl}
&S(2\ell+h)+S(h)&\,\,\,h\ll \ell,\\
&2S(\ell)&\,\,\,h\gg \ell,
\end{array}\,\,
\eea
where $S(l)$ is the entanglement entropy of a strip with width $l$. From the above 
expression and the definition of the mutual information \eqref{IAB}, it is  then clear that in the case of $h\gg \ell$, the mutual information becomes zero, while for  $h\ll \ell$, one finds 
\be\label{MI}
I(\ell,\ell,h)=2S(\ell)-S(2\ell+h)-S(h)\equiv I.
\ee
In what follows we will  consider $h\ll \ell$. Therefore to find the mutual information of two parallel strips depicted in Fig.1, one essentially,
needs to compute the entanglement entropy of three strips\footnote{
If the widths of two parallel strips in Fig.1 are not the same, one should 
compute four entanglement entropies corresponding to $\ell_1, \ell_2, h $ and $\ell_1+\ell_2+h$.} with  widths $h, \ell$ and $2\ell+h$. To make
the paper self contained we have reviewed the computation of holographic  entanglement entropy 
in appendix A. 1.

By making use of the holographic entanglement entropy of a strip in a $d$-dimensional
CFT whose gravity dual is provided by the AdS geometry (see  \eqref{SV}) 
the mutual information for two parallel strips for the vacuum state is found 
\be\label{vac}
{I}_{\text{vac}}=\frac{L^{d-2}c_0}{4G_N}\left(-\frac{2}{\ell^{d-2}}+\frac{1}{(2\ell+h)^{d-2}}+\frac{1}{h^{d-2}}\right),
\ee
where $c_0=\frac{2^{d-2}}{d-2}\left(\sqrt{\pi}\Gamma(\frac{d}{2(d-1)})/\Gamma(\frac{1}{2(d-1)})\right)^{d-1}$. Note that $h\ll \ell$ condition guarantees the positivity of the resultant 
mutual information.

Let us consider the mutual information of the same strips for a thermal state whose gravity dual is provided by an AdS black brane metric
\bea\label{BB}
ds^2=\frac{1}{\rho^2}\bigg(-f(\rho)dt^2+\frac{d\rho^2}{f(\rho)}+d\vec{x}^2\bigg),\hspace*{1cm}f(\rho)=1-m\rho^d.
\eea
The corresponding mutual information may be analytically expressed in certain limits and it is 
illustrative to study it in these limits. For example, one may assume that $\ell\ll \rho_H$ in which 
all entanglement entropies\footnote{Note that since $h\ll \ell$ one also has $h\ll \rho_H$.} involving  in the computation of mutual information, \eqref{MI}, may be expanded as  \eqref{SBH0} leading to
\bea\label{IBH}
{I}_{\text{BH}}={I}_{\text{vac}}
-\frac{L^{d-2}c_1}{2G_N}\; \frac{ (\ell+h)^2}{\rho_H^d},
\eea
where 
\be
c_1=\frac{1}{16(d+1)\sqrt{\pi}}\;\frac{\Gamma(\frac{1}{2(d-1)})^2\Gamma(\frac{1}{d-1})}{\Gamma(\frac{d}{2(d-1)})^2\Gamma(\frac{1}{2}+\frac{1}{d-1})}.
\ee
On the other hand using the holographic renormalization one finds that the dual excited state has non-zero energy which is given by
\bea\label{DE}
\Delta E=\frac{(d-1)L^{d-2}\ell}{16\pi G_N\rho_H^d}.
\eea
Therefore combining equations \eqref{IBH} and \eqref{DE} one arrives at 
\bea\label{FirstM}
\frac{\Delta I}{\Delta E}=-\frac{8\pi c_1}{d-1} \; \ell \left(1+\frac{h}{\ell}\right)^2,\;\;\;\;\;\;\;\;\;\;\;
{\rm with}\;\;\Delta I=I_{\rm BH}-I_{\rm vac}.
\eea
In the light of the first law of entanglement thermodynamics
\cite{{Bhattacharya:2012mi},{Allahbakhshi:2013rda},{Blanco:2013joa},{Wong:2013gua}} 
one may think of the above equation as the first law for mutual information. 
Note that due to the minus sign it is obvious that as one increases the energy
by $\Delta E$ the mutual information decreases by $\Delta I$ such that equation
\eqref{FirstM} holds. In other words it indicates that the mutual information of two static 
regions is maximal when the system is in the  vacuum state.

On the other hand for the case of $h\ll \rho_H \ll \ell$, the corresponding 
entanglement entropy for the region $h$ should be approximated by equation \eqref{SBH0}, while for
those of $\ell$ and $2\ell+h$ one has to use the large entangling region expansion given by equation \eqref{SBH1}. Therefore one arrives at\cite{Fischler:2012uv}
\be\label{MIS}
I=\frac{L^{d-2}}{4G_N}\left(\frac{c_0}{h^{d-2}}-\frac{c_2}{\rho_{H}^{d-2}}-\frac{h}{2\rho_H^{d-1}}-\frac{c_1h^2}{\rho_H^{d}}\right).
\ee
Here $c_2$ is a positive number  which can be found numerically using, {\it e.g.} 
Mathematica (see Appendix A. 1.). For example for $d=3,4$ it is $c_2=0.88,0.33$, respectively.
It is worth noting that although the term containing $c_2$ is subleading in the expression of 
the entanglement entropy at large entangling region, it plays an important role in the 
expression of mutual information and indeed it might be as important as the other terms.  
Finally for $\rho_H \ll h$ and $\rho_H \ll \ell$ the mutual information is identically zero\cite{Fischler:2012uv}.

%%%%%%%%%%%%%%%%%%%%%%%%%%%%%%%%%%%%%%%%%%%%%%%%%%%%%%%%%%%%%%%%%%%%%
%%%%%%%%%%%%%%%%%%%%%%%%%%%%%%%%%%%%%%%%%%%%%%%%%%%%%%%%%%%%%%%%%%%%%

\section{ Time evolution of mutual information}

In this section we shall study the scaling behavior of the holographic mutual information during 
the process of thermalization. We will consider a case where 
the quench occurs in a time interval $\delta t\rightarrow 0$ so that 
the corresponding gravitational description of the process may be provided by 
AdS-Vaidya metric given by \eqref{Vaidya}. 
We will compute the  holographic mutual information for the parallel strips depicted in Fig.1.  We must emphasis that in the following studies the resultant semi-analytic expansion just gives us a piece wise and not a smooth function for mutual information.

Following our discussions in the previous section we will assume $h\ll \ell$ so that to 
find the mutual information, essentially, one needs to
compute holographic entanglement entropy of three strips with widths $ h, \ell$ and $2\ell+h$ in the AdS-Vaidya metric \eqref{Vaidya}.
%backgrounfwhich are denoted by $S_1=S(h),S_2=S(\ell)$ and $S_3=S(2\ell+h$), respectively.
To do so, one should use the covariant proposal of the entanglement entropy which has been reviewed in the appendix A.2.  Note that in the present case 
we will have to deal with three hypersurfaces. For each of these hypersurfaces we denote the crossing point and the turning point by $(\rho_{i\;c},\rho_{i\; t})$ with $i=1, 2, 3$.

%Now the problem is to find the scaling behavior  of the above entanglement entropies as the system evolves
%with time.
 The system has several scales and therefore as time passes one should  look for different behaviors of the 
corresponding entropies in different scales. In the present case where two strips have the same width
there are four  scales given by $\rho_H, h, \ell$ and $2\ell+h$.  As a matter of fact, having noted that $h\ll \ell$, there are four main possibilities for the order of scales as follows
\bea
\rho_H\ll \frac{h}{2}\ll\frac{\ell}{2}<\ell+\frac{h}{2},\;\;\;\;\;\;\;\;\;\;\;\;\frac{h}{2}\ll\rho_H\ll \frac{\ell}{2}<\ell+\frac{h}{2},\nonumber\\
\frac{h}{2}\ll \frac{\ell}{2}<\rho_H<\ell+\frac{h}{2},\;\;\;\;\;\;\;\;\;\;\;\;\frac{h}{2}\ll \frac{\ell}{2}<\ell+\frac{h}{2}\ll \rho_H,
\eea
which we will study them separately.

Note that in all cases for $v<0$ the step function in \eqref{Vaidya} is zero and the dual geometry is an AdS 
solution which is a static background. Therefore the mutual information of the vacuum state before the quench is given by the equation \eqref{vac}. Note also that as one increases the width of 
strips there is an upper limit for the mutual information given by
\be\label{IMaxv}
{I}_{\text{vac}}^{\rm max}=\frac{L^{d-2}}{4G_N}\;\frac{c_0}{h^{d-2}},
\ee
which is the absolute value of the finite part of the entanglement entropy for a strip with the width $h$.
%%%%%%%%%%%%%%%%%%%%%%%%%%%%%%%%%%%%%%%%%%%%%%%%%%%%%%%%%%%%%

\subsection{First case: $\rho_H\ll \frac{h}{2}$}
%%%%%%%%%%%%%%%%%%%%%%%%%%%%%%%%%%%%%%%%%%%%%%%%%%%%%%%%%%%%%%

In this case the width of the all entangling regions involved in the 
computation of the mutual information are much greater than the horizon radius and, consequently, the corresponding co-dimension two hypersurfaces can penetrate the horizon. % and probe the $v<0$ region.  
To study  the time scaling behavior of the mutual information, one may distinguish between five time intervals as stated 
below.

%%%%%%%%%%%%%%111111111%%%%%%%%%%%%%%%%%%%%%%%%%%%%%%
\subsubsection{Early time: $t\ll \rho_H$}

In this time interval, the co-dimension two hypersurfaces in the bulk associated with the entanglement entropies appeared in equation \eqref{MI} cross the null shell almost at same point which is  very close to the boundary, 
\bea
\rho_{1\; c}\approx \rho_{2\;c}\approx \rho_{3\;c}\approx \rho_{c},\hspace*{1cm}\text{and}\hspace*{1cm} \frac{\rho_{c}}{\rho_H}\ll 1.
\eea
Therefore the holographic entanglement entropy for all regions may be well approximated 
by the equation \eqref{SBHET} 
\be
S(l_i)=\frac{L^{d-2}}{4G_N}\left(\frac{1}{(d-2)\epsilon^{d-2}}-\frac{c_0}{l_i^{d-2}}+\frac{t^2}{4\rho_H^d}+
{\cal O} (t^{d+2})\right),\;\;\;\;\;\;\;\;\; i=1,2,3.
\ee
Here and throughout this section we use a notion in which $l_1=h,\,\, l_2=\ell$ and 
$l_3=2\ell+h$. Plugging these expressions 
into the equation \eqref{MI}, one finds
\bea\label{earlytime}
I=\frac{L^{d-2}c_0}{4G_N}\left(-\frac{2}{\ell^{d-2}}+\frac{1}{h^{d-2}}+\frac{1}{(2\ell+h)^{d-2}}\right)+
{\cal O}(t^{2d})=I_{\rm vac}+{\cal O}(t^{2d}).
\eea
One observes that the mutual information starts from its value in the vacuum, $I_{\rm vac}$,  
and remains fixed
up to order of ${\cal O}(t^{2d})$  at the early times.

%%%%%%%%%%%%%%%%1111111111%%%%%%%%%%%%%%%%%%%%%%
\subsubsection{ Steady behavior: $\rho_H\ll t\ll \frac{h}{2}$ \label{3.1.2}}

The system reaches a local equilibrium at $t\sim \rho_H$ when it has ceased  production of thermodynamic entropy, though the entanglement entropy 
still increases. In this time interval all entanglement entropies appearing in  
\eqref{MI} grow linearly with time (see \eqref{SBHL}). In other words, one has
\be
S(l_i)=\frac{L^{d-2}}{4G_N}\left(\frac{1}{(d-2)\epsilon^{d-2}}-\frac{c_0}{l_i^{d-2}}+ \frac{\sqrt{-\tilde{f}(\rho_{i\;m })}}{\rho_{i\;m}^{d-1}}t+\cdots\right), \;\;\;\;\;\;\;\;i=1,2,3.
\ee
where $\tilde{f}(\rho)$ and $\rho_m$ are defined in appendix A.2. The mutual information is then obtained from  \eqref{MI} as follows
\bea
I=I_{\rm vac} +\frac{L^{d-2}}{4G_N}\left(2\frac{\sqrt{-\tilde{f}(\rho_{2\;m}})}{\rho_{2\;m}^{d-1}}-\frac{\sqrt{-\tilde{f}(\rho_{1\; m}})}{\rho_{1\; m}^{d-1}}-\frac{\sqrt{-\tilde{f}(\rho_{3\; m}})}{\rho_{3\; m}^{d-1}}\right)t+\cdots\, .
\eea
Since  we are dealing with the large entangling regions,  the turning points of all hypersurfaces
are large, and therefore from the equation \eqref{largesimp1} one can deduce 
 $\rho_{i\;m}=\rho_m=
(2(d-1)/(d-2))^{1/d} \rho_H$. As a result, the second term in the above equation vanishes leading 
to a constant mutual information in this time interval too. Thus starting from a static solution one gets
almost constant mutual information all the way from $t=0$ to $t\sim \frac{h}{2}$.

%%%%%%%%%%%%%%%%%%1111111111%%%%%%%%%%%%%%%%%%%%

\subsubsection{Linear Growth: $\frac{h}{2}\ll t \ll \frac{\ell}{2}$}

Using \eqref{SBH1} and \eqref{SBHL} one can show that the entanglement entropy 
associated with the entangling region $h$ will be saturated to its equilibrium 
value at $ t\propto \frac{h}{2}-c_2 \rho_H+c_0 \frac{\rho_H^{d-1}}{h^{d-2}}$. 
Therefore in this time interval the  entanglement entropy $S(h)$ is given by  \eqref{SBH1} 
\be
S(h)\approx\frac{L^{d-2}}{4G_N }\left(\frac{1}{(d-2)\epsilon^{d-2}}+\frac{h}{2\rho_H^{d-1}}
-\;\frac{c_2}{\rho_H^{d-2}}\right).
\ee
On the other hand entanglement entropies associated with the entangling regions $\ell$ and $2\ell+h$ 
are still increasing linearly with time. Thus from the equation \eqref{SBHL} one has
\be
S(l_i)=\frac{L^{d-2}}{4G_N}\left(\frac{1}{(d-2)\epsilon^{d-2}}-\frac{c_0}{l_i^{d-2}}+ \frac{\sqrt{-\tilde{f}(\rho_{i\;m })}}{\rho_{i\;m}^{d-1}}t+\cdots\right), \;\;\;\;\;\;\;\;i=2,3.
\ee
Plugging these results into equation \eqref{MI} one finds
\be
{I}=\frac{L^{d-2}}{4G_N}\left[ -\frac{2c_0}{\ell^{d-2}}+\frac{c_0}{(2\ell+h)^{d-2}}-\frac{1}{\rho_{H}^{d-1}}\frac{h}{2}+\frac{c_2}{\rho_{H}^{d-2}}+\left(2 \frac{\sqrt{-\tilde{f}(\rho_{2\;m })}}{\rho_{2\;m}^{d-1}}- \frac{\sqrt{-\tilde{f}(\rho_{3\;m })}}{\rho_{3\;m}^{d-1}}\right)t+\cdots\right],
\ee
which can be recast into the following form
\bea\label{mutuallingro}
I=I_{\rm vac}+\frac{L^{d-2}}{4G_N}\left(\frac{c_2}{\rho_H^{d-2}}-\frac{c_0}{h^{d-2}}\right)+
\frac{L^{d-2}}{4G_N\rho_H^{d-1}}
\left(v_E t-\frac{h}{2}\right),
\eea
where $v_E$ is given by \eqref{VEsimple}. Here we have  used the fact that the entangling regions are large so that $\rho_{i m}=\rho_m$ with $\rho_m$ is
given by  \eqref{largesimp1}.
Note that the above mutual information is positive as long as $\rho_H\ll \frac{h}{2}$ and  $\frac{h}{2}\ll t$. It is also clear that the mutual information in this time interval is always bigger than $I_{\rm vac}$ and grows linearly with time. It is also worth noting that to get a positive mutual information it was crucial to keep
the subleading term $\frac{c_2}{\rho_H^{d-2}}$ in the equation \eqref{SBH1}.

Assuming  to have a linear growth all the way up to $t\sim \frac{\ell}{2}-c_2 \rho_H+c_0\frac{\rho_H^{d-1}}{\ell^{d-2}}$ where the entanglement entropy 
associated with the entangling region $\ell$ saturates to its equilibrium value, the mutual information
reaches  its maximum value during the thermalization process. More precisely setting 
\be\label{TMAX}
v_E \,\,t_{\rm max}^{(1)}\sim \frac{\ell}{2}-c_2 \rho_H+c_0\frac{\rho_H^{d-1}}{\ell^{d-2}},
\ee
one finds
\be\label{IMax1}
I_{\rm max}^{(1)}\approx I_{\rm vac}+\frac{L^{d-2}}{4G_N}\left(\frac{c_0}{\ell^{d-2}}-\frac{c_0}{h^{d-2}}\right)+
\frac{L^{d-2}}{4G_N\rho_H^{d-1}}
\left(\frac{\ell}{2}-\frac{h}{2}\right).
\ee
Here  $t_{\rm max}^{(1)}$ is the time when the mutual information reaches its maximum value
$I_{\rm max}^{(1)}$.

%%%%%%%%%%%%%%%%%%1111111111111%%%%%%%%%%%%%%%%%%%%%%%%%%%
\subsubsection{ Linear decreasing: $\frac{\ell}{2} < t < \ell+\frac{h}{2}$\label{sub3.1.4}}

As we have already mentioned at $t\sim \frac{\ell}{2}-c_2 \rho_H+c_0\frac{\rho_H^{d-1}}{\ell^{d-2}}$ the entanglement entropy $S(\ell)$ saturates to its equilibrium value. Therefore in this time interval both entanglement entropies $S(\ell)$ and $S(h)$ have 
to be approximated by their equilibrium values as follows
\be
S(l_i)\approx\frac{L^{d-2}}{4G_N }\left(\frac{1}{(d-2)\epsilon^{d-2}}+\frac{l_i}{2\rho_H^{d-1}}
-\;\frac{c_2}{\rho_H^{d-2}}\right),\;\;\;\;\;\;\;\;\;\;i=1,2,
\ee
 while the one associated with the entangling region $2\ell+h$ still grows linearly with time (see \eqref{SBHL})
\be
S(2\ell+h)=\frac{L^{d-2}}{4G_N}\left(\frac{1}{(d-2)\epsilon^{d-2}}-\frac{c_0}{(2\ell+h)^{d-2}}+ \frac{\sqrt{-\tilde{f}(\rho_{3\;m })}}{\rho_{3\;m}^{d-1}}t+\cdots\right).
\ee
Therefore, in this case the mutual information is 
\be\label{SD0}
I\approx \frac{L^{d-2}}{4G_N}\left(\frac{c_0}{(2\ell+h)^{d-2}}-\frac{c_2}{\rho_H^{d-2}}+\frac{\ell}{\rho_H^{d-1}}
-\frac{h}{2\rho_H^{d-1}}-\frac{v_E}{\rho_H^{d-1}}t\right),
\ee
which may be simplified as follows  
\be\label{SD}
I\approx I_{\rm max}^{(1)}+\frac{L^{d-2}}{4G_N}\left(\frac{c_0}{\ell^{d-2}}-\frac{c_2}{\rho_H^{d-2}}\right)
+\frac{L^{d-2}}{4G_N\rho_H^{d-1}}\left(\frac{\ell}{2}-v_E t\right).
\ee
From this expression, it is then clear that in this time interval $I<I_{\rm max}^{(1)}$, also note that $I$ declines linearly  with time and 
is positive for $t<\ell+\frac{h}{2}$.
% and also $I<I_{\rm max}^{(1)}$. One observes that the mutual information declines linearly with time in this %interval.

%%%%%%%%%%%%%%%%%%%%%1111111111%%%%%%%%%%%%%%%%%%%%%%%%%%%%%%%%%%%%
\subsubsection{ Saturation}\label{section5.1}

If one waits enough the entanglement entropy associated with the entangling region $2\ell+h$
will also saturate to its equilibrium value at $t\sim \ell+\frac{h}{2}-c_2 \rho_H+c_0\frac{\rho_H^{d-1}}{(2\ell+h)^{d-2}}$. So that the mutual information 
will also saturates to its equilibrium value studied in the previous section.  Of course generally it is not
correct to plug just the equilibrium values of the corresponding entanglement entropies into 
the equation \eqref{MI} to find the mutual information. Indeed if one naively do that, in the 
present case, the resultant mutual information would become negative.
Therefore the mutual information must reach its equilibrium value at a saturation time $t^{(1)}_s<
\ell+\frac{h}{2}-c_2 \rho_H+c_0\frac{\rho_H^{d-1}}{(2\ell+h)^{d-2}}$.

%Moreover the time when the mutual information reaches its equilibrium value is not just the one at
%which the entanglement entropy of $2\ell+h$ 
%saturates to its equlibrium value. In fact the corresponding entanglement entropy saturates at $t \sim %\ell+\frac{h}{2}$ and from equation \eqref{SD} it leads to a negative mutual information.

To find the saturation time and the equilibrium value of mutual information we note that 
at the end of the thermalization process the resultant background will be 
an AdS black brane. On the other hand as we have already mentioned in the previous section 
when both the width of strips $\ell$ and distance between them $h$ are large compared to 
the radius of the horizon namely, $\rho_H\ll \ell$ and $\rho_H\ll h$, the mutual information is zero.
Therefore in the present case one would expect that  the mutual information 
becomes zero at the end of the thermalization process. Using this fact, one may estimate the saturation 
time as follows.

Indeed assuming  the mutual information decreases all the way till it becomes zero, from  \eqref{SD0}, one should set
\be\label{Isat1}
I^{(1)}_{\text{sat}}\approx \frac{L^{d-2}}{4G_N}\left(\frac{c_0}{(2\ell+h)^{d-2}}-\frac{c_2}{\rho_H^{d-2}}+\frac{\ell}{\rho_H^{d-1}}
-\frac{h}{2\rho_H^{d-1}}-\frac{v_E}{\rho_H^{d-1}}t^{(1)}_s\right)=0,
\ee
so that the saturation time reads
\bea\label{tsat}
v_E\,t^{(1)}_s\approx \ell-\frac{h}{2}-c_2\rho_H+\frac{c_0\rho_H^{d-1}}{(2\ell+h)^{d-2}}\approx \ell-\frac{h}{2}-c_2\rho_H,
\eea
which shows that the mutual information saturates long before $\ell+\frac{h}{2}-c_2 \rho_H+c_0\frac{\rho_H^{d-1}}{(2\ell+h)^{d-2}}$ which would be 
the saturation time of the entanglement entropy of a strip with width $2\ell+h$. Indeed this
result is consistent with the numerical results of\cite{{Balasubramanian:2011at},{Allais:2011ys},{Callan:2012ip},{Li:2013sia}}. 

Let us summarize the results of this subsection. We have found that for the case where 
$\rho_H\ll \frac{h}{2}$ the mutual information starts from its value in 
the vacuum and remains almost constant up to $t\sim \frac{h}{2}$ , then it starts
growing with time linearly. It reaches its maximum value at $t_{\text{max}}^{(1)}$ after that 
it decreases linearly with time till it becomes zero at the saturation time which takes place 
approximately at $t_s^{(1)}\sim \ell-\frac{h}{2}-c_2\rho_H$ (See Fig.2).

\begin{figure}[h]
\label{fig:2}
\begin{center}
\begin{tikzpicture}[scale=.7]
\draw[->,ultra thick] (0,0) -- (0,7) node[left] {$I$}; 
\draw[->,ultra thick] (0,0) -- (9.5,0) node[below] {$t$};
\draw[dashed,orange,very thick] (1.25,2) -- (1.25,0);
\draw[blue,very thick] (0,2) -- (1.1,2);
\draw[blue,very thick] (1.39,2.175) arc (-45:-90:.4cm and 0.6cm);
\draw[blue,very thick] (1.4,2.17) -- (3,5);
\draw[blue,very thick] (3.39,5.10) arc (70:130:.4cm and 0.6cm);
\draw[dashed,red,very thick] (3.25,5.1) -- (3.25,0);
\draw[dashed,red,very thick] (3.25,5.12) -- (0,5.12);
\draw[blue,very thick] (3.4,5.1) -- (6,0);
\draw[blue,very thick] (6,0) -- (9.4,0);
\draw[->,brown,very thick] (7,0) .. controls (7.5,2) and (7.9,0.5) .. (8.5,1);
\draw[] (0,2) node[left] {$I^{\text{max}}_{\text{vac}}>I_{\text{vac}}$}; 
\draw[] (0,5.12) node[left] {$I^{(1)}_{\text{max}}$}; 
\draw[] (9.5,1.5) node[left] {\small$I^{(1)}_{\text{sat}}=0$}; 
\draw[] (1.26,0) node[below] {$\frac{h}{2}$}; 
\draw[] (3.30,0) node[below] {\small{$t_{\text{max}}^{(1)}$}}; 
\draw[] (6,0) node[below] {\small{$t_s^{(1)}$}}; 
\end{tikzpicture}
\caption{Schematic behavior of mutual information during the thermalization process for 
$\rho_H\ll \frac{h}{2}$. Here $I_{\rm vac}, I_{\rm vac}^{\rm max}$, $t_{\text{max}}^{(1)}$, $I^{(1)}_{\rm max}$, $I^{(1)}_{\rm sat}$ and $t_s^{(1)}$
are given by equations \eqref{vac}, \eqref{IMaxv}, \eqref{TMAX}, \eqref{IMax1}, \eqref{Isat1} and \eqref{tsat} respectively.}
\end{center}
\end{figure}

%%%%%%%%%%%%%%%%%%%%%%%%%%%%%%%%%%%%%%%%%%%%%%%%%%%%%%%
\subsection{Second case: $\frac{h}{2}\ll\rho_H\ll \frac{\ell}{2}<\ell+\frac{h}{2}$}

In this case, similar to the previous subsection, one can study the behavior of the mutual 
information in five time intervals. We note, however, that  since we have  $\frac{h}{2}\ll\rho_H$ condition, the co-dimension two hypersurface corresponding to the entangling region $h$ cannot probe the $v<0$ region.

%%%%%%%%%%%%%%%%2222222222222222%%%%%%%%%%%%%%%%%%%%%%%%%%%%
\subsubsection{Early time: $t \ll \frac{h}{2}$}
In this time interval the behavior of the entanglement entropies appearing in the equation \eqref{MI} at the early times are  the same as 
that  in the previous case. Thus   in this time interval, the mutual information is essentially given by the equation (\ref{earlytime}), which means  that it  remains constant for $t\ll \frac{h}{2}$.

%%%%%%%%%%%%%%%%%%%%%%222222222222222%%%%%%%%%%%%%%%%%%%%%%%%%%
\subsubsection{ Quadratic growth:  $\frac{h}{2} \ll t\ll\rho_H$\label{quadraticgrowth1}}

In this time interval the co-dimension two hypersurface corresponding to the entangling region $h$
remains all the time in the region of  $v>0$  which is, indeed, a static AdS black brane. Therefore the
corresponding entanglement entropy $S(h)$ reaches its equilibrium value which, in the present case,
is given by equation  \eqref{SBH0}. On the other hand since we are still in the range of $t\ll \frac{\ell}{2}$,
the entanglement entropies associated with the entangling regions $\ell$ and $2\ell+h$ are still at the early times so that they should be approximated by   \eqref{SBHET}. Therefore one gets
\be
 S(h)=S_{\text{vac}}+\frac{L^{d-2}}{4G_N} c_1mh^2,
\ee
and
\be
S(l_i) \approx \frac{L^{d-2}}{4G_N}\left(\frac{1}{(d-2)\epsilon^{d-2}}-\frac{c_0}{l_i^{d-2}}+\frac{t^2}{4 \rho_H^d}+{\cal O}(t^{d+2})\right), \hspace*{1cm}
 \,\, i=2,3.
\ee
Plugging these expressions into equation \eqref{MI} one arrives at
\bea\label{quadraticgrowth}
{I}\approx I_{\rm vac} +\frac{L^{d-2}}{4G_N\rho_H^d}\left(-c_1h^2+\frac{t^2}{4}\right),
\eea
showing that the mutual information has a quadratic growth up to $t\sim \rho_H$.

%%%%%%%%%%%%%%%%%%%2222222222222222%%%%%%%%%%%%%%%%%%%%%%%%%%%%%%%
\subsubsection{Linear growth  $\rho_H \ll t \ll \frac{\ell}{2}$}

In this case  the entanglement entropy $S(h)$ is still given by equation \eqref{SBH0}, while
since the system has reached a local equilibrium  and moreover $\rho_H \ll \frac{\ell}{2}$, the 
equation \eqref{SBHL} should be used to describe the entanglement entropies associated with the
entangling regions  $\ell$ and $2\ell+h$,
\be
S(l_i)\approx\frac{L^{d-2}}{4G_N}\left(\frac{1}{(d-2)\epsilon^{d-2}}-\frac{c_0}{l_i^{d-2}}+ \frac{\sqrt{-\tilde{f}(\rho_{i\,m})}}{\rho_{i\,m}^{d-1}}t+\cdots\right),\hspace*{1cm}\,\,i=2,3.
\ee
This leads to the following expression for mutual information
\bea\label{i2lingrowth}
{I}\approx I_{\rm vac}+\frac{L^{d-2}}{4G_N\rho_H^{d-1}} \left(
v_E t-\frac{c_1}{\rho_H} h^2\right).
\eea
Here, again, we have used the fact that the entangling regions are large so that $\rho_{i\,m}=\rho_{m}$. 
Moreover,  in this time interval, the conditions $h\ll \rho_H$ and $\rho_H\ll t$ guarantee that the resultant mutual information will be positive and bigger than $I_{\rm vac}$.

The linear growth lasts all the way up to 
\bea\label{tMax2}
t^{(2)}_{\rm max}\approx \frac{\ell}{2}-c_2 \rho_H+c_0\frac{\rho_H^{d-1}}{\ell^{d-2}},
\eea
when $S(\ell)$  saturates to its equilibrium value.  By making use  of equation \eqref{TMAX} one may also  estimate the maximum value of the mutual information as follows 
\bea\label{IMax2}
{I}^{(2)}_{\rm max}\approx I_{\rm vac}+\frac{L^{d-2}}{4G_N\rho_H^{d-1}} \left(
\frac{\ell}{2}-c_2 \rho_H+c_0\frac{\rho_H^{d-1}}{\ell^{d-2}}-\frac{c_1h^2}{\rho_H} \right)
\eea

%%%%%%%%%%%%%%%%%%%%%22222222222222222222%%%%%%%%%%%%%%%%%%%%%%

\subsubsection{Linear decreasing: $\frac{\ell}{2}< t< \ell+\frac{h}{2}$}

In this time interval both the entanglement entropies $S(h)$ and $S(\ell)$ are saturated to their
equilibrium values, though because of  their  size of entangling regions,  the corresponding equilibrium values are given by
different expressions. Indeed, although the equilibrium value of $S(h)$ is given by the equation \eqref{SBH0}, for that of $S(\ell)$ one should use \eqref{SBH1}.
The entanglement entropy $S(2\ell+h)$ is still growing with time as \eqref{SBHL}. 
Therefore, the mutual information in this time interval linearly decreases as time goes on and it is given by 
\bea\label{YY}
{I}\approx I^{(2)}_{\rm max} +\frac{L^{d-2}}{4G_N}\left(\frac{c_0}{\ell^{d-2}}-\frac{c_2}{\rho_H^{d-2}}\right)+\frac{L^{d-2}}{4G_N\rho_H^{d-1}}\left(\frac{\ell}{2}-v_E t\right).
\eea
Note that the mutual information is positive and also $I<I^{(2)}_{\rm max}$.

%%%%%%%%%%%%%%%%%%%%%%%%%22222222222222222222222222222222%%%%%%%%%%%%%%%%%

\subsubsection{Saturation}

As we have already mentioned, the final state of our system after a global quench we are 
considering is a thermal state whose gravity dual is provided by an AdS black brane. On the other hand
for this static background the mutual information of two strips depicted in Fig.1 with the
condition $\frac{h}{2}\ll \rho_H\ll \frac{\ell}{2}$ is given by  \eqref{MIS}.  Therefore in the
present case the equilibrium value of the mutual information is 
\bea\label{satcase2}
I^{(2)}_{\text{sat}}\approx \frac{L^{d-2}}{4G_N}\left(\frac{c_0}{h^{d-2}}-\frac{c_2}{\rho_{H}^{d-2}}-\frac{h}{2\rho_H^{d-1}}-\frac{c_1h^2}{\rho_H^{d}}\right).
\eea
It is then possible to estimate the saturation time by assuming that the mutual information decreases
linearly with time till it reaches its equilibrium value \eqref{satcase2}. Indeed equating equations
\eqref{YY} and \eqref{satcase2} one finds 
\be\label{tsat2}
v_E\, t^{(2)}_s\approx \frac{c_0 \rho_H^{d-1}}{(2\ell+h)^{d-2}}+\ell+\frac{h}{2}-c_2\rho_H \approx \ell+\frac{h}{2}-c_2\rho_H,
\ee
Moreover, from  \eqref{satcase2} it is obvious that
\be\label{Isat2}
I^{(2)}_{\text{sat}}=I_{\text{vac}}-\frac{L^{d-2}}{4G_N}\left(\frac{c_2}{\rho_{H}^{d-2}}+\frac{h}{2\rho_{H}^{d-1}}+c_1\frac{h^2}{\rho_{H}^{d}}+\frac{c_0}{(2\ell+h)^{d-2}}-\frac{2c_0}{\ell^{d-2}}\right)
\ee
which shows that in this case with the condition $\frac{h}{2}\ll\rho_H\ll \frac{\ell}{2}$  
the expression in the parentheses is always positive leading to the fact that $I^{(2)}_{\text{sat}}<I_{\text{vac}}$.

Let us summarize the results of this subsection. In fact  we have  found that the mutual information starts from its value in 
the vacuum and remains almost constant up to $t\sim \frac{h}{2}$, then it grows with time quadratically till $t\sim \rho_H$. After that a linear behavior starts and it reaches its maximum value at $t_{\text{max}}^{(2)}$.  Finally 
it decreases linearly with time till it saturates to a constant value at the saturation time which takes place 
approximately at $t_s^{(2)}\sim \ell+\frac{h}{2}-c_2\rho_H$ (See Fig.3).
\begin{figure}[h]
\label{fig:3}
\begin{center}
\begin{tikzpicture}[scale=.7]
\draw[->,ultra thick] (0,0) -- (0,7) node[left] {$I$}; 
\draw[->,ultra thick] (0,0) -- (9.7,0) node[below] {$t$};
\draw[dashed,orange,very thick] (1.1,2) -- (1.1,0);
\draw[blue,very thick] (0,2) -- (1,2);
\draw[blue,very thick] (1,2) parabola (2,4);
\draw[blue,very thick] (2,4) .. controls (2.1,4.3) and (3.8,5.85) .. (4,5.8);
\draw[dashed,purple,very thick] (2,4.05) -- (2,0);
\draw[dashed,red,very thick] (4,5.8) -- (4,0);
\draw[dashed,red,very thick] (4,5.8) -- (0,5.8);
\draw[blue,very thick] (4,5.8) .. controls (4.1,5.8) and (5,5.1) .. (7.7,1);
\draw[blue,very thick] (7.7,1) .. controls (7.8,1) and (8,1) .. (9.5,1);
\draw[dashed,violet,very thick] (0,1) -- (7.7,1);
\draw[dashed,violet,very thick] (7.7,0) -- (7.7,1);
%\draw[step=1cm,gray,very thin] (0,0) grid (9,9);
\draw[] (0,2) node[left] {$I^{\text{max}}_{\text{vac}}>I_{\text{vac}}$}; 
\draw[] (0,5.8) node[left] {$I^{(2)}_{\text{max}}$}; 
\draw[] (0,1) node[left] {$I^{(2)}_{\text{sat}}$}; 
\draw[] (1.1,0) node[below] {$\frac{h}{2}$}; 
\draw[] (2,-0.1) node[below] {\small{$\rho_H$}}; 
\draw[] (4,0) node[below] {\small{$t_{\text{max}}^{(2)}$}}; 
\draw[] (7.7,0) node[below] {\small{$t_s^{(2)}$}}; 
\end{tikzpicture}
\caption{Schematic behavior of mutual information during the thermalization process for 
$\frac{h}{2}\ll \rho_H\ll \frac{\ell}{2}$. Here $I_{\rm vac}, I_{\rm vac}^{\rm max}$, $t_{\text{max}}^{(2)}$, $I^{(2)}_{\rm max}$, $I^{(2)}_{\rm sat}$ and $t_s^{(2)}$ 
are given by equations \eqref{vac}, \eqref{IMaxv}, \eqref{tMax2}, \eqref{IMax2}, \eqref{Isat2} and \eqref{tsat2}, respectively.}

\end{center}
\end{figure}
%%%%%%%%%%%%%%%%%%%%%%%%%%%%%%%%%%%%%%%%%%%%%%%%%%%%%%%%%%%

%%%%%%%%%%%%%%%%%%%%%%%%%%%%%%%%%%%%%%%%%%%%%%%%%%%%%%%
\subsection{Third case: $\frac{h}{2}\ll \frac{\ell}{2}<\rho_H<\ell+\frac{h}{2}$}

In this case the entanglement entropies associated with the entangling regions $h$ and $\ell$ saturate
to their equilibrium values before the system reaches a local equilibrium.  Therefore the corresponding 
co-dimension two hypersurfaces cannot probe the region near and behind the horizon. In other
words, the entanglement entropies $S(h)$ and $S(\ell)$ do not exhibit linear growth, though
$S(2\ell+h)$  could still grow linearly with time before it reaches its equilibrium value.

Actually in this case the behavior of the mutual information for early times is almost the same as 
that in the previous case. Namely it starts from its value in the vacuum and remains fixed up to
$t\sim \frac{h}{2}$ then it begins to grow quadratically with time 
\bea
{I}\approx I_{\rm vac} +\frac{L^{d-2}}{4G_N\rho_H^d}\left(-c_1h^2+\frac{t^2}{4}\right),
\eea
Let us assume that the entanglement entropy associated with the entangling region $\ell$ grows
quadratically with time till it reaches its equilibrium value. Then using the fact that in this case
the corresponding equilibrium value is given by \eqref{SBH0}, one may 
estimate the time when the mutual information becomes maximum as follows
\bea\label{TMAX3}
\frac{t^{(3)}_{\text{max}}}{4\sqrt{c_1}} \sim \frac{\ell}{2}
\eea
by which the maximum value of the mutual information reads 
\bea\label{IMax3}
{I}^{(3)}_{\rm max}\approx I_{\rm vac} +\frac{L^{d-2}c_1}{4G_N\rho_H^d}\left(\ell^2-h^2\right).
\eea
Let us now study the other  time intervals in more details. 
%%%%%%%%%%%%%%%%%%%2222222222222222%%%%%%%%%%%%%%%%%%%%%%%%%%%%%%%
\subsubsection{Quadratic decreasing  $\frac{\ell}{2} < t < \rho_H$}
In this time interval, the entanglement entropies associated with the entangling regions  
$h$ and $\ell$ are saturated to their equilibrium  values   given by (see  \eqref{SBH0})
\be
 S(l_i)=S_{i\;\text{vac}}+\frac{L^{d-2}}{4G_N} c_1ml_i^2,\;\;\;\;\;i=1,2.
\ee
On the other hand since  we are in the regime of $t<\rho_H<\ell+\frac{h}{2}$ the entanglement entropy  $S(2\ell+h)$ is still at the early times and should be  approximated by equation \eqref{SBHET}
 \be
S(2\ell+h)\approx\frac{L^{d-2}}{4G_N}\left(\frac{1}{(d-2)\epsilon^{d-2}}-\frac{c_0}{(2\ell+h)^{d-2}}+
 \frac{t^2}{4\rho_h^d}+\cdots\right).
\ee
Plugging these results into equation \eqref{MI}, one finds
\bea
{I}\approx I_{\rm max}^{(3)}+\frac{L^{d-2}}{4G_N\rho_H^{d}} \left(
c_1\ell^2- \frac{t^2}{4}\right).
\eea
Since $t>t^{(3)}_{\rm max}$ it is clear that  $I<I^{(3)}_{\rm max}$.

%%%%%%%%%%%%%%%%%%%%%3333333333333333333333%%%%%%%%%%%%%%%%%%%%%%

%%%%%%%%%%%%%%%%%%%2222222222222222%%%%%%%%%%%%%%%%%%%%%%%%%%%%%%%
\subsubsection{Linear decreasing  $ \rho_H< t<\ell+\frac{h}{2}$}
In this time interval, the entanglement entropies  $S(h)$ and $S(\ell)$ are the same as that in the previous case. On the other hand  since in this time interval the system is locally equilibrated the entanglement entropy $S(2\ell+h)$ exhibits a linear growth. Therefore one has
\be
 S(l_i)=S_{i\; \text{vac}}+\frac{L^{d-2}}{4G_N} c_1ml_i^2,\;\;\;\;\;i=1,2.
\ee
and 
\be
S(2\ell+h)\approx\frac{L^{d-2}}{4G_N}\left(\frac{1}{(d-2)\epsilon^{d-2}}-\frac{c_0}{(2\ell+h)^{d-2}}+ \frac{\sqrt{-\tilde{f}(\rho_{3\,m})}}{\rho_{3\,m}^{d-1}}t+\cdots\right).
\ee
Plugging these results into equation \eqref{MI}, one finds
\bea\label{SD3}
{I}\approx I^{(3)}_{\rm max}+\frac{L^{d-2}}{4G_N\rho_H^{d}} \left(
c_1 \ell^2-v_E \rho_H t\right).
\eea
Here  we have used the large entangling region approximation for $\rho_{3\;m}$.  Note that in this time interval because of $t < \ell+\frac{h}{2}$, one obtains a positive value for mutual information.

%%%%%%%%%%%%%%%%%%%%%3333333333333333333333%%%%%%%%%%%%%%%%%%%%%%

\subsubsection{Saturation}

At this stage let us consider the situation that takes place after a long time when all the entanglement entropies appeared in \eqref{MI} saturate to their equilibrium values. 
Since the  entanglement entropies $S(h)$ and $S(\ell)$  have already saturated (given by the 
equation  \eqref{SBH0}),  they do not change as the system evolves with time,
though the one associated with entangling region $2\ell+h$ will be saturated whose
equilibrium value is given by  \eqref{SBH1}. This would lead to the following mutual 
information
\bea
I_{\text{sat}}^{(3)}\approx \frac{L^{d-2}}{4G_N}\left(-\frac{2c_0}{\ell^{d-2}}+\frac{c_0}{h^{d-2}}+\frac{c_2}{\rho_{H}^{d-2}}-\frac{2\ell+h}{2\rho_H^{d-1}}+\frac{2c_1\ell^2}{\rho_H^{d}}-\frac{c_1h^2}{\rho_H^{d}}\right),
\eea
which may be recast to the following form
\be\label{satcase3}
I_{\text{sat}}^{(3)}\approx I_{\text{vac}}+\frac{L^{d-2}}{4G_N}\left(\frac{c_2}{\rho_{H}^{d-2}}-\frac{2\ell+h}{2\rho_{H}^{d-1}}+\frac{2c_1\ell^2}{\rho_{H}^{d}}-\frac{c_1h^2}{\rho_{H}^{d}}-\frac{c_0}{(2\ell+h)^{d-2}}\right).
\ee
From this expression it is clear that in the range we are interested in (i.e.  $\frac{h}{2}\ll \frac{\ell}{2}<\rho_H<\ell+\frac{h}{2}$)  the expression in the parentheses is always negative and therefore one gets $I_{\text{sat}}^{(3)}<I_{\text{vac}}$\footnote{In the previous version of this paper, there is a mistake about the sign of the second term in equation \eqref{satcase3} in the desired range of the parameters. Doing a numerical analysis which we will discuss in section 6, we find that in this range of parameters this term is always negative.}.
To estimate the saturation time, with the assumption that the mutual information decreases linearly with 
time, one may equate equations \eqref{satcase3} and \eqref{SD3} to find
\be\label{tsat3}
v_E \,t_s^{(3)}\approx  \ell+\frac{h}{2}-c_2 \rho_H+\frac{c_0 \rho_H^{d-1}}{(2l+h)^{d-2}}\approx \ell+\frac{h}{2}-c_2 \rho_H.
\ee

To summarize the results of this subsection, one has observed that the mutual information starts from its value in 
the vacuum and remains almost constant up to $t\sim \frac{h}{2}$, then it starts
growing with time quadratically till $t_{\text{max}}^{(3)}$. Finally 
it first declines quadratically and then linearly with time till it saturates to a constant value at the saturation time which takes place 
approximately at $t_s^{(3)}\sim \ell+\frac{h}{2}-c_2\rho_H$, (See Fig.4).
\begin{figure}[h]\label{fig:4}
\begin{center}
\begin{tikzpicture}[scale=.7]
\draw[->,ultra thick] (0,0) -- (0,7) node[left] {$I$}; 
\draw[->,ultra thick] (0,0) -- (10,0) node[below] {$t$};
\draw[blue,very thick] (0,1) -- (1,1);
\draw[blue,very thick] (1,1) parabola (2,4);
\draw[blue,very thick] (2,4) .. controls (2.4,6) and (3.9,6.1) .. (5,4);
%\draw[blue,very thick] (2,4) .. controls (2.4,6) .. (5,4);
\draw[blue,very thick] (5,4) .. controls (6,3.25) and (5.05,4.05) .. (8.4,0.8);
\draw[blue,very thick] (8.4,0.8) .. controls (8.4,0.76) .. (9,0.77);
\draw[dashed,orange,very thick] (1,1) -- (1,0);
\draw[dashed,red,very thick] (3.25,5.58) -- (3.25,0);
\draw[dashed,red,very thick] (3.25,5.58) -- (0,5.58);
\draw[dashed,purple,very thick] (5,4) -- (5,0);
\draw[dashed,violet,very thick] (8.4,0.8) -- (8.4,0);
\draw[dashed,violet,very thick] (0,0.8) -- (8.4,0.8);
\draw[] (0,1.5) node[left] {$I^{\text{max}}_{\text{vac}}>I_{\text{vac}}$}; 
\draw[] (0,0.6) node[left] {$I^{(3)}_{\text{sat}}$}; 
\draw[] (0,5.7) node[left] {$I^{(3)}_{\text{max}}$}; 
\draw[] (5,0) node[below] {$\rho_{H}$}; 
\draw[] (1,0) node[below] {$\frac{h}{2}$}; 
\draw[] (3.25,0) node[below] {$t_{\text{max}}^{(3)}$}; 
\draw[] (8.28,0) node[below] {$t_s^{(3)}$}; 
%\draw[blue,very thick] (5,2) -- (7,2);%\draw[step=1cm,gray,very thin] (0,0) grid (11,11);
\end{tikzpicture}
\caption{Schematic behavior of mutual information during the thermalization process for 
$\frac{h}{2}\ll \frac{\ell}{2}<\rho_H$. Here $I_{\rm vac}, I_{\rm vac}^{\rm max}$, $t_{\text{max}}^{(3)}$,$I^{(3)}_{\rm max}$, $I^{(3)}_{\rm sat}$ and $t_s^{(3)}$
are given by equations \eqref{vac}, \eqref{IMaxv}, \eqref{TMAX3}, \eqref{IMax3}, \eqref{satcase3} and \eqref{tsat3}, respectively.}
\end{center}
\end{figure}
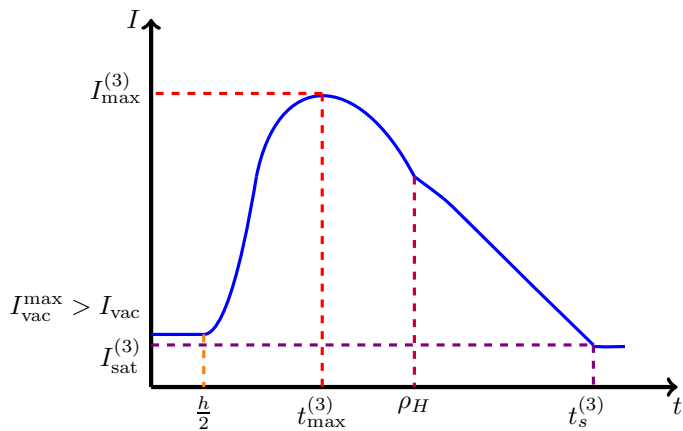

%%%%%%%%%%%%%%%%%%%%%%%%%%%%%%%%%%%%%%%%%%%%%%%%%%%%%%%
\subsection{Fourth case: $\frac{h}{2}\ll \frac{\ell}{2}<\ell+\frac{h}{2}\ll \rho_H$}

In this case due to the fact that all entangling regions $h,\ell$ and $2\ell+h$ are smaller than the
radius of horizon, the corresponding entanglement entropies saturate to their equilibrium values before
the system reaches a local equilibrium. Therefore neither the entanglement entropies nor the mutual
information exhibit linear growth with time during the process  of thermalization.

In fact to study the behavior of the entanglement entropies $S(h), S(\ell)$ and $S(2\ell+h)$ one should use
either the equation \eqref{SBH0}  or \eqref{SBHET} depending on whether they have been saturated or not.
Actually the situation is very similar to the third case studied above.  Namely the mutual information
starts from its value in vacuum and remains fixed up to $t\sim \frac{h}{2}$ when it starts growing
quadratically  with time 
\bea
{I}\approx I_{\rm vac} +\frac{L^{d-2}}{4G_N\rho_H^d}\left(-c_1h^2+\frac{t^2}{4}\right).
\eea
Assuming to have this quadratic growth up to its maximum value given by
\bea\label{IMax4}
{I}^{(4)}_{\rm max}\approx I_{\rm vac} +\frac{L^{d-2}c_1}{4G_N\rho_H^d}\left(\ell^2-h^2\right),
\eea
one may estimate a time when the mutual information becomes maximum 
\bea\label{TMAX4}
\frac{t^{(4)}_{\text{max}}}{4\sqrt{c_1}} \sim \frac{\ell}{2}.
\eea
Then it starts decreasing quadratically with time 
\bea\label{TT2}
{I}\approx I^{(4)}_{\rm max}+\frac{L^{d-2}}{4G_N\rho_H^{d}} \left(
c_1 \ell^2-\frac{t^2}{4}\right),
\eea
which is positive as long as $t<\ell+\frac{h}{2}$. Finally the mutual information reaches its equilibrium value
as system evolves with time. Indeed the saturation takes place, when the values of the entanglement entropies appearing in  \eqref{MI} become that of  an AdS black brane given by \eqref{SBH0}. Therefore from the equation \eqref{MI}, the saturated mutual information is obtained as
\bea\label{Isat4}
{I^{(4)}_{\text{sat}}}={I}_{\text{vac}}
-\frac{L^{d-2}}{2G_N}\;c_1 \frac{ (\ell+h)^2}{\rho_H^d}.
\eea
Assuming to have the quadratic decreasing all the way to the saturation point, one may estimate the
saturation time by equating equations  \eqref{TT2} and \eqref{Isat4} which leads to 
\be\label{tsatcase4}
\frac{t^{(4)}_s}{4\sqrt{c_1}}\approx  \ell+\frac{h}{2}.
\ee
These behaviors are summarized in Fig.5.
\begin{figure}[h]
\label{fig:5}
\begin{center}
\begin{tikzpicture}[scale=.7]
\draw[->,ultra thick] (0.4,-1) -- (0.4,7) node[left] {$I$}; 
\draw[->,ultra thick] (0.4,-1) -- (9,-1) node[below] {$t$};
\draw[blue,very thick] (0.4,1) -- (1,1);
\draw[blue,very thick] (1,1) parabola (2,4);
\draw[blue,very thick] (2,4) .. controls (2.4,6) and (4,6.5) .. (5,0.2);
\draw[blue,very thick] (5,0.2) .. controls (5.05,0.15) .. (8,0.15);
\draw[dashed,orange,very thick] (1,1) -- (1,-1);
\draw[dashed,red,very thick] (2.85,5.3) -- (2.85,-1);
\draw[dashed,red,very thick] (2.8,5.3) -- (0.4,5.3);
\draw[dashed,violet,very thick] (5,0.15) -- (5,-1);
\draw[dashed,violet,very thick] (0.4,0.15) -- (5.05,0.15);
\draw[] (0.4,1.15) node[left] {$I^{\text{max}}_{\text{vac}}>I_{\text{vac}}$}; 
\draw[] (0.4,0.25) node[left] {$I^{(4)}_{\text{sat}}$}; 
\draw[] (0.4,5.3) node[left] {$I^{(4)}_{\text{max}}$}; 
\draw[] (1,-1) node[below] {$\frac{h}{2}$}; 
\draw[] (2.70,-1) node[below] {$t_{\text{max}}^{(4)}$}; 
\draw[] (5.0,-1) node[below] {$t_s^{(4)}$}; 
%\draw[blue,very thick] (5,2) -- (7,2);
%\draw[step=1cm,gray,very thin] (0,0) grid (11,11);
\end{tikzpicture}
\caption{Schematic behavior of mutual information during the thermalization process for 
$l_i\ll\rho_H$. Here $I_{\rm vac}, I_{\rm vac}^{\rm max}$, $I^{(4)}_{\rm max}$, $t_{\text{max}}^{(4)}$, $I^{(4)}_{\rm sat}$ and $t_s^{(4)}$
are given by equations \eqref{vac}, \eqref{IMaxv}, \eqref{IMax4}, \eqref{TMAX4}, \eqref{Isat4} and \eqref{tsatcase4}, respectively.}
\end{center}
\end{figure}

%%%%%%%%%%%%%%%%%%%%%%%%%%%%%%%%%%%%%%%%%%%%%%%%%%%%%%%%%%%%%%%%%%%%%%%%%
\section{$n$-partite information for static backgrounds}

In this section by making use of the AdS/CFT correspondence we will study $n$-partite information of 
a subsystem consists of $n$ disjoint regions $A_i,\; i=1,\cdots, n$ in a $d$-dimensional
CFT for the vacuum and thermal states whose gravity duals are provided by  AdS and AdS
black brane geometries, respectively. The $n$ disjoint regions are given by
$n$ parallel infinite strips of equal width $\ell$ separated by $n-1$ regions of width $h$, as
depicted in Fig.6.
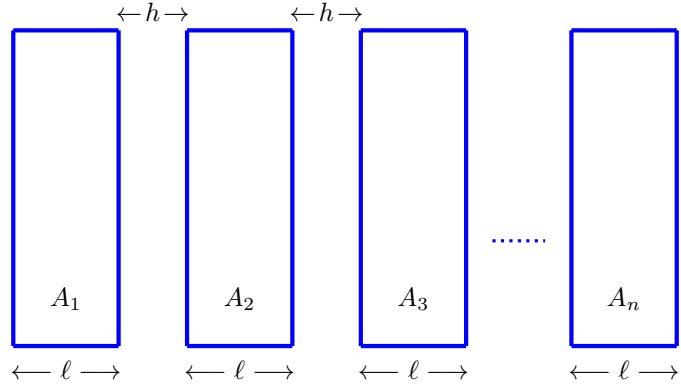
\begin{figure}[h]
\label{tripartiteregions}
\begin{center}
\begin{tikzpicture}[scale=.7]
\draw[ultra thick,blue] (-0.5,0) -- (-0.5,6);
\draw[ultra thick,blue] (-0.5,0) -- (1.5,0);
\draw[ultra thick,blue] (1.5,0) -- (1.5,6);
\draw[ultra thick,blue] (-0.5,6) -- (1.5,6);

\draw[ultra thick,blue] (2.8,0) -- (2.8,6);
\draw[ultra thick,blue] (2.8,0) -- (4.8,0);
\draw[ultra thick,blue] (4.8,0) -- (4.8,6);
\draw[ultra thick,blue] (2.8,6) -- (4.8,6);

\draw[ultra thick,blue] (6.1,0) -- (6.1,6);
\draw[ultra thick,blue] (6.1,0) -- (8.1,0);
\draw[ultra thick,blue] (8.1,0) -- (8.1,6);
\draw[ultra thick,blue] (6.1,6) -- (8.1,6);

\draw[dotted,blue,very thick] (8.6,2) -- (9.6,2);

\draw[ultra thick,blue] (10.1,0) -- (10.1,6);
\draw[ultra thick,blue] (10.1,0) -- (12.1,0);
\draw[ultra thick,blue] (12.1,0) -- (12.1,6);
\draw[ultra thick,blue] (10.1,6) -- (12.1,6);

\draw[] (0.5,0.5) node[above] {$A_1$}; 
\draw[] (3.8,0.5) node[above] {$A_2$}; 
\draw[] (7.1,0.5) node[above] {$A_3$}; 
\draw[] (11.1,0.5) node[above] {$A_n$}; 
\draw[] (0.5,-0.1) node[below] {$\longleftarrow \ell  \longrightarrow$}; 
\draw[] (3.8,-0.1) node[below] {$\longleftarrow \ell  \longrightarrow$}; 
\draw[] (7.1,-0.1) node[below] {$\longleftarrow \ell  \longrightarrow$}; 
\draw[] (11.1,-0.1) node[below] {$\longleftarrow \ell  \longrightarrow$}; 
\draw[] (2.15,6) node[above] {$\leftarrow \! h \! \rightarrow$}; 
\draw[] (5.45,6) node[above] {$\leftarrow \! h \! \rightarrow$}; 
\end{tikzpicture}
\caption{$n$ disjoint entangling regions $A_i, i=1,\cdots, n$  for computing $n$-partite information.}
\end{center}
\end{figure}

Following our discussions in the introduction we shall define the $n$-partite information as 
follows \cite{Hayden:2011ag}
\be\label{np1}
I^{[n]}(A_{\{i\}})=\sum_{i=1}^nS(A_i)-\sum_{i<j}^n S(A_i\cup A_j)+\sum_{i<j<k}^n S(A_i\cup A_j\cup A_k)
-\cdots\cdots -(-1)^n S(A_1\cup A_2\cup\cdots\cup A_n).
\ee
The main subtlety in evaluating  the above quantity is the computation of entanglement entropy of
union of subsystem. As  we have already mentioned in the previous section in 
order to compute the holographic mutual information there are two possibilities to get minimal 
surface in the bulk associated to the entanglement entropy of the union $S(A\cup B)$. In the present case
where we are dealing with parallel strips with $h\ll \ell$, taking the  minimal surface leads to
\bea
S({A_i\cup A_{i+j}})=\Bigg\{ \begin{array}{rcl}
&S(2\ell+h)+S(h)&\,\,\,j=1,\\
&2S(\ell)&\,\,\,j>1,
\end{array}\, .
\eea
Similarly for the union of three regions one uses
\bea
S({A_i\cup A_{i+j}\cup A_{i+j+k}})=\Bigg\{ \begin{array}{rcl}
&S(3\ell+2h)+2S(h)&\,\,\,j=1,k=1\\
&S(2\ell+h)+S(\ell)+S(h)&\,\,\,j=1,k>1,\; {\rm or}\; j>1, k=1,\\
&3S(\ell)&\,\,\,j>1,k>1
\end{array}\, ,
\eea
and  more generally for arbitrary integer numbers $k, m$ and  $j>1$  one has
\bea
&&S(A_i\cup A_{i+1} \cdots \cup A_{i+k}\cup A_{i+k+j}\cup A_{i+k+j+1}\cdots
\cup A_{i+k+j+m})\\ \nonumber &&\;\;\;\;\;\;\;\;\;\;\;\;\;\;\;\;\;\;\;\;\;\;\;\;\;\;\;=
S(A_i\cup A_{i+1}\cdots \cup A_{i+k})+S( A_{j}\cup A_{j+1} \cdots
\cup A_{j+m})\\ \nonumber
 &&\;\;\;\;\;\;\;\;\;\;\;\;\;\;\;\;\;\;\;\;\;\;\;\;\;\;\;
=S(k\ell+(k-1)h)+(k-1)S(h)+S(m\ell+(m-1)h)+(m-1)S(h).
\eea 
By making use of these expressions, the equation \eqref{np1}  evaluated for the system depicted
in Fig. 6 may be simplified significantly as follows
\bea\label{np}
I^{[n]}(A_{\{i\}})&=&(-1)^{n}\bigg[2S\bigg((n-1)\ell+(n-2)h\bigg)-S\bigg(n\ell+(n-1)h\bigg)-S\bigg((n-2)\ell+(n-3)h\bigg)\bigg]\cr
&\equiv& (-1)^n {\tilde I}^{[n]}.
\eea
Interestingly enough,  one observes that among various co-dimension two hypersurfaces only three of them corresponding to $n\ell+(n-1)h,\;(n-1)\ell+(n-2)h$ and $(n-2)\ell+(n-3)h$ contribute to the $n$-partite information\footnote{If one replaces $n-3$ by $|n-3|$, this equation reduces to the mutual information for 
$n=2$.}. Therefore in order to compute the $n$-partite information one should essentially redo the same computations we have done for the  mutual information. In what follows using the AdS/CFT correspondence we will compute ${\tilde I}^{[n]}$ which as we will see it is always positive. In other words the 
holographic $n$-partite information has definite sign: for even $n$ it is positive and for odd $n$ it is
negative.

Let us consider the holographic $n$-partite information for the vacuum state of a CFT whose gravity 
dual is given by an AdS background. Indeed from the equation \eqref{SV} one finds
\bea\label{Jvac}
{\tilde I}^{[n]}_{\text{vac}}=\frac{L^{d-2}c_0}{4G_N}\left(-\frac{2}{((n-1)\ell+(n-2)h)^{d-2}}
+\frac{1}{(n\ell+(n-1)h)^{d-2}}+\frac{1}{((n-2)\ell+(n-3)h)^{d-2}}\right).
\eea
Using a numerical calculation one can show that for fixed $\frac{h}{\ell}$ the above quantity for all values of
$d$ and $n$ is positive and approaches zero in large $\ell$ limit.

For a thermal case whose gravity dual is provided by an AdS black brane geometry, 
and in the limit of 
$\ell\ll \rho_H$, utilizing the equation  \eqref{SBH0} one arrives at 
\bea\label{JBH}
{\tilde I}^{[n]}_{\text{BH}}={\tilde I}^{[n]}_{\text{vac}}-\frac{L^{d-2}}{2G_N}c_1\frac{(\ell+h)^2}{\rho_H^d}.
\eea
On the other hand, by making use of the equation \eqref{DE} one finds
\bea\label{jFirstM}
\frac{\Delta{\tilde I}^{[n]}}{\Delta E}=\frac{8\pi c_1}{d-1}\; \ell \left(1+\frac{h}{\ell}\right)^2,
\eea
 This relation shows that by increasing the temperature, the $n$-partite information is increased (decreased) for $n$ even (odd).

On the other hand for the case of $\rho_H\ll \ell$ since all entangling regions appearing in the 
definition of  $n$-partite information \eqref{np} contains  a factor of $\ell$, the corresponding 
entanglement entropy should be approximated by the equation \eqref{SBH1}. We note, however, that 
since 
\be
-2[(n-1)\ell+(n-2)h]+[(n-2)\ell+(n-3)h]+ [n\ell+(n-1)h]=0,
\ee
the resultant  $n$-partite information vanishes.

%%%%%%%%%%%%%%%%%%%%%%%%%%%444444444444444444444%%%%%%%%%%%%%%%%%%%%%%%%%%%

\section{Time evolution of $n$-partite information}

In this section we would like to study the scaling behavior of  $n$-partite information during the process of thermalization. This is indeed a generalization of the mutual 
information studied in the section three. Again, the thermalization process we are considering  is holographically 
modelled by the  AdS-Vaidya metric \eqref{Vaidya}. Therefore one, essentially, needs
to study different scaling behaviors of three entanglement entropies appearing in the
$n$-partite information \eqref{np} in the AdS-Vaidya metric. To do so, we will utilize
the results reviewed in the appendix A.2.

In general for the system we are considering there are four time scales given by the radius of the
horizon $\rho_H$ and three entangling regions appearing in equation \eqref{np} which are
 $(n-2)\ell+(n-3)h,\; (n-1)\ell+(n-2)h$ and $n\ell+(n-1)h$.
Since we are interested in $h\ll \ell$ situation, one may recognize
four possibilities for the order of these scales as follows
\bea
2\rho_H\ll {(n-2)}\ell+{(n-3)}h<{(n-1)}\ell+{(n-2)}h<{n}\ell+{(n-1)}h,\nonumber\\
{(n-2)}\ell+{(n-3)}h<2\rho_H<{(n-1)}\ell+{(n-2)}h<{n}\ell+{(n-1)}h,\nonumber\\
{(n-2)}\ell+{(n-3)}h<{(n-1)}\ell+{(n-2)}h<2\rho_H<{n}\ell+{(n-1)}h,\nonumber\\
{(n-2)}\ell+{(n-3)}h<{(n-1)}\ell+{(n-2)}h<{n}\ell+{(n-1)}h<2\rho_H,
\eea
which could be studied separately. We note, however, that since the situation is very 
similar to the mutual information, one would expect to get the same qualitative behaviors
for the $n$-partite information. In what follows we will explore the first case listed above in more detail and
will briefly present the results of other cases.

\subsection{First case: $2\rho_H\ll {(n-2)}\ell+{(n-3)}h$}
%%%%%%%%%%%%%%%%%%%%%%%%%%%%%%%%%%%%%%%%%%%%%%%%%%%%%%%%%%%%%%

In this case since all entangling regions involved in the computations of $n$-partite information
are larger than the radius of horizon, the corresponding co-dimension two hypersurfaces in the 
bulk would have a chance to penetrate the horizon.  Indeed in this case there are five time intervals
in which the $n$-partite information behaves differently. We will study these intervals  separately. It is worth noting 
that before the thermalization process, the system is in the vacuum state and therefore the corresponding
$n$-partite information is given by the equation \eqref{Jvac}.

%%%%%%%%%%%%%%111111111%%%%%%%%%%%%%%%%%%%%%%%%%%%%%%
\subsubsection{Early time: $t\ll \rho_H$}

At the early times all co-dimension  two hypersurfaces cross the null shell almost at the same point 
that is
very close to the boundary. Therefore all entanglement entropies appearing in 
the $n$-partite information
\eqref{np} should be approximated by the equation \eqref{SBHET}. Thus at the early times one finds
\bea\label{jearly}
{\tilde I}^{[n]}={\tilde I}^{[n]}_{\rm vac}+{\cal O}(t^{2d}),
\eea
showing that  the $n$-partite information starts from $I^{[n]}_{\rm vac}$ in the vacuum and remains fixed
up to order of  ${\cal O}(t^{2d})$.

%%%%%%%%%%%%%%%%1111111111%%%%%%%%%%%%%%%%%%%%%%
\subsubsection{ Steady behavior: $\rho_H\ll t\ll \frac{(n-2)}{2}\ell+\frac{(n-3)}{2}h$}

The system reaches a local equilibrium at $t\sim \rho_H$ after which it does not
produce thermal entropy, though the entanglement entropies associated with the entangling regions
appearing in the $n$-partite information still increase with time. Actually since all the entangling regions
are larger than the radius of horizon,  the corresponding  entanglement entropies grow
linearly with time (see \eqref{SBHL}). Therefore one finds
\bea
{\tilde I}^{[n]}={\tilde I}^{[n]}_{\rm vac} +\frac{L^{d-2}}{4G_N}\left(2\frac{\sqrt{-\tilde{f}(\rho_{2\;m}})}{\rho_{2\;m}^{d-1}}-\frac{\sqrt{-\tilde{f}(\rho_{1\; m}})}{\rho_{1\; m}^{d-1}}-\frac{\sqrt{-\tilde{f}(\rho_{3\; m}})}{\rho_{3\; m}^{d-1}}\right)t+\cdots\, .
\eea
Here $\tilde{f}(\rho)$ and $\rho_m$ are defined in appendix A.2, and $\rho_{i\;m}$ for $i=1,2,3$ are 
associated with entangling regions $(n-2)\ell+(n-3)h,\; (n-1)\ell+(n-2)h$ and $n\ell+(n-1)h$,
respectively. Since  we are dealing with the large entangling regions,  the turning points of all hypersurfaces
are large, and therefore from  \eqref{largesimp1} one can deduce  $\rho_{i\;m}=\rho_m=
(2(d-1)/(d-2))^{1/d} \rho_H$. As a result, the second term in the above equation vanishes leading 
to a constant $n$-partite  information in this time interval too.

%%%%%%%%%%%%%%%%%%1111111111%%%%%%%%%%%%%%%%%%%%

\subsubsection{Linear growth : $\frac{(n-2)}{2}\ell+\frac{(n-3)}{2}h< t < \frac{(n-1)}{2}\ell+\frac{(n-2)}{2}h$}

In this time interval the entanglement entropy associated with the entangling 
region $(n-2)\ell+(n-3)h$
is saturated to its equilibrium value given by  \eqref{SBH1}, though 
the others are still growing linearly with time as  \eqref{SBHL}. Plugging these 
results into the equation \eqref{np} one arrives at
\be
{\tilde I}^{[n]}={\tilde I}^{[n]}_{\rm vac}+\frac{L^{d-2}}{4G_N}\left(\frac{c_2}{\rho_H^{d-2}}-\frac{c_0}{((n-2)\ell+(n-3)h)^{d-2}}\right)+
\frac{L^{d-2}}{4G_N\rho_H^{d-1}}
\left(v_E t-\frac{(n-2)\ell+(n-3)h}{2}\right),
\ee
where $v_E=\frac{\sqrt{d/(d-2)}}{(\frac{2(d-1)}{(d-2)})^{(d-1)/d}}$. Note also that in order to find the above 
expression we have used the large entangling region limit by which
 $\rho_{i m}=\rho_m$ with $\rho_m$ is given by  \eqref{largesimp1}. 

From the above equation it is clear that in this time interval ${\tilde I}^{[n]}$ is 
bigger than its value in the vacuum ${\tilde I}^{[n]}_{\rm vac}$, and grows linearly with time.
Actually the linear growth  continues until the entanglement entropy $S((n-1)\ell+(n-2)h)$ saturates to its equilibrium value at which ${\tilde I}^{[n]}$ reaches its maximum at 
\be\label{tmin}
v_E \,\,t_{\rm max}^{[n](1)}\sim \frac{(n-1)\ell+(n-2)h}{2}-c_2\rho_H+c_0\frac{\rho_H^{d-1}}{((n-1)\ell+(n-2)h)^{d-2}},
\ee
with the maximum value given by
\be\label{Jmin11}
{\tilde I}_{\rm max}^{[n]\;(1)}\approx {\tilde I}^{[n]}_{\rm vac}+\frac{L^{d-2}}{4G_N}\left[
\frac{\ell+h}{2\rho_H^{d-1}}-\frac{c_0}{((n-2)\ell+(n-3)h)^{d-2}}+\frac{c_0}{((n-1)\ell+(n-2)h)^{d-2}}\right].
\ee

It is important to note that since ${\tilde I}^{[n]}$ is positive the actual sign of the $n$-partite information 
is given by the prefactor $(-1)^n$ in the equation \eqref{np}. Therefore for odd $n$ the $n$-partite 
information has, actually, a minimum, though for even $n$ it has a maximum.

%%%%%%%%%%%%%%%%%%1111111111111%%%%%%%%%%%%%%%%%%%%%%%%%%%
\subsubsection{ Linear decreasing : $\frac{(n-1)}{2}\ell+\frac{(n-2)}{2}h< t < \frac{n}{2}\ell+\frac{(n-1)}{2}h$\label{sub6.1.4}}

In this case the first two entanglement entropies associated with the 
entangling regions  $(n-2)\ell+(n-3)h$ and $(n-1)\ell+(n-2)h$  are saturated while the last one 
is still increasing linearly with time. Therefore from equations \eqref{SBH1},  \eqref{SBHL} and 
\eqref{np} one finds
 \be\label{JSD}
{\tilde I}^{[n]}\approx {\tilde I}_{\rm max}^{[n](1)}+\frac{L^{d-2}}{4G_N}\left(\frac{c_0}{((n-1)\ell+(n-2)h)^{d-2}}-\frac{c_2}{\rho_H^{d-2}}\right)
+\frac{L^{d-2}}{4G_N\rho_H^{d-1}}\left(\frac{n-1}{2}\ell+\frac{n-2}{2}h-v_E t\right).
\ee

%%%%%%%%%%%%%%%%%%%%%1111111111%%%%%%%%%%%%%%%%%%%%%%%%%%%%%%%%%%%%
\subsubsection{ Saturation}\label{section6.1}

In the present case where $\rho_H\ll \ell$, as we have seen in the previous section all 
entanglement entropies saturate to their equilibrium values and the $n$-partite information 
becomes zero. Then assuming to have linear decreasing all the way to the saturation, one 
may estimate the saturation time by setting the equation \eqref{JSD} to zero,
\be\label{Jsat1}
{\tilde I}^{[n](1)}_{\text{sat}}\approx \frac{L^{d-2}}{4G_N}\left[\frac{c_0}{(n\ell+(n-1)h)^{d-2}}-\frac{c_2}{\rho_H^{d-2}}+\frac{n\ell+(n-1)h}{2\rho_H^{d-1}}
-v_E\frac{t_s}{\rho_H^{d-1}}\right]=0,
\ee
which can be solved for saturation time
\bea\label{Jtsat}
v_E t^{[n](1)}_s\approx \frac{n}{2}\ell+\frac{n-1}{2}h-c_2\rho_H+\frac{c_0\rho_H^{d-1}}{(n\ell+(n-1)h)^{d-2}}\approx \frac{n}{2}\ell+\frac{n-1}{2}h-c_2\rho_H.
\eea
It is worth noting that the $n$-partite information saturates before $\frac{n}{2}\ell+\frac{n-1}{2}h-c_2\rho_H+c_0\frac{\rho_H^{d-1}}{(\frac{n}{2}\ell+\frac{n-1}{2}h)^{d-2}}$
which is essentially the time when entanglement entropy $S(n\ell+(h-1) h)$ saturates to its
equilibrium value.  

As a result, we found that in the case of
$\rho_H\ll l_i$, the quantity ${\tilde I}^{[n]}$ starts from its value in 
the vacuum and remains almost constant up to $t\sim \frac{n-2}{2}\ell+\frac{n-3}{2}h$, then it 
grows linearly  with time till it reaches its maximum value at $t_{\text{max}}^{[n](1)}$. After that 
it decreases linearly with time till it becomes zero at the saturation time given by 
$t_s^{[n](1)}\sim \frac{n}{2}\ell+\frac{n-1}{2}h-c_2\rho_H$. One observes that 
the quantity ${\tilde I}^{[n]}$ has the same behavior 
as the mutual information, though scaling behaviors occur at different time scales. 
Note that the to find the actual value of the  $n$-partite information,  the factor of $(-1)^n$ should also be taken into account. Therefore although the behavior should be
the same, the $n$-partite information is either negative (for odd $n$) or positive (for even $n$). To illustrate
the situation we have summarized the results in Fig.7 for tripartite information (Note that in this case because $n$ is odd we have $t_{\text{min}}, I_{\text{min}}$ instead of $t_{\text{max}}, I_{\text{max}}$).
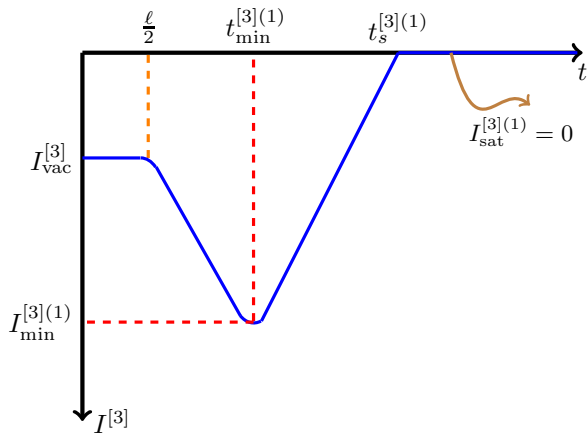
\begin{figure}[h]
\begin{center}
\begin{tikzpicture}[scale=.7]
\draw[->,ultra thick] (0,0) -- (0,-7) node[right] {$I^{[3]}$}; 
\draw[->,ultra thick] (0,0) -- (9.5,0) node[below] {$t$};
\draw[dashed,orange,very thick] (1.25,-2) -- (1.25,0);
\draw[blue,very thick] (0,-2) -- (1.1,-2);
\draw[blue,very thick] (1.39,-2.175) arc (45:90:.4cm and 0.6cm);
\draw[blue,very thick] (1.4,-2.17) -- (3,-5);
\draw[blue,very thick] (3.39,-5.10) arc (-70:-130:.4cm and 0.6cm);
\draw[dashed,red,very thick] (3.25,-5.1) -- (3.25,0);
\draw[dashed,red,very thick] (3.25,-5.12) -- (0,-5.12);
\draw[blue,very thick] (3.4,-5.1) -- (6,0);
\draw[blue,very thick] (6,0) -- (9.4,0);
\draw[->,brown,very thick] (7,0) .. controls (7.5,-2) and (7.9,-0.5) .. (8.5,-1);
\draw[] (0,-2) node[left] {$I^{[3]}_{\text{vac}}$}; 
\draw[] (0,-5.12) node[left] {$I^{[3](1)}_{\text{min}}$}; 
\draw[] (9.5,-1.5) node[left] {\small$I^{[3](1)}_{\text{sat}}=0$}; 
\draw[] (1.26,0) node[above] {$\frac{\ell}{2}$}; 
\draw[] (3.30,0) node[above] {$t_{\text{min}}^{[3](1)}$}; 
\draw[] (6.0,0) node[above] {$t_s^{[3](1)}$}; 
\end{tikzpicture}
\caption{Schematic behavior of tripartite information during the thermalization process for the case $\rho_H\ll \frac{\ell}{2}$. Here $I^{[3]}_{\rm vac}$,  $I^{[3](1)}_{\rm min}$ and $I^{[3](1)}_{\rm sat}$,
up to a minus sign, 
are given by equations \eqref{Jvac}, \eqref{Jmin11} and \eqref{Jsat1}, respectively.}
\end{center}\label{fig7}
\end{figure}

%%%%%%%%%%%%%%%%%%%%%%%%%%%%%%%%%%%%%%%%%%%%%%%%%%%%%%%

\subsection{Other cases}

Since for the model we are considering the $n$-partite information (or more precisely the quantity
${\tilde I}^{[n]}$) has the same structure as the mutual information ( three entanglement entropies
have to be computed), the behavior of ${\tilde I}^{[n]}$ should be the same as that of 
the mutual information. Indeed we have explicitly shown this in the previous subsection 
for the case where all entangling regions are bigger than radius of the horizon. 
Having reached to this conclusion in what follows we just briefly present the results of other 
cases.

\subsubsection{Second case: $\frac{(n-2)}{2}\ell+\frac{(n-3)}{2}h<\rho_H<\frac{(n-1)}{2}\ell+\frac{(n-2)}{2}h<\frac{n}{2}\ell+\frac{(n-1)}{2}h$}

In this case ${\tilde I}^{[n]}$ starts from its value and remains constant at the  early times till 
the first entanglement entropy associated with the entangling region $(n-2)\ell+(n-3) h$ saturates.
Then one gets a quadratic growth as follows
\bea\label{jquadratic}
{\tilde I}^{[n]}\approx {\tilde I}^{[n]}_{\rm vac} +\frac{L^{d-2}}{4G_N\rho_H^d}\left(\frac{t^2}{4}-c_1\big((n-2)\ell+(n-3)h\big)^2\right),
\eea
in the time interval $\frac{(n-2)}{2}\ell+\frac{(n-3)}{2}h < t<\rho_H$. When system reaches 
a local equilibrium one gets linear growth. Indeed for time interval $\rho_H < t < \frac{(n-1)}{2}\ell+\frac{(n-2)}{2}h$ one has linear growth, while  for
 $\frac{(n-1)}{2}\ell+\frac{(n-2)}{2}h<t<\frac{n}{2}\ell+\frac{(n-1)}{2}h$ one gets linear decreasing
with time. Therefore it has a maximum value given by  
\bea\label{jmin2}
{\tilde I}^{[n](2)}_{\rm max}\approx {\tilde I}^{[n]}_{\rm vac}+\frac{L^{d-2}}{4G_N\rho_H^{d-1}} \left(\frac{(n-1)\ell+(n-2)h}{2}-c_2 \rho_H+\frac{c_0\rho_H^{d-1}}{((n-1)\ell+(n-2)h)^{d-2}}
-\frac{c_1((n-2)\ell+(n-3)h)^2}{\rho_H}\right).
\eea
Finally it saturates at
\be
v_E\, t^{[n](2)}_s\approx \frac{c_0 \rho_H^{d-1}}{(n\ell+(n-1)h)^{d-2}}+\frac{n}{2}\ell+\frac{n-1}{2}h-c_2\rho_H \approx \frac{n}{2}\ell+\frac{n-1}{2}h-c_2\rho_H,
\ee
to its equilibrium value given by 
\bea\label{jsat2}
{\tilde I}^{[n](2)}_{\text{sat}}={\tilde I}^{[n]}_{\text{vac}}&+&\frac{L^{d-2}}{4G_N\rho_H^{d-1}}\Big(\frac{(n-2)\ell+(n-3)h}{2}-c_2\rho_{H}-\frac{c_1((n-2)\ell+(n-3)h)^2}{\rho_{H}}\nonumber\\
&-&\frac{c_0\rho_H^{d-1}}{(n\ell+(n-1)h)^{d-2}}+\frac{2c_0\rho_H^{d-1}}{((n-1)\ell+(n-2)h)^{d-2}}\Big)
\eea
showing that  ${\tilde I}^{[n]}_{\text{vac}}>{\tilde I}^{[n](2)}_{\text{sat}}>0$. We have depicted the qualitative behavior of tripartite information in this case in Fig.8.
\begin{figure}[h]
\begin{center}
\begin{tikzpicture}[scale=.7]
\draw[->,ultra thick] (0,0) -- (0,-7) node[right] {$I^{[3]}$}; 
\draw[->,ultra thick] (0,0) -- (9.7,0) node[below] {$t$};
\draw[dashed,orange,very thick] (1.1,-2) -- (1.1,0);
\draw[blue,very thick] (0,-2) -- (1,-2);
\draw[blue,very thick] (1,-2) parabola (2,-4);
\draw[blue,very thick] (2,-4) .. controls (2.1,-4.3) and (3.8,-5.85) .. (4,-5.8);
\draw[dashed,purple,very thick] (2,-4.05) -- (2,0);
\draw[dashed,red,very thick] (4,-5.8) -- (4,0);
\draw[dashed,red,very thick] (4,-5.8) -- (0,-5.8);
\draw[blue,very thick] (4,-5.8) .. controls (4.1,-5.8) and (5,-5.1) .. (7.7,-1);
\draw[blue,very thick] (7.7,-1) .. controls (7.8,-1) and (8,-1) .. (9.5,-1);
\draw[dashed,violet,very thick] (0,-1) -- (7.7,-1);
\draw[dashed,violet,very thick] (7.7,0) -- (7.7,-1);
%\draw[step=1cm,gray,very thin] (0,0) grid (9,9);
\draw[] (0,-2) node[left] {$I^{[3]}_{\text{vac}}$}; 
\draw[] (0,-5.8) node[left] {$I^{[3](2)}_{\text{min}}$}; 
\draw[] (0,-1) node[left] {$I^{[3](2)}_{\text{sat}}$}; 
\draw[] (1.1,0) node[above] {$\frac{\ell}{2}$}; 
\draw[] (2,0.1) node[above] {\small{$\rho_H$}}; 
\draw[] (4,0) node[above] {$t_{\text{min}}^{[3](2)}$}; 
\draw[] (7.7,0) node[above] {$t_{s}^{[3](2)}$}; 
\end{tikzpicture}
\caption{Schematic behavior of tripartite information during the thermalization process for 
$\frac{\ell}{2}< \rho_H< \ell+\frac{h}{2}$. Here $I^{[n]}_{\rm vac}$,  $I^{[3](2)}_{\rm min}$ and $I^{[3](2)}_{\rm sat}$, up to a minus sign, 
are given by equations \eqref{Jvac}, \eqref{jmin2} and \eqref{jsat2}, respectively.}

\end{center}
\end{figure}
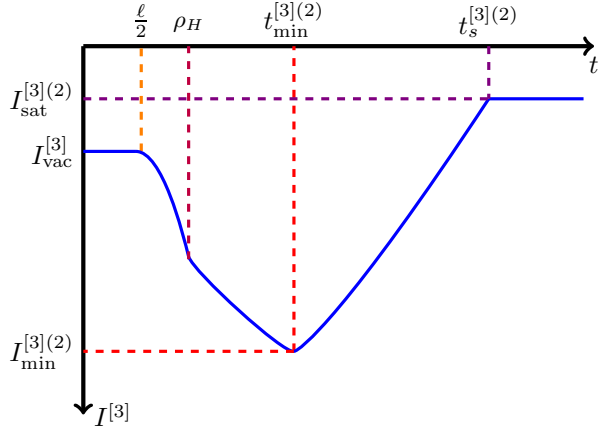

%%%%%%%%%%%%%%%%%%%%%%%%%%%%%%%%%%%%%%%%%%%%%%%%%%%%%%%%%%%

%%%%%%%%%%%%%%%%%%%%%%%%%%%%%%%%%%%%%%%%%%%%%%%%%%%%%%%
\subsubsection{Third case: $\frac{(n-1)}{2}\ell+\frac{(n-2)}{2}h<\rho_H<\frac{n}{2}\ell+\frac{(n-1)}{2}h$}

In this case the situation is exactly the same as the third case of the mutual information. Namely the 
quantity ${\tilde I}^{[n]}$ starts from its value in vacuum and remains fixed at the early times. Then it 
grows quadratically with time till reaches a maximum after that it decreases quadratically and then 
linearly with time up to the saturation point. The maximum and saturation  values are given by
\bea\label{jmin3}
{\tilde I}^{[n](3)}_{\rm max}\approx {\tilde I}^{[n]}_{\rm vac}+\frac{L^{d-2}c_1}{4G_N\rho_H^d}\left((n-1)\ell+(n-2)h)^2-((n-2)\ell+(n-3)h)^2\right),
\eea
and 
\bea\label{jsatcase3}
{\tilde I}_{\text{sat}}^{[n](3)}={\tilde I}^{[n]}_{\text{vac}}+\frac{L^{d-2}}{4G_N\rho_H^{d-1}}&\bigg(&c_2\rho_{H}-\frac{n\ell+(n-1)h}{2}+\frac{2c_1((n-1)\ell+(n-2)h)^2}{\rho_{H}}-\frac{c_1((n-2)\ell+(n-3)h)^2}{\rho_{H}}\cr
&-&\frac{c_0\rho_H^{d-1}}{(n\ell+(n-1)h)^{d-2}}\bigg).
\eea
The corresponding saturation time is 
\be
v_E \,t_s^{[n](3)}\approx  \frac{n}{2}\ell+\frac{n-1}{2}h+\frac{c_0 \rho_H^{d-1}}{(n\ell+(n-1)h)^{d-2}}-c_2 \rho_H\approx \frac{n}{2}\ell+\frac{n-1}{2}h-c_2 \rho_H.
\ee
The situation for tripartite information is depicted in Fig.9.
\begin{figure}\label{fig:44}[h]
\begin{center}
\begin{tikzpicture}[scale=.7]
\draw[->,ultra thick] (0,0) -- (0,-7) node[right] {$I^{[3]}$}; 
\draw[->,ultra thick] (0,0) -- (10,0) node[below] {$t$};
\draw[blue,very thick] (0,-1) -- (1,-1);
\draw[blue,very thick] (1,-1) parabola (2,-4);
\draw[blue,very thick] (2,-4) .. controls (2.4,-6) and (3.9,-6.1) .. (5,-4);
%\draw[blue,very thick] (2,-4) .. controls (2.4,-6) .. (5,-4);
\draw[blue,very thick] (5,-4) .. controls (6,-3.25) and (5.05,-4.05) .. (8.4,-0.8);
\draw[blue,very thick] (8.4,-0.8) .. controls (8.4,-0.76) .. (9,-0.77);
\draw[dashed,orange,very thick] (1,-1) -- (1,0);
\draw[dashed,red,very thick] (3.25,-5.58) -- (3.25,0);
\draw[dashed,red,very thick] (3.25,-5.58) -- (0,-5.58);
\draw[dashed,purple,very thick] (5,-4) -- (5,0);
\draw[dashed,violet,very thick] (8.4,-0.8) -- (8.4,0);
\draw[dashed,violet,very thick] (0,-0.8) -- (8.4,-0.8);
\draw[] (0,-1.5) node[left] {$I^{[3]}_{\text{vac}}$}; 
\draw[] (0,-0.6) node[left] {$I^{[3](3)}_{\text{sat}}$}; 
\draw[] (0,-5.7) node[left] {$I^{[3](3)}_{\text{min}}$}; 
\draw[] (5,0) node[above] {$\rho_{H}$}; 
\draw[] (1,0) node[above] {$\frac{\ell}{2}$}; 
\draw[] (3.25,0) node[above] {$t_{\text{min}}^{[3](3)}$}; 
\draw[] (8.28,0) node[above] {$t_s^{[3](3)}$}; 
%\draw[blue,very thick] (5,-2) -- (7,-2);%\draw[step=1cm,gray,very thin] (0,0) grid (11,-11);
\end{tikzpicture}
\caption{Schematic behavior of tripartite information during the thermalization process for 
$\ell+\frac{h}{2}< \rho_H<\frac{3\ell}{2}+h$. Here $I^{[3]}_{\rm vac}$,  $I^{[3](3)}_{\rm min}$ and 
$I^{[3](3)}_{\rm sat}$, up to a minus sign, 
are given by equations \eqref{Jvac}, \eqref{jmin3} and \eqref{jsatcase3}, respectively.}
\end{center}
\end{figure}
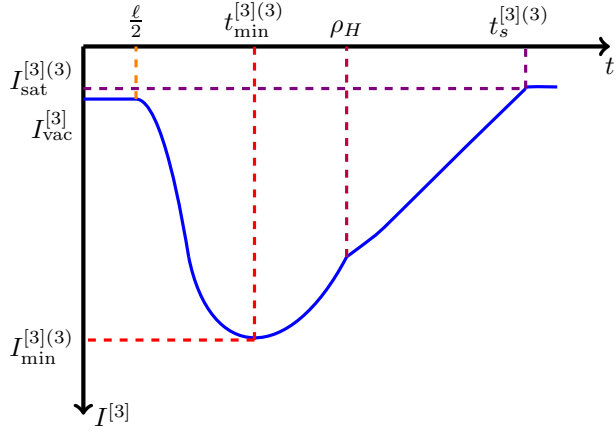
%%%%%%%%%%%%%%%%%%%%%%%%%%%%%%%%%%%%%%%%%%%%%%%%%%%%

%%%%%%%%%%%%%%%%%%%%%%%%%%%%%%%%%%%%%%%%%%%%%%%%%%%%%%%
\subsubsection{Fourth case: $\frac{n}{2}\ell+\frac{n-1}{2}h\ll \rho_H$}

In this case which all  entangling regions involving in the computation of the $n$-partite information
\eqref{np} are smaller than the radius of the horizon, the corresponding entanglement entropies saturate to their equilibrium values before the system reaches a local 
equilibrium. Therefore during the process of thermalization the $n$-partite information does not
exhibit linear growth. Indeed ${\tilde I}^{[n]}$ starts from its value at the vacuum and remains fixed at the early time. 
Then it grows quadratically with time and then decreases quadratically till it reaches its equilibrium value.
The maximum occurs at 
\bea
\frac{t^{[n](4)}_{\text{max}}}{4\sqrt{c_1}} \sim \frac{(n-1)\ell+(n-2)h}{2},
\eea
with the value of 
\bea\label{jmin4}
{\tilde I}^{[n](4)}_{\rm max}\approx {\tilde I}^{[n]}_{\rm vac}+\frac{L^{d-2}c_1}{4G_N\rho_H^d}\left((n-1)\ell+(n-2)h)^2-((n-2)\ell+(n-3)h)^2\right).
\eea
Finally the equilibrium value and the corresponding saturation time are  given by 
\bea\label{jsat4}
{\tilde I}^{[n](4)}_{\text{sat}}={\tilde I}^{[n]}_{\text{vac}}
-\frac{L^{d-2}}{2G_N}\;c_1 \frac{ (\ell+h)^2}{\rho_H^d},\;\;\;\;\;\;\;\;\;
\frac{t^{(4)}_s}{4\sqrt{c_1}}\approx  \frac{n}{2}\ell+\frac{n-1}{2}h.
\eea
The situation is illustrated in Fig.10 for tripartite information.
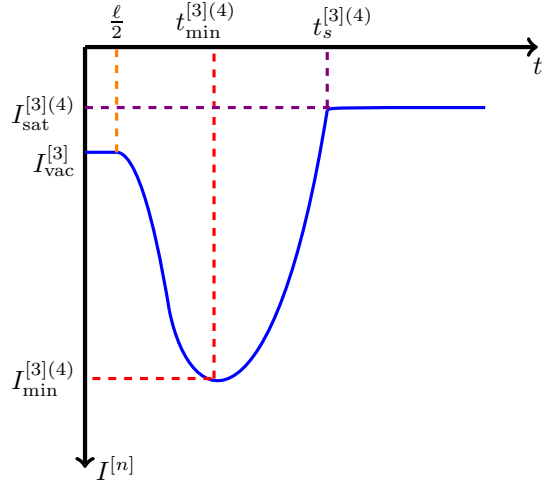
\begin{figure}[h]
\begin{center}
\begin{tikzpicture}[scale=.7]
\draw[->,ultra thick] (0.4,1) -- (0.4,-7) node[right] {$I^{[n]}$}; 
\draw[->,ultra thick] (0.4,1) -- (9,1) node[below] {$t$};
\draw[blue,very thick] (0.4,-1) -- (1,-1);
\draw[blue,very thick] (1,-1) parabola (2,-4);
\draw[blue,very thick] (2,-4) .. controls (2.4,-6) and (4,-6.5) .. (5,-0.2);
\draw[blue,very thick] (5,-0.2) .. controls (5.05,-0.15) .. (8,-0.15);
\draw[dashed,orange,very thick] (1,-1) -- (1,1);
\draw[dashed,red,very thick] (2.85,-5.3) -- (2.85,1);
\draw[dashed,red,very thick] (2.8,-5.3) -- (0.4,-5.3);
\draw[dashed,violet,very thick] (5,-0.15) -- (5,1);
\draw[dashed,violet,very thick] (0.4,-0.15) -- (5.05,-0.15);
\draw[] (0.4,-1.15) node[left] {$I^{[3]}_{\text{vac}}$}; 
\draw[] (0.4,-0.25) node[left] {$I^{[3](4)}_{\text{sat}}$}; 
\draw[] (0.4,-5.3) node[left] {$I^{[3](4)}_{\text{min}}$}; 
\draw[] (1,1) node[above] {$\frac{\ell}{2}$}; 
\draw[] (2.70,1) node[above] {$t_{\text{min}}^{[3](4)}$}; 
\draw[] (5.3,1) node[above] {$t_{s}^{[3](4)}$}; 
%\draw[blue,very thick] (5,2) -- (7,2);
%\draw[step=1cm,gray,very thin] (0,0) grid (11,11);
\end{tikzpicture}
\caption{Schematic behavior of tripartite information during the thermalization process for 
$\ell_i\ll\rho_H$. Here $I^{[3]}_{\rm vac}$, $I^{[3](4)}_{\rm min}$ and $I^{[3](4)}_{\rm sat}$ 
are given by equations \eqref{Jvac}, \eqref{jmin4} and \eqref{jsat4}, respectively.}
\end{center}
\end{figure}

%%%%%%%%%%%%%%%%%%%%%%%%%%%%%%%%%%%%%%%%%%%%%%%%%%%%%%%%%%%%%%%%
\section{Numerical results}

So far following \cite{{Liu:2013iza},{Liu:2013qca}} we have analytically studied the behavior of
$n$-partite information in a process of thermalization with certain assumptions.
In order to examine our results in this section we will study the behavior of $n$-partite information 
numerically. In particular we will mainly focus on the mutual information and 3-partite information
in more details and then briefly study 4-partite and 5-partite information.

It is worth mentioning that although  such a numerical analysis has been already done in {\it e.g.} \cite{{Balasubramanian:2011at},{Allais:2011ys}}, in what follows our main 
interest is to explore  various scaling regimes  we have obtained in the  previous sections.
This could be used to  examine the validity of our assumptions, approximations and results. 

\begin{figure}[h!]
\centering
\includegraphics[scale=0.9]{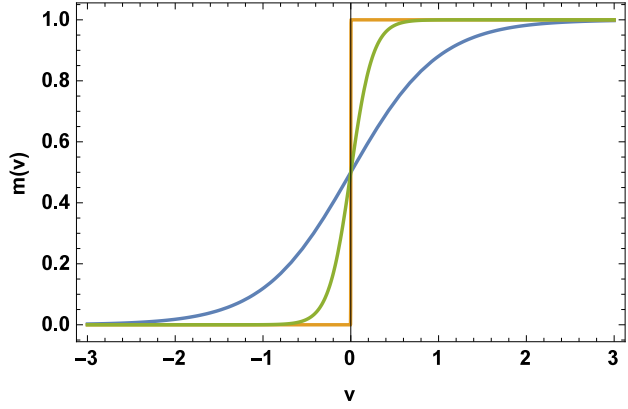}
\caption{The profile of the mass function for $\rho_H=1$ and $a=1\text{(blue)}, 0.25\text{(green)}, 0.001\text{(orange)}$, the latter one is very close to the step function behavior.}
\label{fig:mv}
\end{figure}

To be concrete  we will consider  a 2+1 dimensional boundary theory, however, the result
could be  extended to higher dimensions.  In this case the area of the extremal surface using \eqref{area00} and setting $L=1$ reads
\be\label{area4d}
A=\frac{1}{2}\int_{-\ell/2}^{\ell/2} dx\; \frac{\sqrt{1-2v'\rho'- v'^2 f(\rho,v)}}{\rho^{2}}\equiv \frac{1}{2}\int_{-\ell/2}^{\ell/2} dx\; \frac{\mathcal{L}}{\rho^{2}},
\ee
The minimization condition leads to the following equations of motion
\bea\label{eomfourdim}
\partial_x(\rho'+v'f)&=&\frac{v'^2}{2}\partial_v f\nonumber\\
\rho v''-2\mathcal{L}^2&=&\frac{ v'^2}{2}\rho \partial_{\rho} f,
\eea
which should be solved with the following  boundary conditions  
\begin{eqnarray}\label{bdyfourdim}
\rho(\frac{\ell}{2})&=&0,\;\;\;\;\;\;\;\;v(\frac{\ell}{2})=t,\;\;\;\;\;\;\;\;\nonumber\\
\rho(0)&=&\rho_t,\;\;\;\;\;\;\;v(0)=v_t.
\end{eqnarray} 
In order to study the equations  numerically one should approximate the theta function 
appearing in $f$ with a smooth analytic function. Actually in what follows we will consider the following 
function (see for example \cite{AbajoArrastia:2010yt})
\bea
f(\rho,v)=1-m(v)\rho^d,\;\;\;\;m(v)=\frac{m_0}{2}\left(1+\tanh \frac{v}{a}\right),
\eea
where $m_0$ is a measure of the horizon radius for the final static black-brane geometry, i.e. $\rho_H=m_0^{-1/2}$ and $a$ is the parameter that controls  the thickness of the null shell. In the limit of $a\rightarrow 0$ this profile coincides with the step function, as  illustrated in Fig.\ref{fig:mv}.

\begin{figure}[h]
\centering
\begin{subfigure}
\centering
\includegraphics[scale=.8]{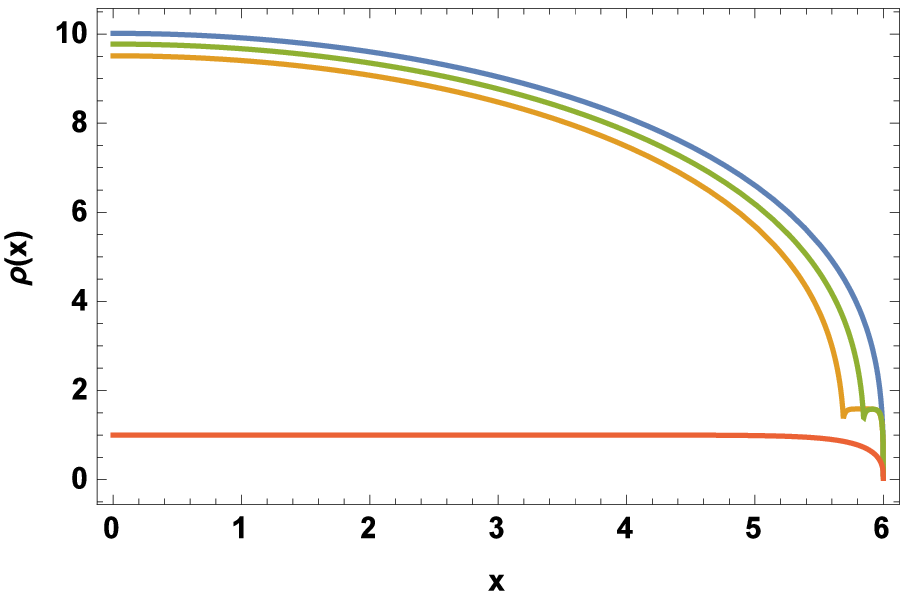}
\end{subfigure}
\begin{subfigure}
\centering
\includegraphics[scale=.8]{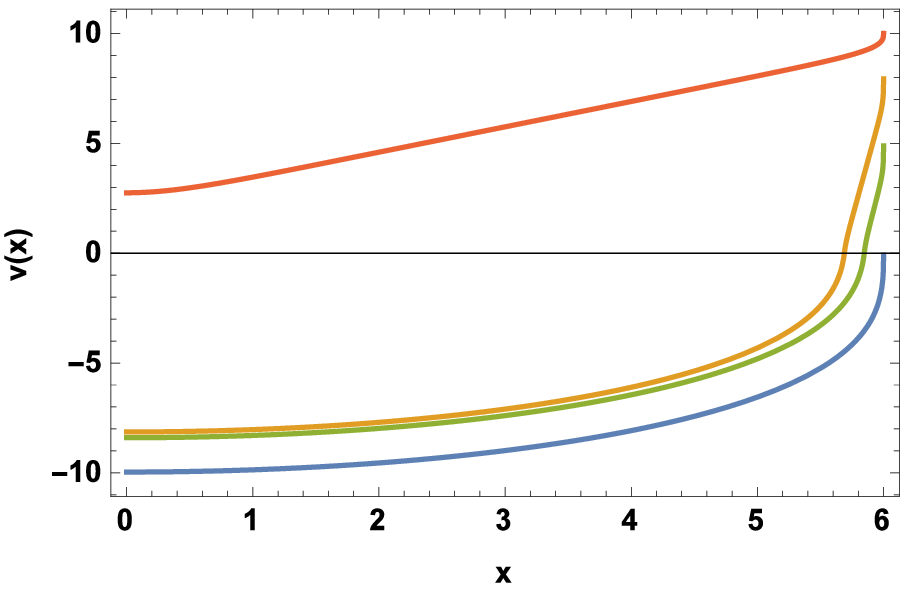}
\end{subfigure}
\caption{The profile of the extremal surface for a strip with $\ell=12$ at thin shell limit $a=0.001$ with $\rho_H=1$ for different boundary times: $t=0\text{(blue)}, 5\text{(green)}, 8\text{(orange)}, 10\text{(red)}$.}
\label{fig:rhoxvx}
\end{figure}

To find the profile of the extremal surface numerically one should solve equations \eqref{eomfourdim} with boundary conditions \eqref{bdyfourdim} using {\it e.g.}  shooting method. For explicit examples
we have plotted different  extremal surfaces for different boundary times in the thin shell limit for a strip entangling region with $\ell=12$ in Fig.\ref{fig:rhoxvx}.

\begin{figure}[h]
\centering
\begin{subfigure}
\centering
\includegraphics[scale=.8]{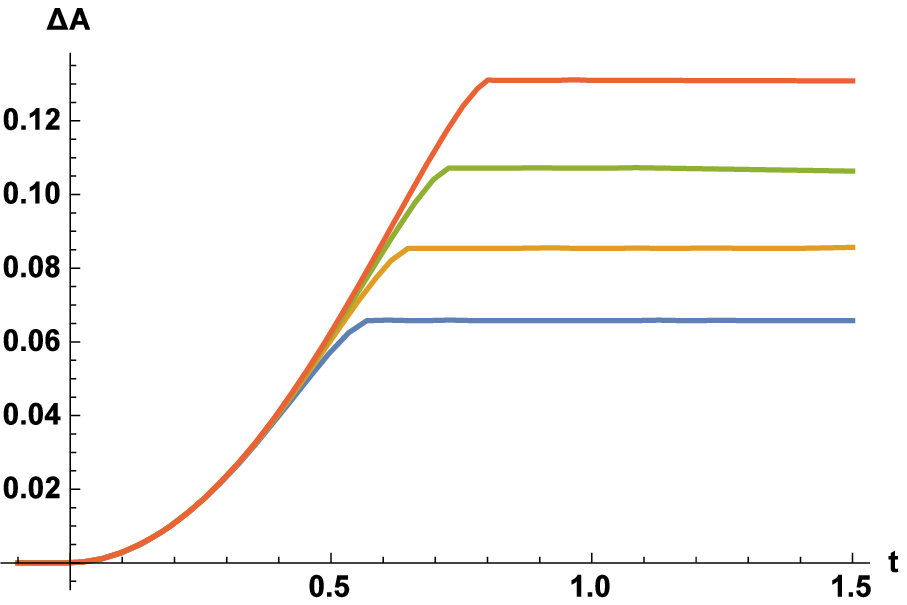}
\end{subfigure}
\begin{subfigure}
\centering
\includegraphics[scale=.8]{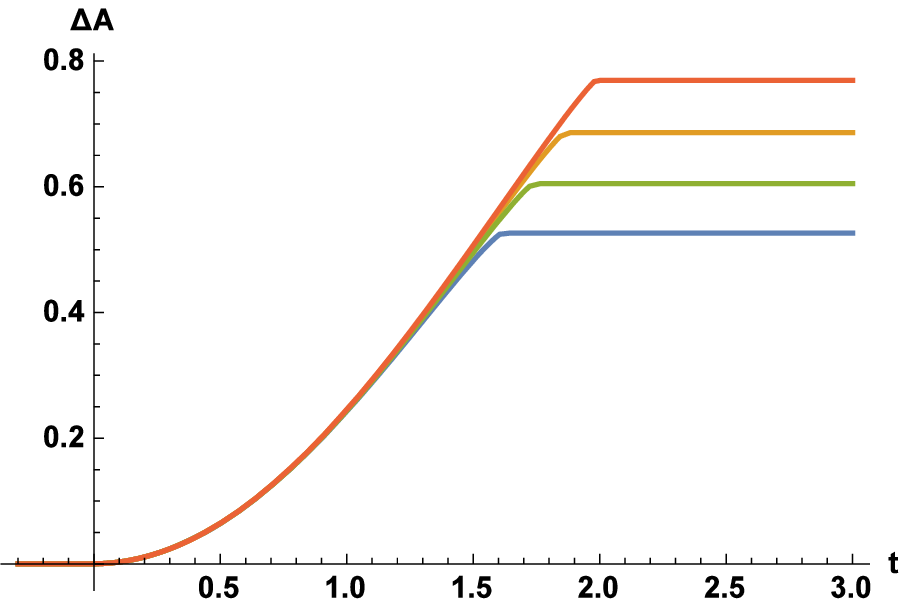}
\end{subfigure}
\caption{Evolution of the regularized area of the minimal surface for $a=0.001$ and $\rho_H=1$. \textit{Left plot}: The small entangling regions  for $\ell= 0.7, 0.8, 0.9, 1$ from bottom to top. \textit{Right plot}: The large entangling regions for $\ell=2.2, 2.4, 2.6, 2.8$ from bottom to top.}
\label{fig:st}
\end{figure}

Having found the extremal surface numerically one could plug the  profile of the extremal surface  into \eqref{area4d} to read the  area of the extremal surface  as a function of boundary time.  
It is, however, important to note that due to the large volume limit, the area  \eqref{area4d} is divergent  and needs to be regularized by introducing a UV cut-off at $\rho=\epsilon$.
 In this case the finite part of the area is given by
\begin{eqnarray}
A_{\text{reg.}}=\int_{0}^{\ell/2-\delta} dx\; \frac{\rho_t^2}{\rho(x)^{4}}-\frac{1}{\epsilon},\;\;\;\;\;\rho(\ell/2-\delta)=\epsilon.
\end{eqnarray} 
Here we have used the conservation law, $\rho^2\mathcal{L}=\rho_t^2$, to simplify the 
final expression. Evolution of the area of extremal surface for a strip in a thin shell limit for the 
large ($\ell >\rho_H$) and small ($\ell<\rho_H$)  entangling regions is depicted in   
Fig.\ref{fig:st}.\footnote{Actually when one considers the large entangling region, one  
must be careful about the swallow tail problem\cite{Albash:2010mv}. We would like to thank P. Fonda for  a discussion on this point.}  Actually in this figure we have plotted $\Delta A$ defined by
\be
\Delta A=A-A_{\rm AdS}=A_{\rm reg.}+\frac{c_0}{\ell}.
\ee

Note that the actual value of the entanglement entropy has an extra factor of $(4G_N)^{-1}$
in front of the area, though in this note we will neglect this factor.
%In the next sections we will study the evolution of the mutual and tripartite information numerically. 

\subsection{Mutual information}

In this section by making use of the numerical results of the holographic entanglement
entropy we will numerically explore  different scaling regimes of the holographic mutual
information. To do so a non-trivial task is how to compute the entanglement entropy of a union
of two subsystems.  As we have already mentioned  there are two minimal surfaces associated with the entanglement entropy $S(A_1\cup A_2)$  (see Fig.14)
\bea\label{SAUB2}
S({A_1\cup A_2})=\Bigg\{ \begin{array}{rcl}
&S(2\ell+h)+S(h)\equiv S_{\text{con.}}(h,\ell)&\,\,\,\,\,h\ll \ell,\\
&2S(\ell)\equiv S_{\text{dis.}}(\ell)&\,\,\,\,\,h\gg \ell,
\end{array}\,\,
\eea
\begin{figure}[h]
\label{0}
\begin{center}
\begin{tikzpicture}[scale=.7]
\draw[ultra thick,black] (0,0) -- (2,0);
\draw[ultra thick,black] (2.5,0) -- (4.5,0);
\draw[ultra thick,blue] (2,0) arc (0:180:1cm);
\draw[ultra thick,blue] (4.5,0) arc (0:180:1cm);
\draw[ultra thick,black] (10,0) -- (12,0);
\draw[ultra thick,black] (12.5,0) -- (14.5,0);
\draw[ultra thick,blue] (14.5,0) arc (0:180:2.25cm);
\draw[ultra thick,blue] (12.5,0) arc (0:180:0.25cm);
%\draw[step=1cm,gray,very thin] (0,-10) grid (25,25);
\draw[] (-1,0.1) node[left] {$S_{\text{dis.}}:$}; 
\draw[] (9,0.1) node[left] {$S_{\text{con.}}:$}; 
\draw[] (1,0) node[below] {$A_1$}; 
\draw[] (3.5,0) node[below] {$A_2$}; 
\draw[] (11,0) node[below] {$A_1$}; 
\draw[] (13.5,0) node[below] {$A_2$}; 
\end{tikzpicture}
\caption{Two different configurations for computing $S(A_1\cup A_2)$. }
\end{center}
\end{figure}
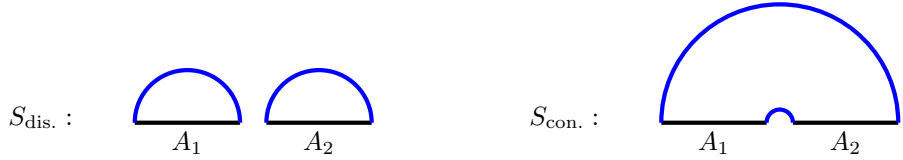
indicating that there is a transition between connected and disconnected configurations as 
one increases  $\frac{h}{\ell}$. Note also that the value of the disconnected configuration is independent of $h$.

Indeed it is easy to find the transition point between these two configurations which can be
done by solving the $S_{\text{con.}}(h,\ell)=S_{\text{dis.}}(\ell)$ for $h$. In particular for a four
dimensional  AdS background in which the corresponding expression 
for the entanglement entropy is given by equation \eqref{SV}  for $d=3$ one finds that the transition occurs at $h=\frac{1}{2} \left(\sqrt{5}-1\right) \ell \sim 0.618\,\ell$. This means that for the vacuum state and
for $h<0.618\,\ell$ one must consider the connected configuration where the  resulting mutual information will be  a finite positive number, though  for $h>0.618\,\ell$, the disconnected 
configuration is favored  and the resulting mutual information is zero.
Since the results we have presented in the previous sections depend crucially on the 
assumption of whether the connected or disconnected configurations are favored in what follows for 
all cases we will compute the evolution of $S(A\cup B)$ too. 

To proceed with the numerical computations we will set $\rho_H=1$ and $a=0.001$.
 Having collected all information and the procedure of our numerical method, let us present 
 our numerical results for the mutual information for all scaling cases we have considered in section 3. 
 
\subsubsection{First case}
\begin{figure}
\centering
\begin{subfigure}
\centering
\includegraphics[scale=.8]{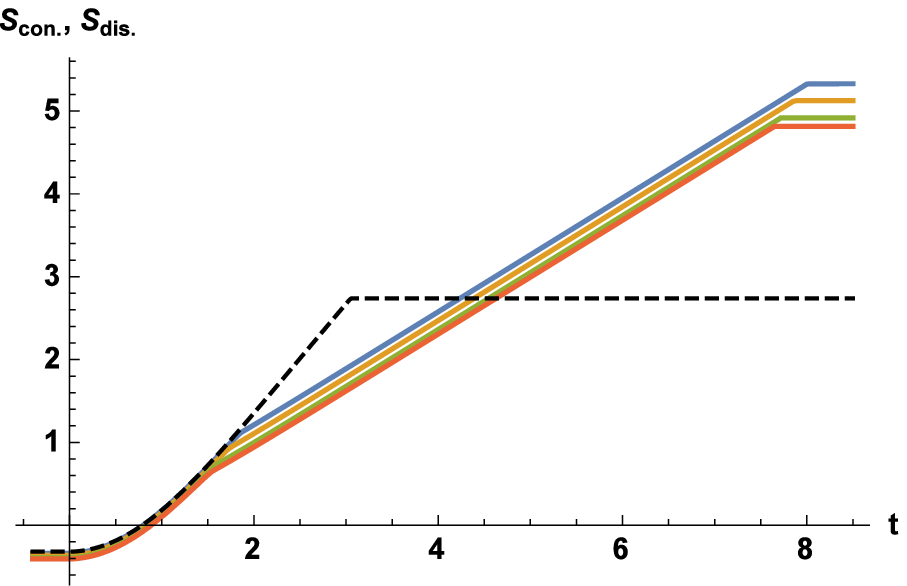}
\end{subfigure}
\begin{subfigure}
\centering
\includegraphics[scale=.8]{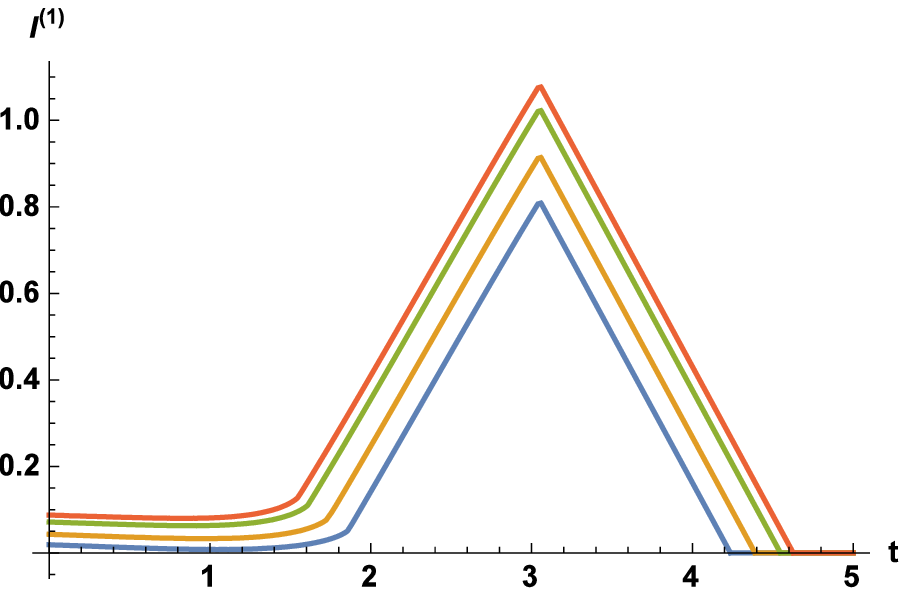}
\end{subfigure}
\caption{\textit{Left plot}: Evolution of entanglement entropy for the connected and disconnected configurations for $\ell=4.5$ and $h=2.1, 2.2, 2.4, 2.6$ from bottom to top. The dashed curve is for disconnected configuration which is independent of $h$. This plot shows that in this range of the parameters the disconnected configuration has the minimal area after the saturation. \textit{Right plot}: Evolution of the mutual information for the same value of fixed $\ell$ and different $h$ now decreasing from bottom to top.}
\label{fig:mutual-1}
\end{figure}

For this case we will fix the width of strips to be $\ell=4.5$ which is larger than the radius of horizon
$\rho_H=1$. Then we will consider different values for $h=2.1,2.2,2.4,2,6$. Note that 
for these cases we have the condition $h<0.618\,\ell$ and therefore the mutual information 
for the vacuum state (AdS geometry) is non-zero. The numerical results are given in Fig.\ref{fig:mutual-1}.
As one observes the numerical results are in good agreement with the analytical 
results we have obtained in the section 3. In particular the left plot in this figure shows that saturation of the mutual information (which takes place at the crossing point of the dashed curve with others) happens long before the saturation of the HEEs. Also according to \eqref{tsat} as we increase the separation between the strips the saturation time decreases.

\begin{figure}[h]
\centering
\begin{subfigure}
\centering
\includegraphics[scale=.8]{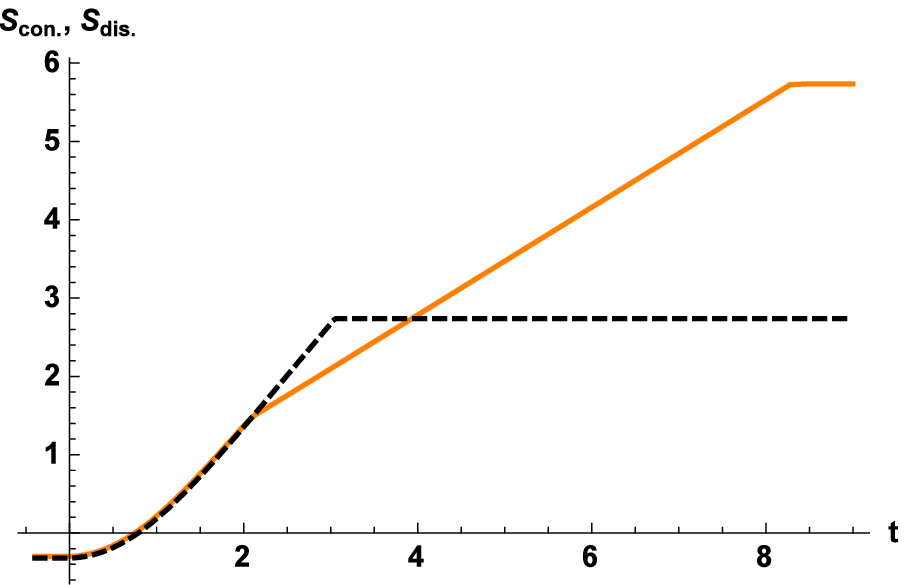}
\end{subfigure}
\begin{subfigure}
\centering
\includegraphics[scale=.8]{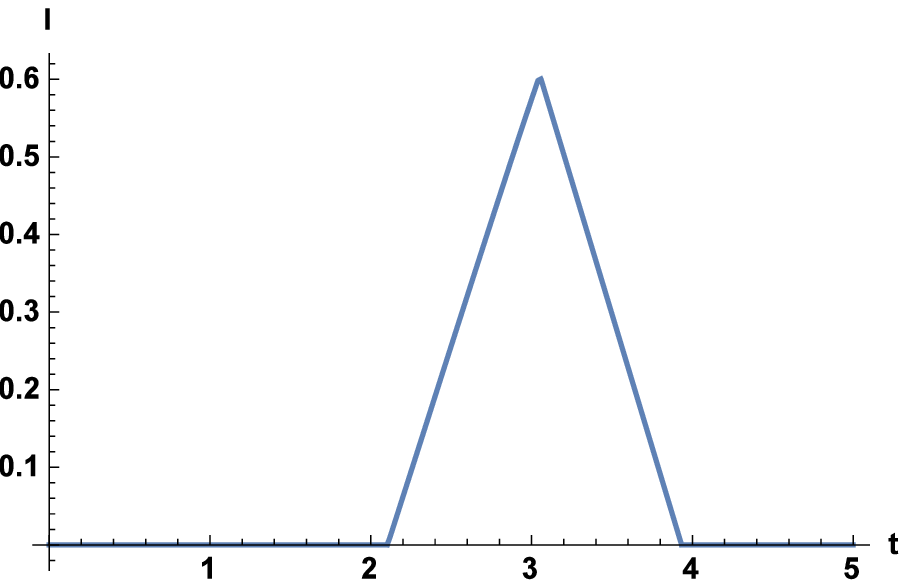}
\end{subfigure}
\caption{\textit{Left plot}: Evolution of entanglement entropy for the connected (orange curve) and disconnected (dashed curve) configurations. At early time the disconnected configuration has the minimal area which leads to a zero initial value for the mutual information. \textit{Right plot}: Evolution of the corresponding mutual information. Here we set $h=3$ and $\ell=4.5$.}
\label{fig:mutual-2}
\end{figure}

On the other hand  for  $h=3$ one has  $h>0.618\,\ell$ so that the mutual information in the vacuum 
state is zero. The corresponding behavior is shown in  Fig.\ref{fig:mutual-2}.

%this case the such that consider
%As we mentioned in section 3.1 we expect that in this case the length of all the entangling regions %are greater than the horizon size and the schematic behavior of the evolution of the mutual %information is similar to Fig.2. 
%The evolution of the mutual information in this case which produced numerically is depicted in Fig.\ref{fig:mutual-1}. Note that in this figure we always consider  to have a finite mutual information at the beginning of the evolution in parallel with our assumptions in section 3.1. Also in this range of the parameters the disconnected configuration has the minimal area after the saturation and the mutual information saturates to zero. On the other hand if we consider $0.618\,\ell<h$, the initial value for the mutual information becomes zero, shows this situation. 

\subsubsection{Second case}
\begin{figure}[h]
\centering
\begin{subfigure}
\centering
\includegraphics[scale=.8]{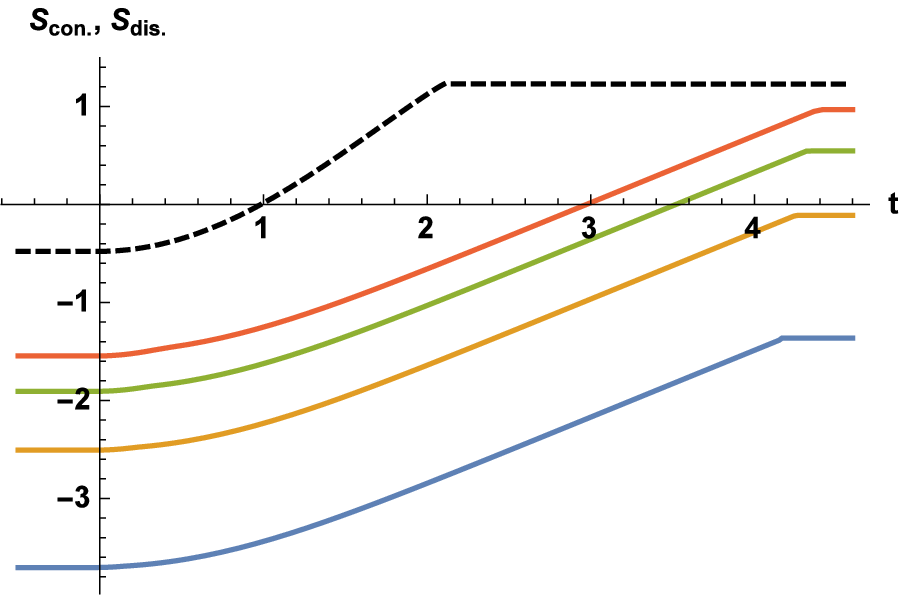}
\end{subfigure}
\begin{subfigure}
\centering
\includegraphics[scale=.8]{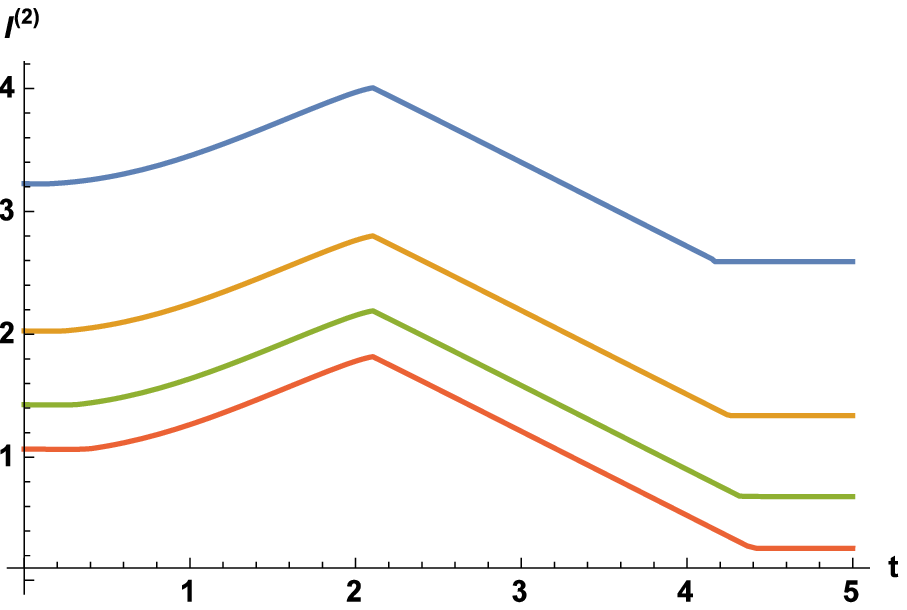}
\end{subfigure}
\caption{\textit{Left plot}: Evolution of entanglement entropy for the connected and disconnected configurations for $\ell=3$ and $h=0.2, 0.3, 0.4, 0.5$ from bottom to top. The dashed curve is for disconnected configuration which is independent of $h$. This plot shows that in this range of the parameters the connected configuration always has the minimal area. \textit{Right plot}: Evolution of the mutual information for the same value of fixed $\ell$ and different $h$ now decreasing from bottom to top.}
\label{fig:mutual-4}
\end{figure}

To study the second case we set $\ell=3$ and consider  $h=0.2, 0.3, 0.4, 0.5$ to make sure that 
the condition $\frac{h}{2}<\rho_H< {\frac{\ell}{2}}$ is satisfied. The numerical results for this case 
are presented in Fig.\ref{fig:mutual-4}. It is worth to mention that in this case, the numerical results indicate that the saturation value is independent of $\ell$ which is in agreement with that corresponding analytical results (see \eqref{satcase2}).

%\begin{figure}
%\centering
%\begin{subfigure}
%\centering
%\includegraphics[scale=.8]{mutual-8}
%\end{subfigure}
%\begin{subfigure}
%\centering
%\includegraphics[scale=.8]{mutual-9}
%\end{subfigure}
%\caption{\textit{Left plot}: Evolution of entanglement entropy for the connected and disconnected configurations for $\ell=2.4$ and $h=0.6, 0.8, 1, 1.1$ from bottom to top. The dashed curve is for disconnected configuration which is independent of $h$. This plot shows that in this range of the parameters the disconnected configuration always has the minimal area after the saturation. \textit{Right plot}: Evolution of the mutual information for the same value of fixed $\ell$ and different $h$ now decreasing from bottom to top. Here we set $a=0.001$ and $\rho_H=1$.}
%\label{fig:mutual-5}
%\end{figure}
\subsubsection{Third case}
\begin{figure}[h]
\centering
\begin{subfigure}
\centering
\includegraphics[scale=.8]{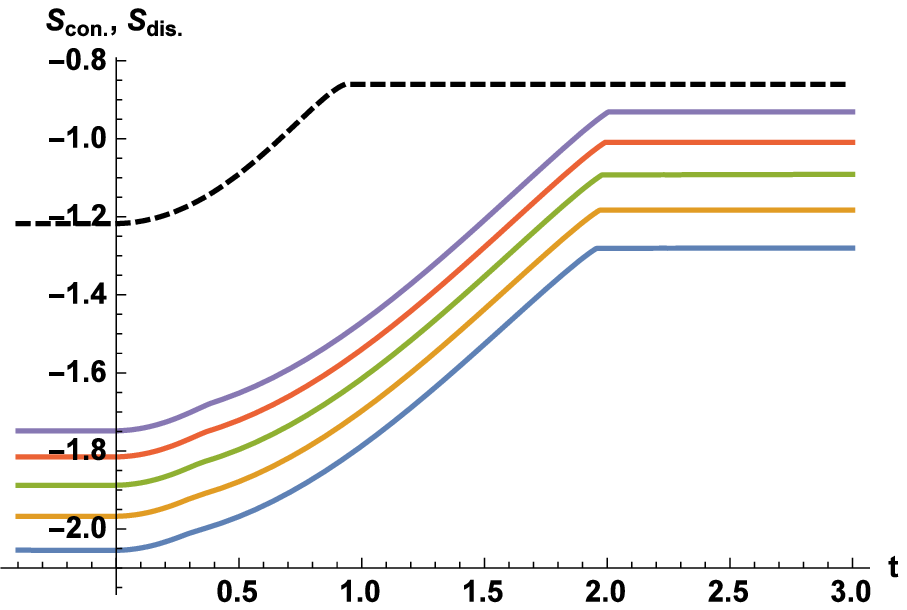}
\end{subfigure}
\begin{subfigure}
\centering
\includegraphics[scale=.8]{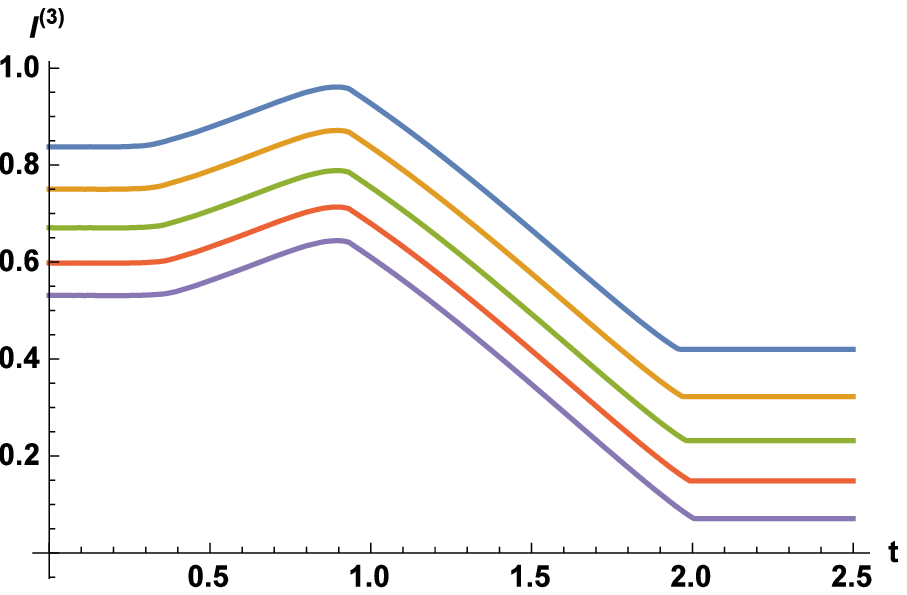}
\end{subfigure}
\caption{\textit{Left plot}: Evolution of entanglement entropy for the connected and disconnected configurations for $\ell=1.18$ and $h=0.4, 0.42, \cdots, 0.48$ from bottom to top. The dashed curve is for disconnected configuration which is independent of $h$. This plot shows that in this range of the parameters the connected configuration always has the minimal area. \textit{Right plot}: Evolution of the mutual information for the same value of fixed $\ell$ and different $h$ now decreasing from bottom to top.}
\label{fig:mutual-6}
\end{figure}

To examine the third case we set $\ell=1.18$ and let $h=0.4, 0.42, \cdots, 0.48$. For these values of 
$\ell$ and $h$ we have plotted the numerical results for the connected and disconnected configurations as well as the  evolution of the mutual information in  Fig.\ref{fig:mutual-6}. Again 
the results are in a good agreement with that discussed in section 3.3.
\subsubsection{Fourth case}
\begin{figure}[h]
\centering
\begin{subfigure}
\centering
\includegraphics[scale=.8]{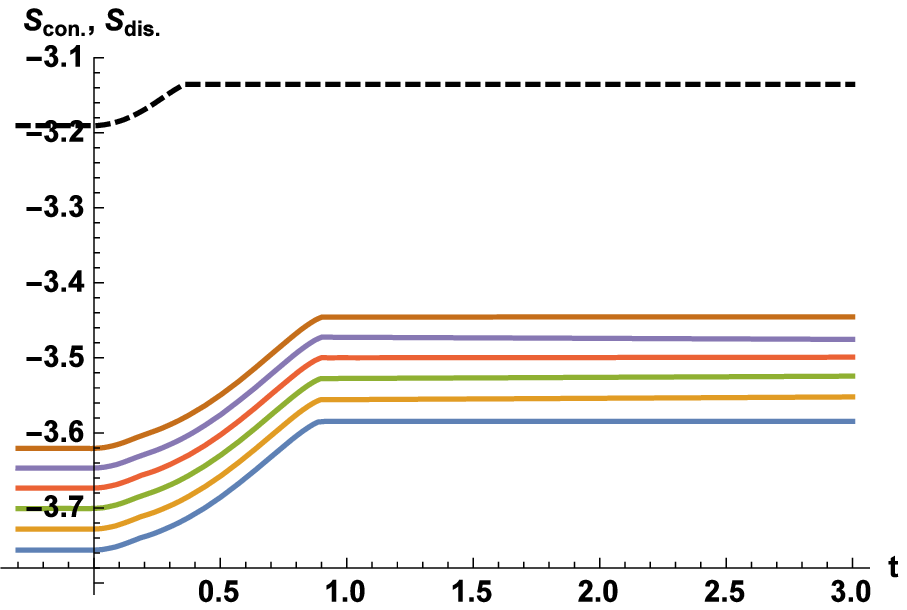}
\end{subfigure}
\begin{subfigure}
\centering
\includegraphics[scale=.8]{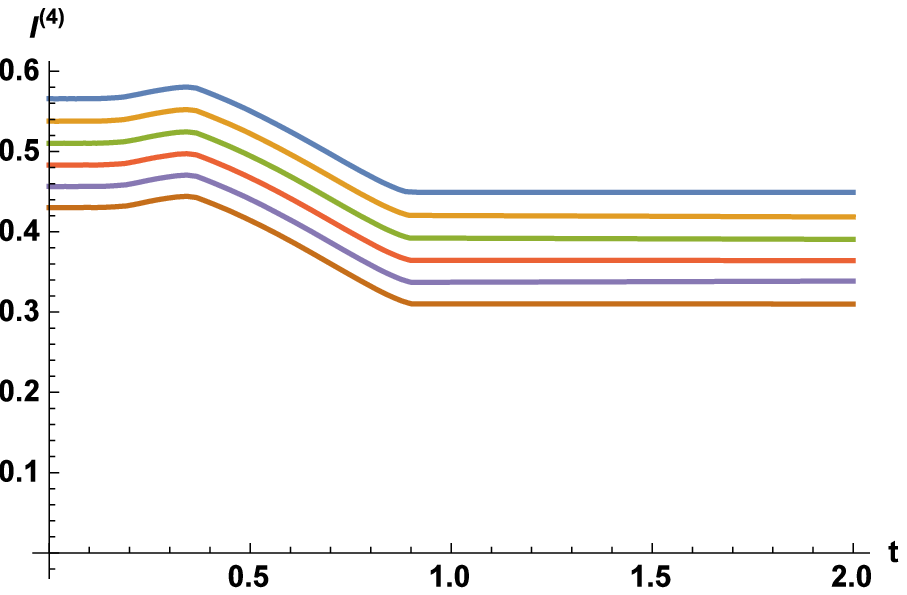}
\end{subfigure}
\caption{\textit{Left plot}: Evolution of entanglement entropy for the connected and disconnected configurations for $\ell=0.45$ and $h=0.23, 0.232,..., 0.24$ from bottom to top. The dashed curve is for disconnected configuration which is independent of $h$. This plot shows that in this range of the parameters the connected configuration always has the minimal area. \textit{Right plot}: Evolution of the mutual information for the same value of fixed $\ell$ and different $h$ now decreasing from bottom to top.}
\label{fig:mutual-7}
\end{figure}

For this case we consider $\ell=0.45$ and $h=0.23, 0.232,..., 0.24$ and the corresponding 
numerical results are depicted in Fig.\ref{fig:mutual-7}. These plots should be compared with 
the Fig.5 in subsection 3.4.

%%%%%%%%%%%%%%%%%%%%%%%
\subsection{3-partite information}
To further examine our analytical results, in this subsection, we will numerically study the 
behavior of 3-partite information during the process of thermalization.
The corresponding system consists of three parallel strips with width $\ell$ separated by 
distances $h$ as drawing in figure 20. 
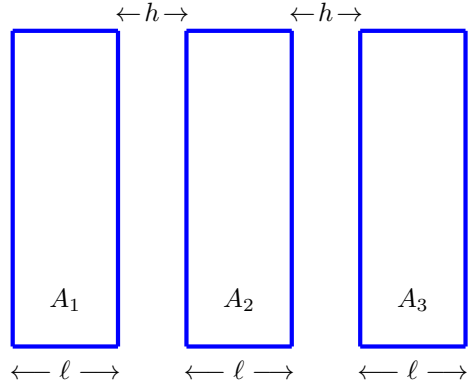
\begin{figure}[h]
\begin{center}
\begin{tikzpicture}[scale=.7]
\draw[ultra thick,blue] (-0.5,0) -- (-0.5,6);
\draw[ultra thick,blue] (-0.5,0) -- (1.5,0);
\draw[ultra thick,blue] (1.5,0) -- (1.5,6);
\draw[ultra thick,blue] (-0.5,6) -- (1.5,6);

\draw[ultra thick,blue] (2.8,0) -- (2.8,6);
\draw[ultra thick,blue] (2.8,0) -- (4.8,0);
\draw[ultra thick,blue] (4.8,0) -- (4.8,6);
\draw[ultra thick,blue] (2.8,6) -- (4.8,6);

\draw[ultra thick,blue] (6.1,0) -- (6.1,6);
\draw[ultra thick,blue] (6.1,0) -- (8.1,0);
\draw[ultra thick,blue] (8.1,0) -- (8.1,6);
\draw[ultra thick,blue] (6.1,6) -- (8.1,6);

\draw[] (0.5,0.5) node[above] {$A_1$}; 
\draw[] (3.8,0.5) node[above] {$A_2$}; 
\draw[] (7.1,0.5) node[above] {$A_3$}; 
\draw[] (0.5,-0.1) node[below] {$\longleftarrow \ell  \longrightarrow$}; 
\draw[] (3.8,-0.1) node[below] {$\longleftarrow \ell  \longrightarrow$}; 
\draw[] (7.1,-0.1) node[below] {$\longleftarrow \ell  \longrightarrow$}; 

\draw[] (2.15,6) node[above] {$\leftarrow \! h \! \rightarrow$}; 
\draw[] (5.45,6) node[above] {$\leftarrow \! h \! \rightarrow$}; 
\end{tikzpicture}
\caption{Three disjoint entangling regions for computing tripartite information.}
\end{center}
\label{fig:3partitef}
\end{figure}

The corresponding 3-partite information is given by
\bea\label{tripartite}
I^{[3]}(A_1,A_2,A_3)=S(A_1)+S(A_2)+S(A_3)-S({A_1\cup A_2})-S({A_1\cup A_3})-S({A_2\cup A_3})+S({A_1\cup A_2}\cup A_3).
\eea
As we have already mentioned in the section 5 the main subtlety in evaluating the 3-partite 
is the way we compute the entanglement entropy of union of subsystems. In order to compute 
these quantities let us review the assumptions which led to a simple expression given in 
equation \eqref{np} for $h\ll \ell$.

Actually for the union of two subsystem one may consider different configurations for the extremal surfaces as depicted in Fig.21\footnote{Note that Fig.21 and Fig.22 are schematic, in $d=3$ the extremal surface for strip entangling region even in the vacuum state is not semicircle. Also for these mixed configurations there exist other suboptimal configurations with respect to these configurations and we do not consider them \cite{Allais:2011ys}.}.
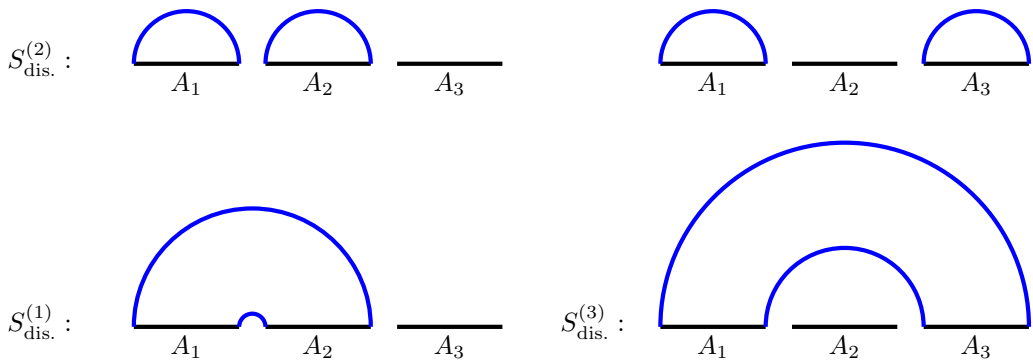
\begin{figure}[h]\label{1}
\begin{center}
\begin{tikzpicture}[scale=.7]
\draw[ultra thick,black] (0,0) -- (2,0);
\draw[ultra thick,black] (2.5,0) -- (4.5,0);
\draw[ultra thick,black] (5,0) -- (7,0);
\draw[ultra thick,blue] (2,0) arc (0:180:1cm);
\draw[ultra thick,blue] (4.5,0) arc (0:180:1cm);
\draw[ultra thick,black] (0,-5) -- (2,-5);
\draw[ultra thick,black] (2.5,-5) -- (4.5,-5);
\draw[ultra thick,black] (5,-5) -- (7,-5);
\draw[ultra thick,blue] (4.5,-5) arc (0:180:2.25cm);
\draw[ultra thick,blue] (2.5,-5) arc (0:180:0.25cm);
%\draw[step=1cm,gray,very thin] (0,-10) grid (25,25);
\draw[] (-1,0.1) node[left] {$S^{(2)}_{\text{dis.}}:$}; 
\draw[] (-1,0.1-5) node[left] {$S^{(1)}_{\text{dis.}}:$}; 
\draw[] (9.5,0.1-5) node[left] {$S^{(3)}_{\text{dis.}}:$}; 
\draw[] (1,0) node[below] {$A_1$}; 
\draw[] (3.5,0) node[below] {$A_2$}; 
\draw[] (6,0) node[below] {$A_3$}; 
\draw[] (11,0) node[below] {$A_1$}; 
\draw[] (13.5,0) node[below] {$A_2$}; 
\draw[] (16,0) node[below] {$A_3$}; 
\draw[ultra thick,black] (10,0) -- (12,0);
\draw[ultra thick,black] (12.5,0) -- (14.5,0);
\draw[ultra thick,black] (15,0) -- (17,0);
\draw[ultra thick,blue] (12,0) arc (0:180:1cm);
\draw[ultra thick,blue] (17,0) arc (0:180:1cm);
\draw[ultra thick,black] (10,-5) -- (12,-5);
\draw[ultra thick,black] (12.5,-5) -- (14.5,-5);
\draw[ultra thick,black] (15,-5) -- (17,-5);
\draw[ultra thick,blue] (17,-5) arc (0:180:3.5cm);
\draw[ultra thick,blue] (15,-5) arc (0:180:1.5cm);
\draw[] (1,-5) node[below] {$A_1$}; 
\draw[] (3.5,-5) node[below] {$A_2$}; 
\draw[] (6,-5) node[below] {$A_3$}; 
\draw[] (11,-5) node[below] {$A_1$}; 
\draw[] (13.5,-5) node[below] {$A_2$}; 
\draw[] (16,-5) node[below] {$A_3$}; 
\end{tikzpicture}
\caption{Schematic  configuration of hypersurfaces for computing $S(A_i\cup A_j)$. In this case two of them have the same contribution to the holographic entanglement entropy $S^{(2)}_{\text{dis.}}$.}
\end{center}
\end{figure}

More precisely one has
\bea
S({A_2\cup A_i})=\Bigg\{ \begin{array}{rcl}
&S(2\ell+h)+S(h)\equiv S^{(1)}_{\text{dis.}}(h,\ell)&\,\,\,\,\,h\ll \ell,\\
&2S(\ell)\equiv S^{(2)}_{\text{dis.}}(\ell)&\,\,\,\,\,h\gg \ell,
\end{array}\,\,\;\;\;\;\;i=1 \;\text{or}\; 3
\eea
and 
\bea
S({A_1\cup A_3})=\Bigg\{ \begin{array}{rcl}
&2S(\ell)\equiv S^{(2)}_{\text{dis.}}(\ell)&\,\,\,\,\,h\ll \ell,\\
&S(3\ell+2h)+S(\ell+2h)\equiv S^{(3)}_{\text{dis.}}(h,\ell)&\,\,\,\,\,h\gg \ell,
\end{array}\,\,\;.
\eea

Similarly one may also study the  union of three subsystems where we could
have different  configurations for the extremal surface as given in Fig.22.
\begin{figure}[h]\label{2}
\begin{center}
\begin{tikzpicture}[scale=.7]
\draw[ultra thick,black] (0,0) -- (2,0);
\draw[ultra thick,black] (2.5,0) -- (4.5,0);
\draw[ultra thick,black] (5,0) -- (7,0);
\draw[ultra thick,blue] (2.5,0) arc (0:180:0.25cm);
\draw[ultra thick,blue] (5,0) arc (0:180:0.25cm);
\draw[ultra thick,blue] (7,0) arc (0:180:3.5cm);
\draw[ultra thick,black] (0,-5) -- (2,-5);
\draw[ultra thick,black] (2.5,-5) -- (4.5,-5);
\draw[ultra thick,black] (5,-5) -- (7,-5);
\draw[ultra thick,blue] (4.5,-5) arc (0:180:2.25cm);
\draw[ultra thick,blue] (2.5,-5) arc (0:180:0.25cm);
\draw[ultra thick,blue] (7,-5) arc (0:180:1cm);
%\draw[step=1cm,gray,very thin] (0,-10) grid (25,25);
\draw[] (-1,0.1) node[left] {$S_{\text{con.}}:$}; 
\draw[] (-1,0.1-5) node[left] {$S^{(4)}_{\text{dis.}}:$}; 
\draw[] (9.5,0.1-5) node[left] {$S^{(6)}_{\text{dis.}}:$}; 
\draw[] (9.5,0.1) node[left] {$S^{(5)}_{\text{dis.}}:$}; 
\draw[] (1,0) node[below] {$A_1$}; 
\draw[] (3.5,0) node[below] {$A_2$}; 
\draw[] (6,0) node[below] {$A_3$}; 
\draw[] (11,0) node[below] {$A_1$}; 
\draw[] (13.5,0) node[below] {$A_2$}; 
\draw[] (16,0) node[below] {$A_3$}; 
\draw[ultra thick,black] (10,0) -- (12,0);
\draw[ultra thick,black] (12.5,0) -- (14.5,0);
\draw[ultra thick,black] (15,0) -- (17,0);
\draw[ultra thick,blue] (12,0) arc (0:180:1cm);
\draw[ultra thick,blue] (14.5,0) arc (0:180:1cm);
\draw[ultra thick,blue] (17,0) arc (0:180:1cm);
\draw[ultra thick,black] (10,-5) -- (12,-5);
\draw[ultra thick,black] (12.5,-5) -- (14.5,-5);
\draw[ultra thick,black] (15,-5) -- (17,-5);
\draw[ultra thick,blue] (17,-5) arc (0:180:3.5cm);
\draw[ultra thick,blue] (14.5,-5) arc (0:180:1cm);
\draw[ultra thick,blue] (15,-5) arc (0:180:1.5cm);
\draw[] (1,-5) node[below] {$A_1$}; 
\draw[] (3.5,-5) node[below] {$A_2$}; 
\draw[] (6,-5) node[below] {$A_3$}; 
\draw[] (11,-5) node[below] {$A_1$}; 
\draw[] (13.5,-5) node[below] {$A_2$}; 
\draw[] (16,-5) node[below] {$A_3$}; 
\end{tikzpicture}
\caption{Four different configurations  for computing $S(A_1\cup A_2\cup A_3)$. }
\end{center}
\end{figure}
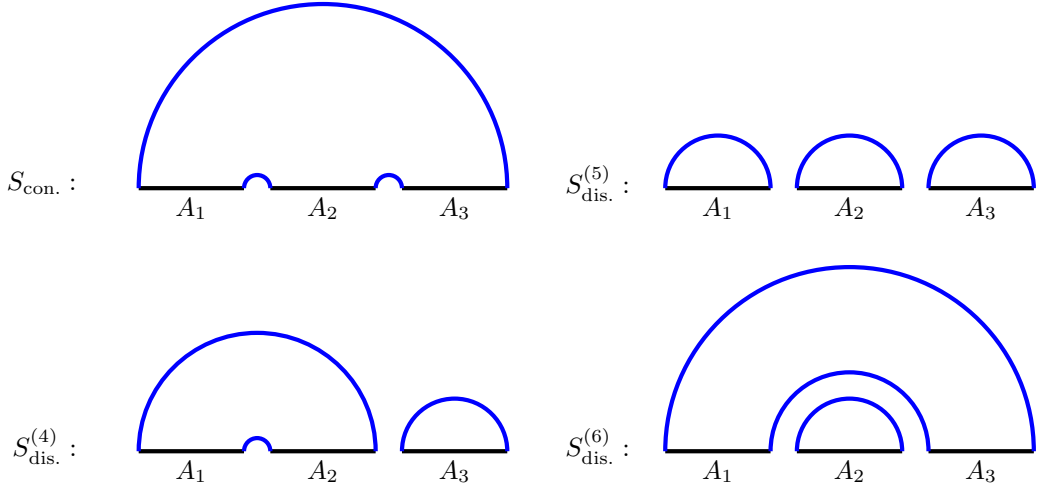

Mathematically these configurations can be translated into the following expressions for the 
union of three strips $A_1, A_2$ and $A_3$
\bea
S({A_1\cup A_2\cup A_3})=\Bigg\{ \begin{array}{rcl}
&S(3\ell+2h)+2S(h)\equiv S_{\text{con.}}(h,\ell)&\,\,\,\,\,h\ll \ell,\\
&S(2\ell+h)+S(\ell)+S(h)\equiv S^{(4)}_{\text{dis.}}(\ell)&\,\,\,\,\,\\
&3S(\ell)\equiv S^{(5)}_{\text{dis.}}(h,\ell)&\,\,\,\,\,h\gg \ell,\\
&S(3\ell+2h)+S(\ell+2h)+S(\ell)\equiv S^{(6)}_{\text{dis.}}(h,\ell)&\,\,\,\,\,
\end{array}\,\,\;.
\eea

\begin{figure}[h]
\centering
\begin{subfigure}
\centering
\includegraphics[scale=.8]{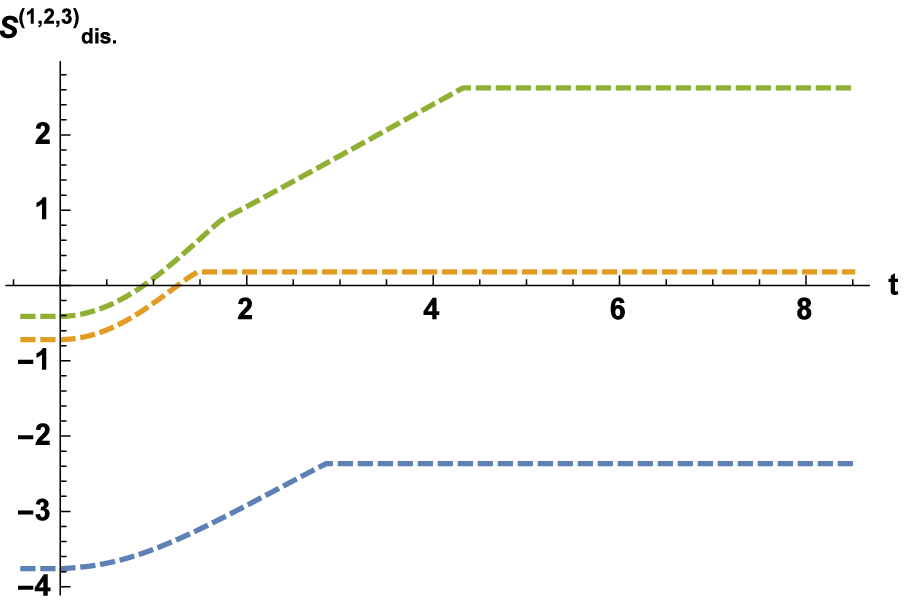}
\end{subfigure}
\begin{subfigure}
\centering
\includegraphics[scale=.8]{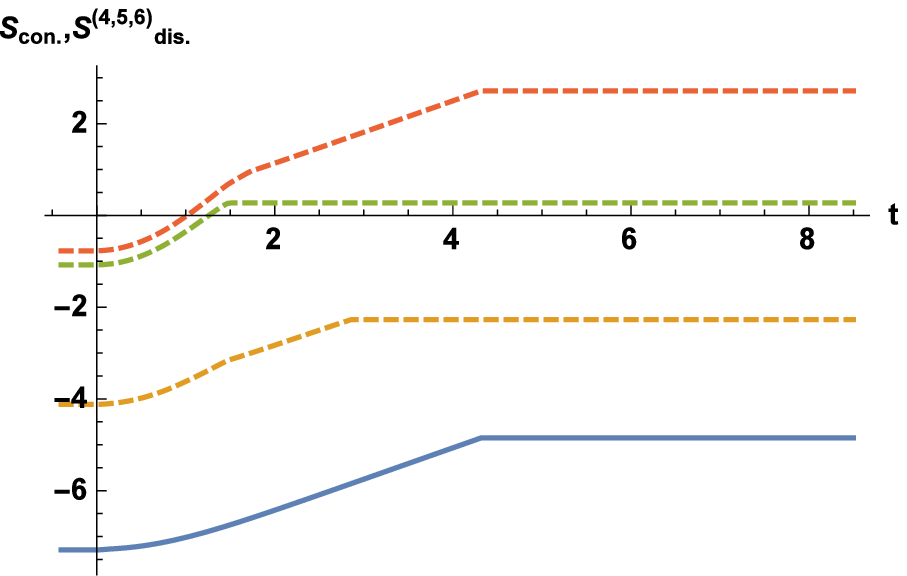}
\end{subfigure}
\caption{Evolution of entanglement entropy for the configurations which correspond to the computation of tripartite information for $h=0.2$ and $\ell=2$. \textit{Left plot}: $S^{(1)}_{\text{dis.}}, S^{(2)}_{\text{dis.}}$ and $S^{(3)}_{\text{dis.}}$ from bottom to top. \textit{Right plot}: $S_{\text{con.}}, S^{(4)}_{\text{dis.}}, S^{(5)}_{\text{dis.}}$ and $S^{(6)}_{\text{dis.}}$ from bottom to top. These plots show that in this range of the parameters the conditions \eqref{assum} satisfied.}
\label{fig:3partite-1}
\end{figure}

\begin{figure}[h]
\centering
\begin{subfigure}
\centering
\includegraphics[scale=.8]{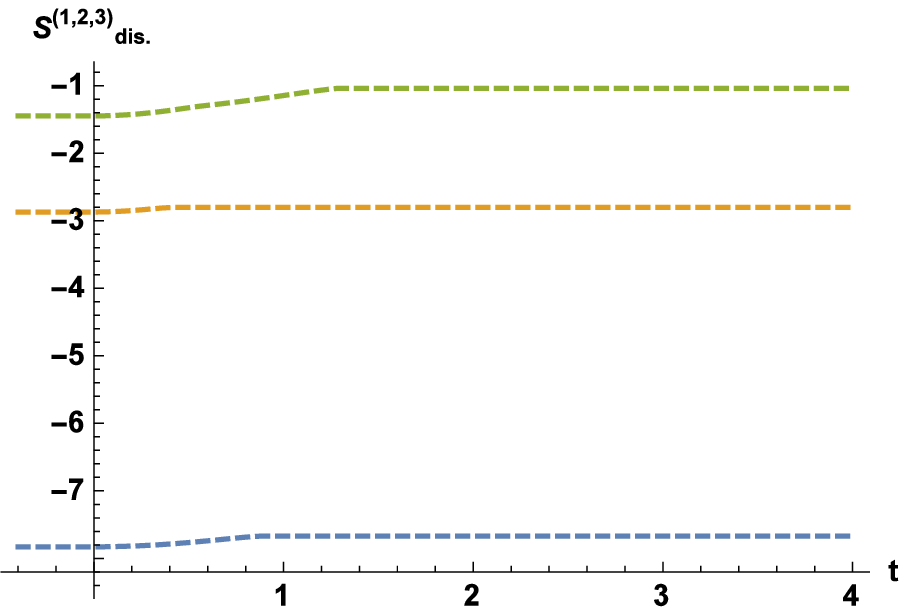}
\end{subfigure}
\begin{subfigure}
\centering
\includegraphics[scale=.8]{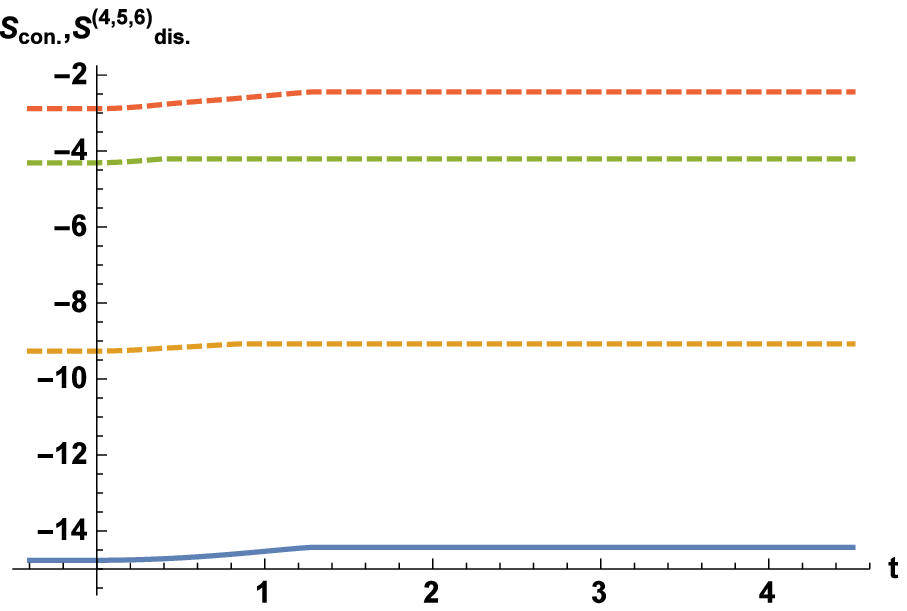}
\end{subfigure}
\caption{Evolution of entanglement entropy for the configurations which correspond to the computation of tripartite information for $h=0.1$ and $\ell=0.5$. \textit{Left plot}: $S^{(1)}_{\text{dis.}}, S^{(2)}_{\text{dis.}}$ and $S^{(3)}_{\text{dis.}}$ from bottom to top. \textit{Right plot}: $S_{\text{con.}}, S^{(4)}_{\text{dis.}}, S^{(5)}_{\text{dis.}}$ and $S^{(6)}_{\text{dis.}}$ from bottom to top. These plots show that in this range of the parameters the conditions \eqref{assum} satisfied.}
\label{fig:3partite-2}
\end{figure}

Putting these results into \eqref{tripartite} and in the limit of  $h\ll \ell$  one arrives at 
\bea\label{limittripartite}
I^{[3]}(\ell,h)=S(3\ell+2h)-2S(2\ell+h)+S(\ell),\;\;\;\;\;\;\;\;h\ll \ell
\eea
which is the same as  \eqref{np} for $n=3$. Indeed to get the above expression for the 3-partite we have assumed that
\bea\label{assum}
S^{(1)}_{\text{dis.}}<S^{(2)}_{\text{dis.}}<S^{(3)}_{\text{dis.}},\;\;\;\;\text{and}\;\;\;\;S_{\text{con.}}<\min\left(S^{(4)}_{\text{dis.}},S^{(5)}_{\text{dis.}},S^{(6)}_{\text{dis.}}\right),\;\;\;\;\;\;\;h\ll \ell. 
\eea
Therefore to proceed with evaluating the 3-partite information, it is crucial to see in what extend 
our assumptions is reliable. Of course, in general, it is not an easy task to prove the above inequalities
during the thermalization process even numerically. Nevertheless we have provided some
numerical examples in figures \ref{fig:3partite-1} and \ref{fig:3partite-2}
showing that in the desired range of parameters these inequalities are indeed hold.

Having explored the subtlety  we are encountering when we are going to compute 3-partite information, in the rest of this subsection we numerically study  different scaling behaviors of 
3-partite  information during the process of thermalization. Actually for all cases which  
we would like study, one should first,   numerically,  check whether the 
conditions \eqref{assum} are satisfied. If the conditions were satisfied, then one can  evaluate 
3-partite information using the expression  \eqref{limittripartite}. Indeed we have done these 
considerations and the results are as follows.

\subsubsection{First case}
\begin{figure}[h]
\centering
\includegraphics[scale=0.9]{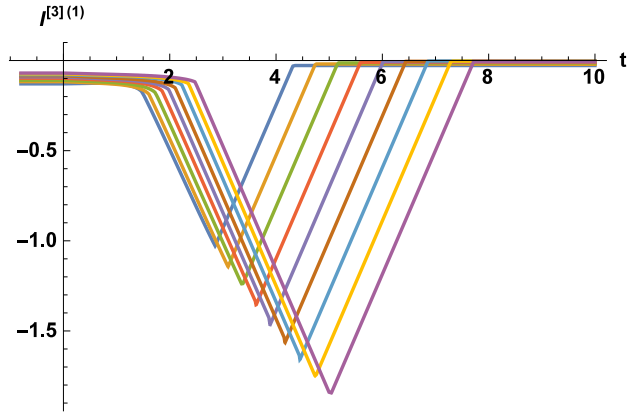}
\caption{Evolution of tripartite information for $h=0.2$ and $\ell=2, 2.2,..., 3.6$ increasing from left to right.}
\label{fig:3partite-3}
\end{figure}

In this case to meet the condition $2 \rho_H<\ell<2\ell+h$,  we will set $h=0.2$ and will
consider different values for the width of strips as $\ell=2,2.2,\cdots, 3.6$. For these values the behavior of the 3-partite information is shown in Fig.\ref{fig:3partite-3}. It is worth mentioning that for each case we have 
numerically checked that the conditions \eqref{assum} are, indeed, satisfied. 

From this figure it is clear that the saturation value of the 3-partite information tends to zero as one increases the size of the entangling regions, which this is in agreement with our analytic results in section 5. Also one can check that the saturation time increases when we increase the strip width as we expect form equation \eqref{Jtsat}.

\subsubsection{Second case}
\begin{figure}[h]
\centering
\includegraphics[scale=0.9]{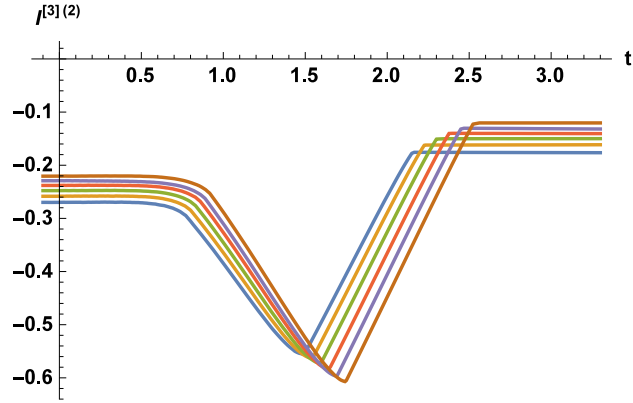}
\caption{Evolution of tripartite information for $h=0.1$ and $\ell=0.96, 1,..., 1.16$ increasing from left to right.}
\label{fig:3partite-4}
\end{figure}

To study this case we will fix $h=0.1$ and consider different values for the width of strips as
$\ell=0.96,1,\cdots,1.16$. It is clear that for these values one has $\ell<2\rho_H<2\ell+h$. For 
these values the  behavior of 3-partite is depicted in Fig.\ref{fig:3partite-4}. Note for all 
results appearing in this figure we have numerically checked the  conditions \eqref{assum} too.

\subsubsection{Third case}
\begin{figure}[h]
\centering
\includegraphics[scale=0.9]{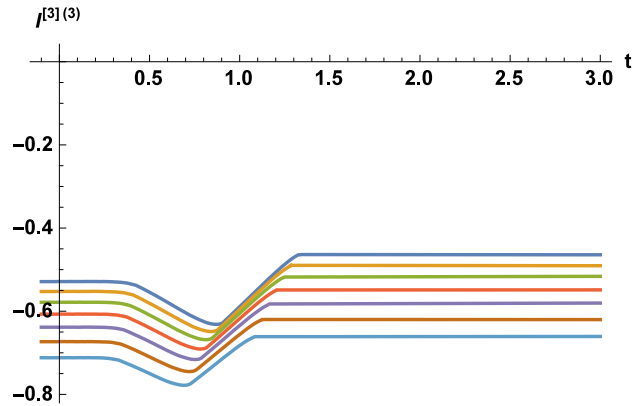}
\caption{Evolution of tripartite information for $h=0.1$ and $\ell=0.4, 0.42,..., 0.52$ increasing from bottom to top.}
\label{fig:3partite-5}
\end{figure}

In this case to maintain the condition $2\ell+h<2\rho_H$ we will set $h=0.1$ and will study 
the behavior of the 3-partite information for different values of the width of strips given by
$\ell=0.4,0.42,\cdots, 0.52$. The corresponding results are given in Fig.\ref{fig:3partite-5}.
Again, the conditions \eqref{assum} are satisfied for our results.
\subsubsection{Fourth case}
\begin{figure}[h]
\centering
\includegraphics[scale=0.9]{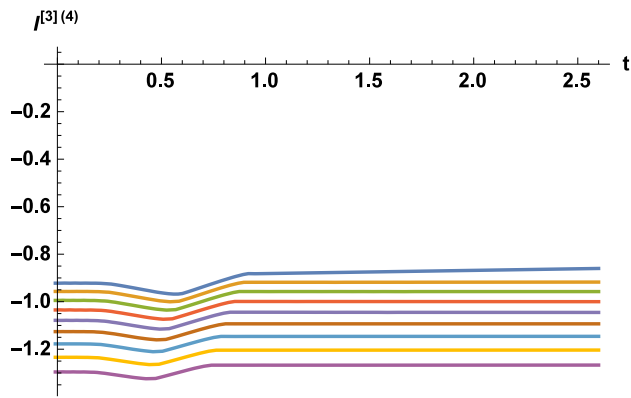}
\caption{Evolution of tripartite information for $h=0.1$ and $\ell=0.24, 0.25,..., 0.32$ increasing from bottom to top.}
\label{fig:3partite-6}
\end{figure}

The last case corresponds to the situation where all entangling regions appearing in the 
expression of 3-partite information are smaller than the radius of horizon. More 
precisely one has $3\ell+2h<2\rho_H$. In order to numerically study 3-partite information
in this region we will set $h=0.1$  and will evaluate the 3-partite for different widths $\ell=
0.24,0.25,\cdots, 0.32$. The results are given in Fig.\ref{fig:3partite-6}. For this case
the conditions \eqref{assum} are also numerically checked. 
%%%%%%%%%%%%%%%%%%%%%%%%%%%%%%%%%%%%%%%%%%%%%%%%%%%%%%%%%%%%%%
%%%%%%%%%%%%%%%%%%%%%%%%%%%%%%%%%%%%%%%%%%%%%%%%%%%%%%%%%%%%%%%%
\subsection{4-partite information}

To extend the presented numerical computations beyond that which has been  already considered in the 
literature in this subsection we will consider 4-partite information. Following our previous discussions
the main point is to explore different configurations we may find for computing the 
entanglement entropy of union of different subsystems. 

\begin{figure}[h!] 
\centering
\begin{tikzpicture}[scale=.55]
\draw[ultra thick,black] (0,0) -- (1,0);
\draw[ultra thick,black] (1.5,0) -- (2.5,0);
\draw[ultra thick,black] (3,0) -- (4,0);
\draw[ultra thick,black] (4.5,0) -- (5.5,0);
\draw[ultra thick,blue] (1.5,0) arc (0:180:0.25cm);
\draw[ultra thick,blue] (2.5,0) arc (0:180:1.25cm);
\draw[] (2.75,0) node[below] {$S_{{\rm dis.}}^{(1)}$}; 

\draw[ultra thick,black] (6.5,0) -- (7.5,0);
\draw[ultra thick,black] (8,0) -- (9,0);
\draw[ultra thick,black] (9.5,0) -- (10.5,0);
\draw[ultra thick,black] (11,0) -- (12,0);
\draw[ultra thick,blue] (7.5,0) arc (0:180:0.5cm);
\draw[ultra thick,blue] (9,0) arc (0:180:0.5cm);
\draw[] (9.25,0) node[below] {$S_{{\rm dis.}}^{(2)}$}; 

\draw[ultra thick,black] (13,0) -- (14,0);
\draw[ultra thick,black] (14.5,0) -- (15.5,0);
\draw[ultra thick,black] (16,0) -- (17,0);
\draw[ultra thick,black] (17.5,0) -- (18.5,0);
\draw[ultra thick,blue] (17,0) arc (0:180:2cm);
\draw[ultra thick,blue] (16,0) arc (0:180:1cm);
\draw[] (15.75,0) node[below] {$S_{{\rm dis.}}^{(3)}$}; 

\draw[ultra thick,black] (19.5,0) -- (20.5,0);
\draw[ultra thick,black] (21,0) -- (22,0);
\draw[ultra thick,black] (22.5,0) -- (23.5,0);
\draw[ultra thick,black] (24,0) -- (25,0);
\draw[ultra thick,blue] (25,0) arc (0:180:2.75cm);
\draw[ultra thick,blue] (24,0) arc (0:180:1.75cm);
\draw[] (22.25,0) node[below] {$S_{{\rm dis.}}^{(4)}$};

\draw[ultra thick,black] (0,-4) -- (1,-4);
\draw[ultra thick,black] (1.5,-4) -- (2.5,-4);
\draw[ultra thick,black] (3,-4) -- (4,-4);
\draw[ultra thick,black] (4.5,-4) -- (5.5,-4);
\draw[ultra thick,blue] (4,-4) arc (0:180:2cm);
\draw[ultra thick,blue] (1.5,-4) arc (0:180:.25cm);
\draw[ultra thick,blue] (3,-4) arc (0:180:.25cm);
\draw[] (2.75,-4) node[below] {$S_{{\rm dis.}}^{(5)}$}; 

\draw[ultra thick,black] (6.5,-4) -- (7.5,-4);
\draw[ultra thick,black] (8,-4) -- (9,-4);
\draw[ultra thick,black] (9.5,-4) -- (10.5,-4);
\draw[ultra thick,black] (11,-4) -- (12,-4);
\draw[ultra thick,blue] (7.5,-4) arc (0:180:.5cm);
\draw[ultra thick,blue] (10.5,-4) arc (0:180:1.25cm);
\draw[ultra thick,blue] (9.5,-4) arc (0:180:.25cm);
\draw[] (9.25,-4) node[below] {$S_{{\rm dis.}}^{(6)}$}; 

\draw[ultra thick,black] (13,-4) -- (14,-4);
\draw[ultra thick,black] (14.5,-4) -- (15.5,-4);
\draw[ultra thick,black] (16,-4) -- (17,-4);
\draw[ultra thick,black] (17.5,-4) -- (18.5,-4);
\draw[ultra thick,blue] (14,-4) arc (0:180:.5cm);
\draw[ultra thick,blue] (15.5,-4) arc (0:180:.5cm);
\draw[ultra thick,blue] (17,-4) arc (0:180:.5cm);
\draw[] (15.75,-4) node[below] {$S_{{\rm dis.}}^{(7)}$}; 

\draw[ultra thick,black] (19.5,-4) -- (20.5,-4);
\draw[ultra thick,black] (21,-4) -- (22,-4);
\draw[ultra thick,black] (22.5,-4) -- (23.5,-4);
\draw[ultra thick,black] (24,-4) -- (25,-4);
\draw[ultra thick,blue] (23.5,-4) arc (0:180:2cm);
\draw[ultra thick,blue] (22.5,-4) arc (0:180:1cm);
\draw[ultra thick,blue] (22,-4) arc (0:180:0.5cm);
\draw[] (22.25,-4) node[below] {$S_{{\rm dis.}}^{(8)}$}; 

\draw[ultra thick,black] (0,-8) -- (1,-8);
\draw[ultra thick,black] (1.5,-8) -- (2.5,-8);
\draw[ultra thick,black] (3,-8) -- (4,-8);
\draw[ultra thick,black] (4.5,-8) -- (5.5,-8);
\draw[ultra thick,blue] (5.5,-8) arc (0:180:2.75cm);
\draw[ultra thick,blue] (4.5,-8) arc (0:180:1.75cm);
\draw[ultra thick,blue] (2.5,-8) arc (0:180:.5cm);
\draw[] (2.75,-8) node[below] {$S_{{\rm dis.}}^{(9)}$}; 

\draw[ultra thick,black] (6.5,-8) -- (7.5,-8);
\draw[ultra thick,black] (8,-8) -- (9,-8);
\draw[ultra thick,black] (9.5,-8) -- (10.5,-8);
\draw[ultra thick,black] (11,-8) -- (12,-8);
%\draw[ultra thick,blue] (9,-8) arc (0:180:.5cm);
\draw[ultra thick,blue] (9.5,-8) arc (0:180:1cm);
\draw[ultra thick,blue] (11,-8) arc (0:180:.25cm);
\draw[ultra thick,blue] (12,-8) arc (0:180:2.75cm);
\draw[] (9.25,-8) node[below] {$S_{{\rm dis.}}^{(10)}$};

\draw[ultra thick,black] (13,-8) -- (14,-8);
\draw[ultra thick,black] (14.5,-8) -- (15.5,-8);
\draw[ultra thick,black] (16,-8) -- (17,-8);
\draw[ultra thick,black] (17.5,-8) -- (18.5,-8);
\draw[ultra thick,blue] (14,-8) arc (0:180:.5cm);
\draw[ultra thick,blue] (16,-8) arc (0:180:.25cm);
\draw[ultra thick,blue] (17.5,-8) arc (0:180:.25cm);
\draw[ultra thick,blue] (18.5,-8) arc (0:180:2cm);
\draw[] (15.75,-8) node[below] {$S_{{\rm dis.}}^{(11)}$}; 

\draw[ultra thick,black] (19.5,-8) -- (20.5,-8);
\draw[ultra thick,black] (21,-8) -- (22,-8);
\draw[ultra thick,black] (22.5,-8) -- (23.5,-8);
\draw[ultra thick,black] (24,-8) -- (25,-8);
\draw[ultra thick,blue] (20.5,-8) arc (0:180:.5cm);
\draw[ultra thick,blue] (23.5,-8) arc (0:180:.5cm);
\draw[ultra thick,blue] (25,-8) arc (0:180:2cm);
\draw[ultra thick,blue] (24,-8) arc (0:180:1cm);
\draw[] (22.25,-8) node[below] {$S_{{\rm dis.}}^{(12)}$}; 

\draw[ultra thick,black] (0,-12) -- (1,-12);
\draw[ultra thick,black] (1.5,-12) -- (2.5,-12);
\draw[ultra thick,black] (3,-12) -- (4,-12);
\draw[ultra thick,black] (4.5,-12) -- (5.5,-12);
\draw[ultra thick,blue] (1,-12) arc (0:180:.5cm);
\draw[ultra thick,blue] (2.5,-12) arc (0:180:.5cm);
\draw[ultra thick,blue] (4.5,-12) arc (0:180:.25cm);
\draw[ultra thick,blue] (5.5,-12) arc (0:180:1.25cm);
\draw[] (2.75,-12) node[below] {$S_{{\rm dis.}}^{(13)}$}; 

\draw[ultra thick,black] (6.5,-12) -- (7.5,-12);
\draw[ultra thick,black] (8,-12) -- (9,-12);
\draw[ultra thick,black] (9.5,-12) -- (10.5,-12);
\draw[ultra thick,black] (11,-12) -- (12,-12);
\draw[ultra thick,blue] (8,-12) arc (0:180:.25cm);
\draw[ultra thick,blue] (9,-12) arc (0:180:1.25cm);
\draw[ultra thick,blue] (11,-12) arc (0:180:.25cm);
\draw[ultra thick,blue] (12,-12) arc (0:180:1.25cm);
\draw[] (9.25,-12) node[below] {$S_{{\rm dis.}}^{(14)}$}; 

\draw[ultra thick,black] (13,-12) -- (14,-12);
\draw[ultra thick,black] (14.5,-12) -- (15.5,-12);
\draw[ultra thick,black] (16,-12) -- (17,-12);
\draw[ultra thick,black] (17.5,-12) -- (18.5,-12);
\draw[ultra thick,blue] (17,-12) arc (0:180:1.25cm);
\draw[ultra thick,blue] (16,-12) arc (0:180:.25cm);
\draw[ultra thick,blue] (17.5,-12) arc (0:180:1.75cm);
\draw[ultra thick,blue] (18.5,-12) arc (0:180:2.75cm);
\draw[] (15.75,-12) node[below] {$S_{{\rm dis.}}^{(15)}$}; 

\draw[ultra thick,black] (19.5,-12) -- (20.5,-12);
\draw[ultra thick,black] (21,-12) -- (22,-12);
\draw[ultra thick,black] (22.5,-12) -- (23.5,-12);
\draw[ultra thick,black] (24,-12) -- (25,-12);
\draw[ultra thick,blue] (22,-12) arc (0:180:.5cm);
\draw[ultra thick,blue] (23.5,-12) arc (0:180:.5cm);
\draw[ultra thick,blue] (25,-12) arc (0:180:2.75cm);
\draw[ultra thick,blue] (24,-12) arc (0:180:1.75cm);
\draw[] (22.25,-12) node[below] {$S_{{\rm dis.}}^{(16)}$};

\draw[ultra thick,black] (3,-16) -- (4,-16);
\draw[ultra thick,black] (4.5,-16) -- (5.5,-16);
\draw[ultra thick,black] (6,-16) -- (7,-16);
\draw[ultra thick,black] (7.5,-16) -- (8.5,-16);
\draw[ultra thick,blue] (4,-16) arc (0:180:.5cm);
\draw[ultra thick,blue] (5.5,-16) arc (0:180:.5cm);
\draw[ultra thick,blue] (7,-16) arc (0:180:.5cm);
\draw[ultra thick,blue] (8.5,-16) arc (0:180:.5cm);
\draw[] (5.75,-16) node[below] {$S_{{\rm dis.}}^{(17)}$};

\draw[ultra thick,black] (9.5,-16) -- (10.5,-16);
\draw[ultra thick,black] (11,-16) -- (12,-16);
\draw[ultra thick,black] (12.5,-16) -- (13.5,-16);
\draw[ultra thick,black] (14,-16) -- (15,-16);
\draw[ultra thick,blue] (12,-16) arc (0:180:.5cm);
\draw[ultra thick,blue] (12.5,-16) arc (0:180:1cm);
\draw[ultra thick,blue] (14,-16) arc (0:180:.25cm);
\draw[ultra thick,blue] (15,-16) arc (0:180:2.75cm);
\draw[] (12.25,-16) node[below] {$S_{{\rm dis.}}^{(18)}$}; 

\draw[ultra thick,black] (16,-16) -- (17,-16);
\draw[ultra thick,black] (17.5,-16) -- (18.5,-16);
\draw[ultra thick,black] (19,-16) -- (20,-16);
\draw[ultra thick,black] (20.5,-16) -- (21.5,-16);
\draw[ultra thick,blue] (17.5,-16) arc (0:180:.25cm);
\draw[ultra thick,blue] (19,-16) arc (0:180:.25cm);
\draw[ultra thick,blue] (20.5,-16) arc (0:180:.25cm);
\draw[ultra thick,blue] (21.5,-16) arc (0:180:2.75cm);
\draw[] (18.75,-16) node[below] {$S_{{\rm con.}}$}; 
\end{tikzpicture}
\caption{Different independent and nonintersecting configurations for computing $S(A_i\cup A_j), S(A_i\cup A_j\cup A_k)$ with $i, j, k=1,\cdots, 4$ and $S(A_1\cup A_2\cup A_3\cup A_4)$. Here we neglect other configurations which have the same area.}
\label{fig:4partiteconfig}
\end{figure}
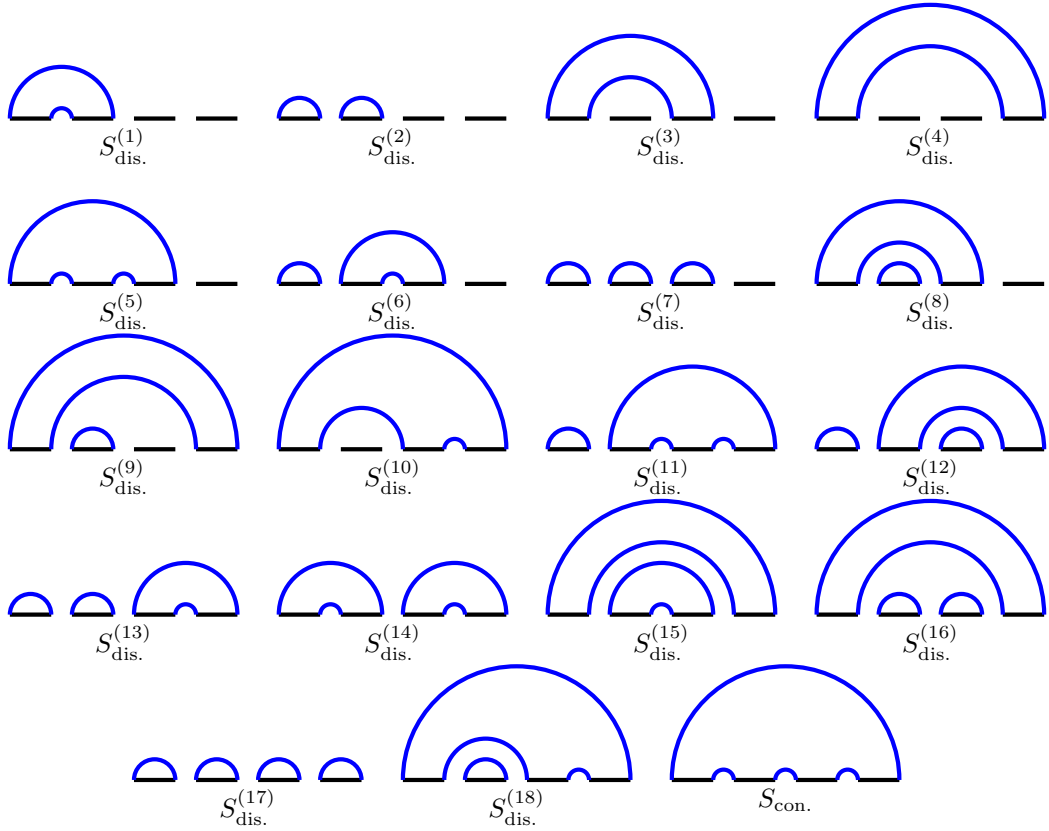

Indeed in the present case the different independent and nonintersecting configurations one should 
study are those appearing in $S(A_i\cup A_j)$, $S(A_i\cup A_j\cup A_k)$ with $i, j, k=1,\cdots, 4$ and $S(A_1\cup A_2\cup A_3\cup A_4)$. The corresponding configurations are illustrated in Fig.\ref{fig:4partiteconfig}.  
Note that in this figure we have just shown  independent configurations  with
different areas.  Each case might have other configurations which could be obtained by 
a permutation from the ones depicted in the Fig.\ref{fig:4partiteconfig}.  For example the cases 
of  $S_{{\rm dis.}}^{(1)}$ and $S_{{\rm dis.}}^{(2)}$ have three and six other configurations, respectively,  which can be obtained by permutations, thought all of them have  the same area.  Indeed for a  generic $S(A_1\cup \cdots \cup A_n)$  entanglement entropy one has  $(2n-1)!!$ inequivalent  configurations.
Nevertheless  having worked with a symmetric entangling regions, the number of 
inequivalent configurations reduce significantly. In particular in the present case we just 
need to study  19 independent configurations one of which is  connected and the others are disconnected, as shown in Fig.\ref{fig:4partiteconfig}. 

\begin{figure}[h!]
\centering
\begin{subfigure}
\centering
\includegraphics[scale=.55]{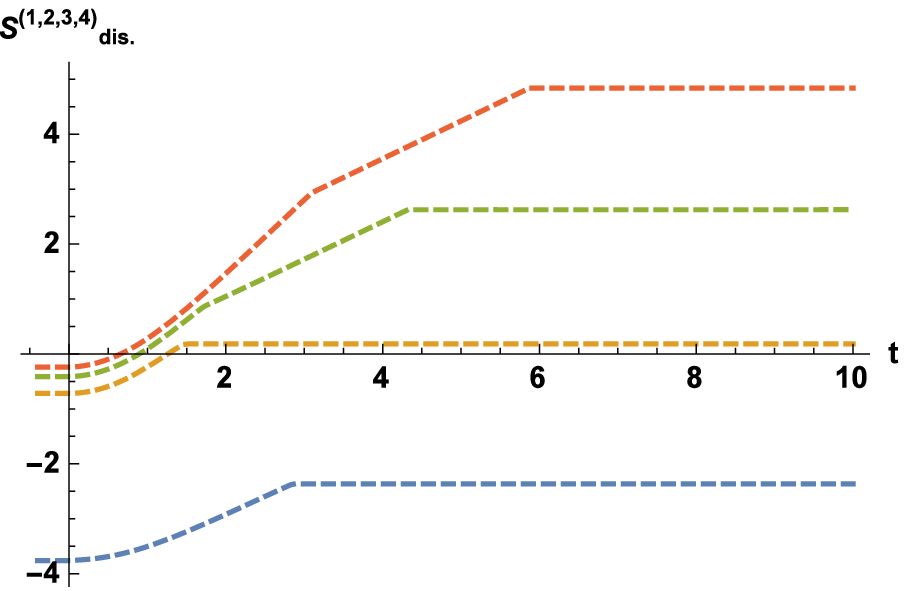}
\end{subfigure}
\begin{subfigure}
\centering
\includegraphics[scale=.55]{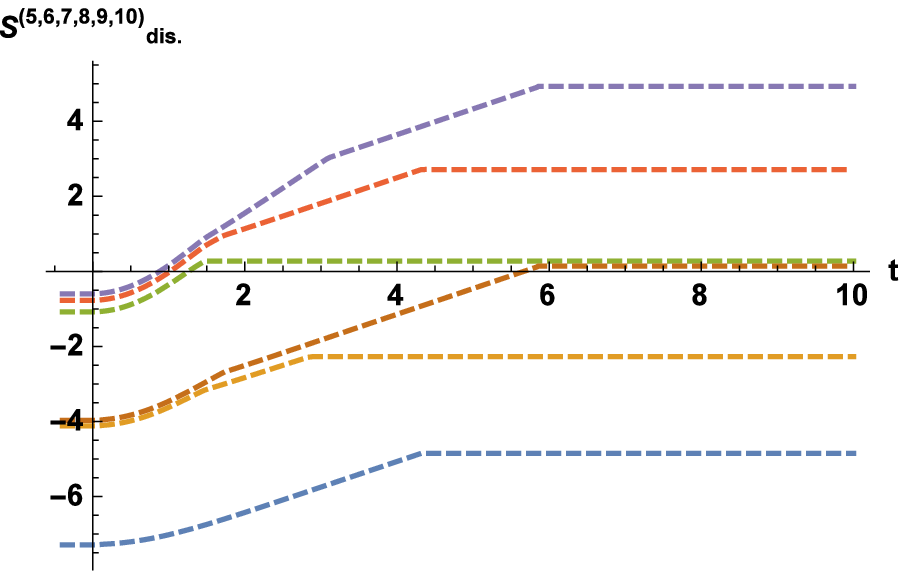}
\end{subfigure}
\begin{subfigure}
\centering
\includegraphics[scale=.55]{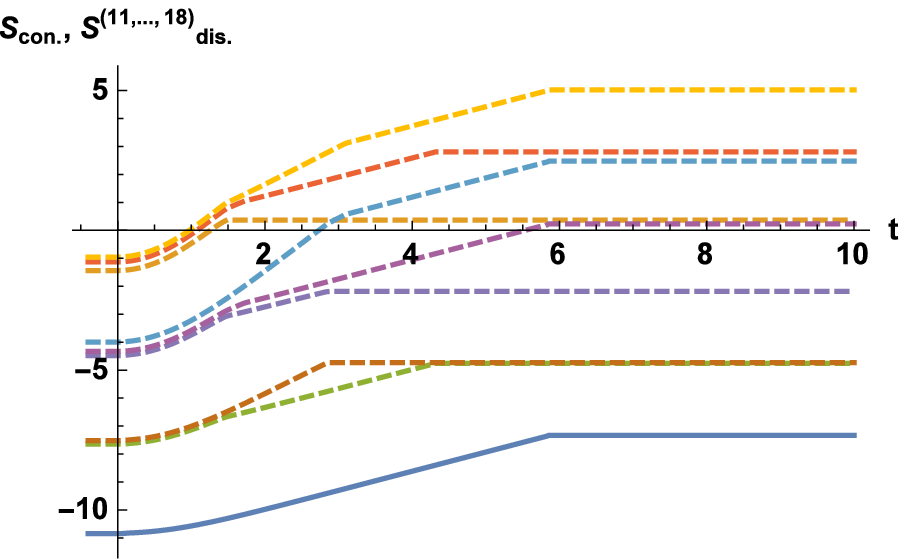}
\end{subfigure}
\caption{Evolution of entanglement entropy for the configurations which correspond to the computation of 4-partite information for $h=0.2$ and $\ell=2$. \textit{Left plot}: $S^{(1)}_{\text{dis.}}, \cdots, S^{(4)}_{\text{dis.}}$ from bottom to top. \textit{Middle plot}: $S^{(5)}_{\text{dis.}},\cdots,  S^{(10)}_{\text{dis.}}$ from bottom to top. \textit{Right plot}: $S_{\text{con.}}$ and $S^{(11)}_{\text{dis.}}, \cdots, S^{(18)}_{\text{dis.}}$. These plots show that in this range of the parameters the conditions \eqref{assum4} satisfied.}
\label{4partite1}
\end{figure}

Now in order to prove equation \eqref{np} for $n=4$  and in the limit of $h\ll \ell$ the following conditions must be satisfied
\bea\label{assum4}
&&S^{(1)}_{\text{dis.}}<S^{(2)}_{\text{dis.}}<\min\left(S^{(3)}_{\text{dis.}},S^{(4)}_{\text{dis.}}\right),\;\;\;\;\text{and}\;\;\;\;
S^{(5)}_{\text{dis.}}<S^{(6)}_{\text{dis.}}<\min\left(S^{(7)}_{\text{dis.}}, S^{(8)}_{\text{dis.}}, S^{(9)}_{\text{dis.}}, S^{(10)}_{\text{dis.}}\right),\nonumber\\
&&\hspace*{2cm}\;\;\text{and}\;\;\;\;\;\;\hspace*{1cm} S_{\text{con.}}<\min\left(S^{(11)}_{\text{dis.}}, \cdots, S^{(18)}_{\text{dis.}}\right). 
\eea
Actually we have numerically checked that these conditions are satisfied within the range of 
parameters we are interested in. For example Fig.\ref{4partite1} shows numerical results  
for certain values of  $h$ and $\ell$. 

\begin{figure}[h!]
\centering
\begin{subfigure}
\centering
\includegraphics[scale=.75]{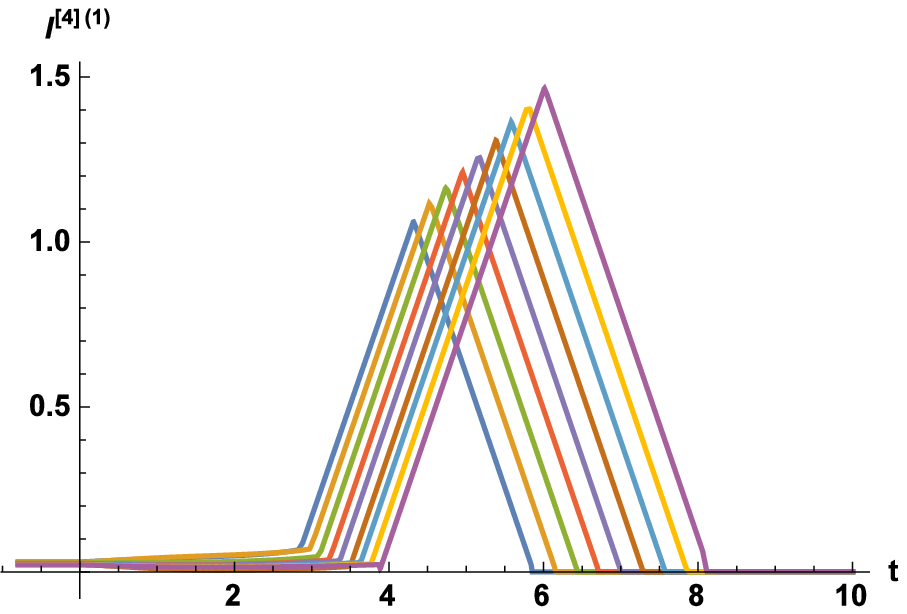}
\end{subfigure}
\begin{subfigure}
\centering
\includegraphics[scale=.75]{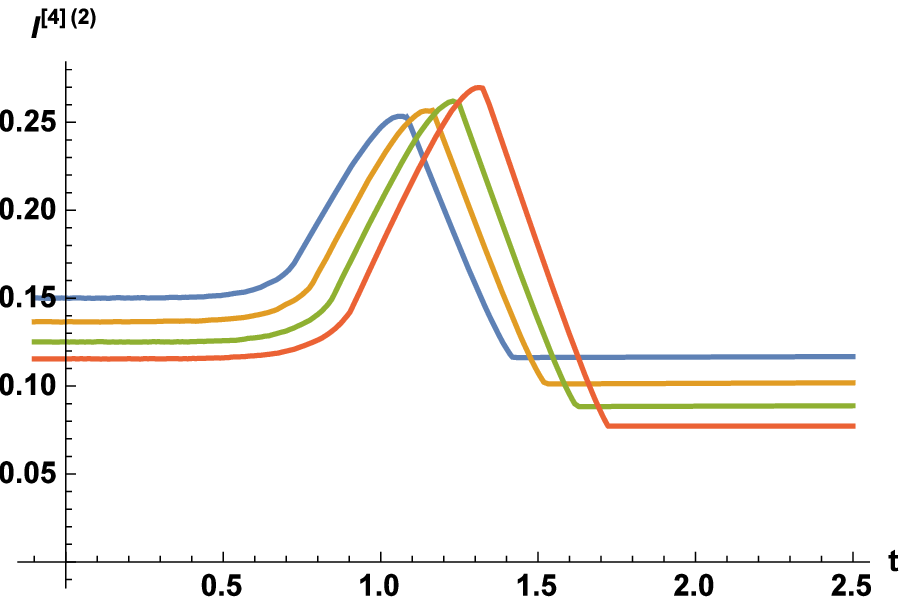}
\end{subfigure}
\begin{subfigure}
\centering
\includegraphics[scale=.75]{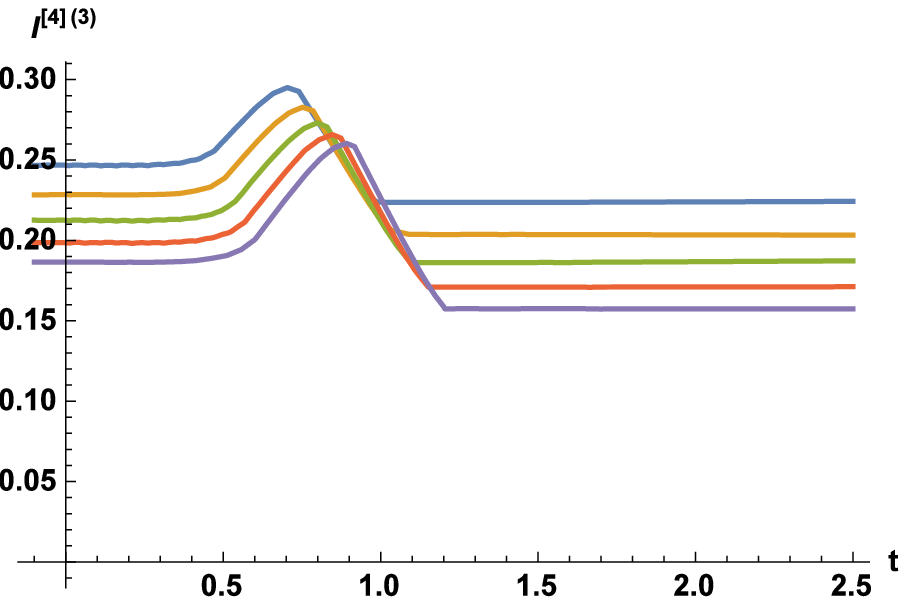}
\end{subfigure}
\begin{subfigure}
\centering
\includegraphics[scale=.75]{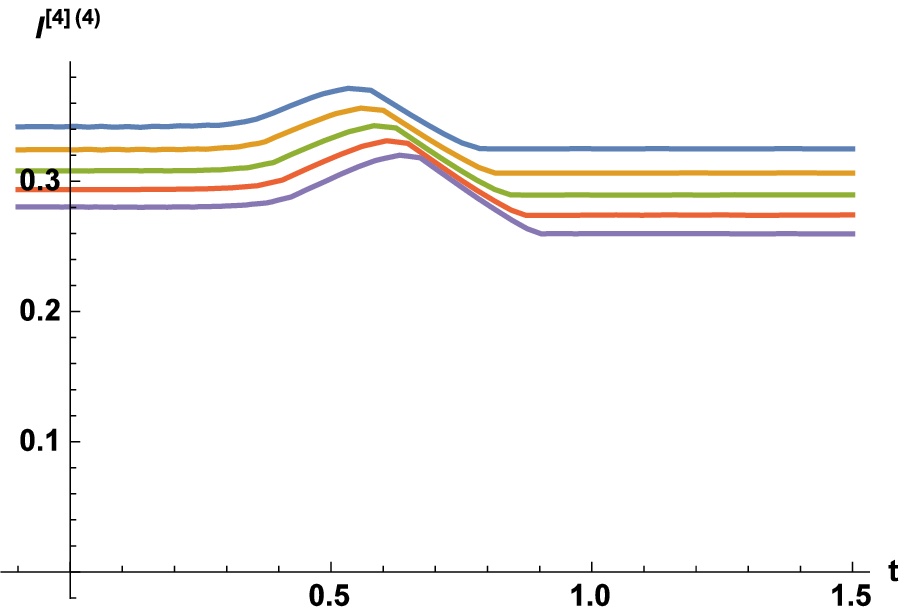}
\end{subfigure}
\caption{Evolution of 4-partite information. \textit{Left up plot}: First case with $h=0.2$ and $\ell=2,\cdots, 2.8$ increasing from left to right. \textit{Right up plot}: Second case with $h=0.1$ and $\ell=0.4,\cdots, 0.52$ increasing from left to right. \textit{Left down plot}: Third case with $h=0.1$ and $\ell=0.24,\cdots, 0.32$ increasing from up to down. \textit{Right down plot}: Fourth case with $h=0.1$ and $\ell=0.17,\cdots, 0.21$ increasing from up to down.}
\label{4partite2}
\end{figure}

Having  check the validity of our assumptions on the minimal configurations one may proceed 
to compute the 4-partite information using equation \eqref{np}.
Of course as we have already mentioned,  due to the relative values of the horizon radius and 
the width of the entangling regions, one recognizes  four different  cases.  Since we have explored 
these possibilities in the cases of mutual information and 3-partite information (see also general argument in section 5) here we just present final numerical results in Fig.\ref{4partite2}.
% for some particular values of $h$ and $\ell$. 

\subsection{5-partite information}

\begin{figure}[h!]
\centering
\begin{subfigure}
\centering
\includegraphics[scale=.75]{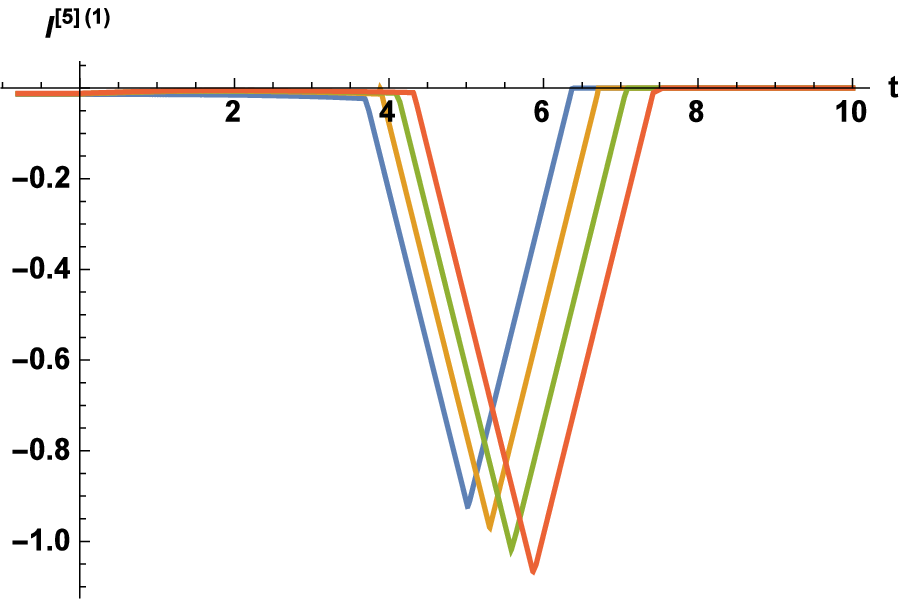}
\end{subfigure}
\begin{subfigure}
\centering
\includegraphics[scale=.75]{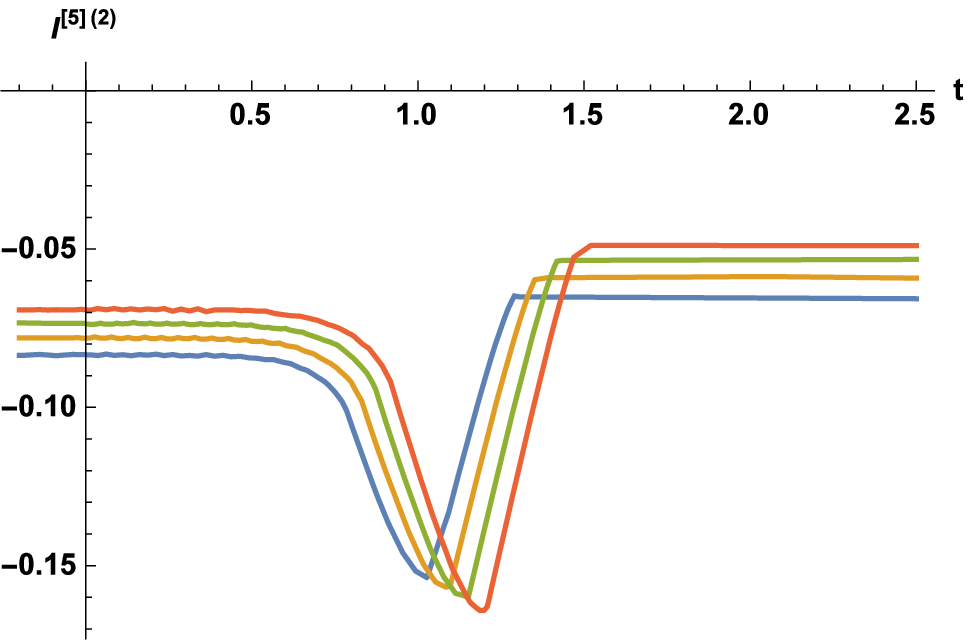}
\end{subfigure}
\caption{Evolution of 5-partite information. \textit{Left plot}: First case with $h=0.2$ and $\ell=1.7,\cdots, 2$ increasing from left to right. \textit{Right plot}: Second case with $h=0.1$ and $\ell=0.26,\cdots, 0.32$ increasing from left to right.}
\label{5partite1}
\end{figure}

Similarly one could also work out the case of 5-partite information. Of course in this case
one must compare different configurations corresponding to $S(A_i\cup A_j)$, $S(A_i\cup A_j\cup A_k) $, $S(A_i\cup A_j\cup A_k\cup A_l)$ with $i, j, k, l=1,\cdots, 5$ and $S(A_1\cup A_2\cup A_3\cup A_4\cup A_5)$.  It is easy to see that in the present case one finds 46 independent nonintersecting configurations with different areas. One of them is  connected and the others are disconnected. Indeed the situation is very similar to what we have done in the previous cases. 
Therefore we just shown the behavior of 5-partite information during the process of thermalization
for the first and second cases (according to the notation in the section 5) in Fig.\ref{5partite1}.

\subsection{Numerics vs. analytic expansions}

In the previous subsections we have numerically studied $n$-partite information (for $n=2,\cdots,5$)  where we have found that the corresponding behaviors qualitatively  are in agreement with the analytical results of sections three and five. To make the comparison clearer, in this subsection, we shall compare the actual numerical values which have been found for different  quantities at distinguished points with those predicted  analytically in the sections three and five. This might help us to better understanding  of limitation and range of the validity of our study.

Table \ref{tablemutual1} presents the numerical results for the special points of the mutual information and the corresponding saturation times for different cases following our notation in the section three. These numerical results are actually obtained by fixing the width $\ell$ while varying $h$
within  its allowed range (see different cases in the section 3). The corresponding results for 3-partite information and those of 4 and 5-partite are, rather 
briefly,  presented in the table \ref{tabletripartite1}  and \ref{tabletripartite4}, respectively.
Note that for these cases we have set $\rho_H=1$ and $a=0.001$. 
\begin{table}[h!]
\centering
\begin{subtable}
\centering
\begin{tabular}{|c|c|c|c|c|c|c|c|c|c|c|}\hline
\multirow{2}{*}{$h$} &  \multicolumn{2}{c|}{ $I^{(1)}_{\rm vac}$ } & \multicolumn{2}{c|}{$I^{(1)}_{\rm max}$} & \multicolumn{2}{c|}{$I^{(1)}_{\rm sat.}$}& \multicolumn{2}{c|}{$t^{(1)}_{\rm max}$}& \multicolumn{2}{c|}{$t^{(1)}_{\rm sat.}$}\\
\hhline{~----------}
        &            Ana. & Num. & Ana. & Num. & Ana. & Num. & Ana. & Num. & Ana. & Num.\\\hline
2.1 & 0.0874& 0.0874      & 1.1051& 1.0771      & 0& 0.0                & 3.2733& 3.0582     & 5.0191& 4.6267       \\ 
        2.2 & 0.0713& 0.0713      & 1.0545& 1.0225      & 0& 0.0                 & 3.2733& 3.0582     & 4.9464& 4.5472       \\ 
        2.4 & 0.0430& 0.0429      & 0.9534& 0.9140      & 0& 0.0                & 3.2733& 3.0582     & 4.8009& 4.3893       \\
\hline 
\end{tabular}
\centering
\end{subtable}
\centering
\begin{subtable}
\centering
\begin{tabular}{|c|c|c|c|c|c|c|c|c|c|c|}
\hline
\multirow{2}{*}{$h$} &  \multicolumn{2}{c|}{ $I^{(2)}_{\rm vac}$ } & \multicolumn{2}{c|}{$I^{(2)}_{\rm max}$} & \multicolumn{2}{c|}{$I^{(2)}_{\rm sat.}$}& \multicolumn{2}{c|}{$t^{(2)}_{\rm max}$}& \multicolumn{2}{c|}{$t^{(2)}_{\rm sat.}$}\\
\hhline{~----------}
        &            Ana. & Num. & Ana. & Num. & Ana. & Num. & Ana. & Num. & Ana. & Num.\\\hline
        0.2  & 3.2261& 3.2256      & 4.0798&4.0054      & 2.6033& 2.5936       & 2.1822& 2.1046      & 4.5099& 4.1683       \\ 
        0.3  & 2.0279& 2.0275      & 2.8749& 2.8007      & 1.3502& 1.3413       & 2.1822& 2.1046      & 4.5827& 4.2636       \\ 
        0.4  & 1.4280& 1.4277      & 2.2654& 2.1910      & 0.6925& 0.6830       & 2.1822& 2.1046      & 4.6554& 4.3588       \\ 
\hline
\end{tabular}
\end{subtable}
\begin{subtable}
\centering
\begin{tabular}{|c|c|c|c|c|c|c|c|c|c|c|}\hline
\multirow{2}{*}{$h$} &  \multicolumn{2}{c|}{ $I^{(3)}_{\rm vac}$ } & \multicolumn{2}{c|}{$I^{(3)}_{\rm max}$} & \multicolumn{2}{c|}{$I^{(3)}_{\rm sat.}$}& \multicolumn{2}{c|}{$t^{(3)}_{\rm max}$}& \multicolumn{2}{c|}{$t^{(3)}_{\rm sat.}$}\\
\hhline{~----------}
        &            Ana. & Num. & Ana. & Num. & Ana. & Num. & Ana. & Num. & Ana. & Num.\\\hline
        0.4  & 0.8379& 0.8376      & 1.0064& 0.9510      & 0.4368& 0.4201       & 0.8728& 0.9467      & 2.0076& 1.9607       \\ 
        0.42 & 0.7506& 0.7503      & 0.9169& 0.8613      & 0.3391& 0.3222       & 0.8728& 0.9467      & 2.0222& 1.9725       \\ 
        0.44 & 0.6710& 0.6707      & 0.8350& 0.7785      & 0.2491& 0.2316       & 0.8728& 0.9467      & 2.0513& 1.9823       \\ 
\hline
\end{tabular}
\end{subtable}
\centering
\begin{subtable}
\centering
\begin{tabular}{|c|c|c|c|c|c|c|c|c|c|c|}\hline
\multirow{2}{*}{$h$} &  \multicolumn{2}{c|}{ $I^{(4)}_{\rm vac}$ } & \multicolumn{2}{c|}{$I^{(4)}_{\rm max}$} & \multicolumn{2}{c|}{$I^{(4)}_{\rm sat.}$}& \multicolumn{2}{c|}{$t^{(4)}_{\rm max}$}& \multicolumn{2}{c|}{$t^{(4)}_{\rm sat.}$}\\
\hhline{~----------}
        &            Ana. & Num. & Ana. & Num. & Ana. & Num. & Ana. & Num. & Ana. & Num.\\\hline
        0.23  & 0.5658& 0.5657      & 0.5863& 0.5788      & 0.4393& 0.4491       & 0.3328& 0.3661      & 0.8358& 0.9193       \\ 
         0.234 & 0.5102& 0.5102      & 0.5304& 0.5229      & 0.3822& 0.3921       & 0.3328& 0.3661      & 0.8387& 0.9222       \\ 
         0.238 & 0.4564& 0.4564      & 0.4764& 0.4690      & 0.3269& 0.3370       & 0.3328& 0.3661      & 0.8417& 0.9251       \\ 
 %       0.24  & 0.4302& 0.4302      & 0.4500& 0.4426      & 0.3000& 0.3101       & 0.3328& 0.3661      %& 0.8432& 0.9265       \\
\hline
\end{tabular}
\end{subtable}
\caption{Comparing specific values of mutual information: $I^{(1)}$ with $\ell=4.5$ (first table),
$I^{(2)}$ with $\ell=3$ (second table), $I^{(3)}$ with $\ell=1.18$ (third table) and $I^{(4)}$ with $\ell=0.45$ (last table).}
\label{tablemutual1}
\end{table}
\begin{table}[h!]
%\centering
%\begin{subtable}
%\centering
%\begin{tabular}{|c|c|c|c|c|c|c|c|c|c|c|}\hline
%\multirow{2}{*}{$\ell$} &  \multicolumn{2}{c|}{ $I^{[3](1)}_{\rm vac}$ } & \multicolumn{2}{c|}{$I^{[3]
%(1)}_{\rm min}$} & \multicolumn{2}{c|}{$I^{[3](1)}_{\rm sat.}$}& \multicolumn{2}{c|}{$t^{[3](1)}_{\rm 
%min}$}& \multicolumn{2}{c|}{$t^{[3](1)}_{\rm sat.}$}\\
%\hhline{~----------}
%        &            Ana. & Num. & Ana. & Num. & Ana. & Num. & Ana. & Num. & Ana. & Num.\\\hline
 %       2.2    & -0.1167& -0.1168       & -0.9130& -1.1359       & 0& $\mathcal{O}(10^{-2})$ & 2.2928& %3.1085         & 3.8116& 4.7578          \\ 
 %       2.4    & -0.1064& -0.1092       & -1.0380& -1.2372       & 0& $\mathcal{O}(10^{-2})$ & 2.5656& %3.3772         & 4.2481& 5.1767          \\ 
 %       2.6    & -0.0977& -0.0981       & -1.1591& -1.3558       & 0& $\mathcal{O}(10^{-2})$ & 2.8411& 3.6248         & 4.6845& 5.6481          \\ 
 %       2.8    & -0.0904& -0.0904       & -1.2770& -1.4663       & 0& $\mathcal{O}(10^{-3})$ & 3.1188& %3.8874         & 5.1210& 6.0445          \\ 
%\hline
%\end{tabular}
%\end{subtable}
\centering
\begin{subtable}
\centering
\begin{tabular}{|c|c|c|c|c|c|c|c|c|c|c|}\hline
\multirow{2}{*}{$\ell$} &  \multicolumn{2}{c|}{ $I^{[3](2)}_{\rm vac}$ } & \multicolumn{2}{c|}{$I^{[3](2)}_{\rm min}$} & \multicolumn{2}{c|}{$I^{[3](2)}_{\rm sat.}$}&  \multicolumn{2}{c|}{$t^{[3](2)}_{\rm min.}$}&  \multicolumn{2}{c|}{$t^{[3](2)}_{\rm sat.}$}\\
\hhline{~----------}
        &            Ana. & Num. & Ana. & Num. & Ana. & Num. & Ana. & Num. & Ana. & Num.\\\hline
        1.04   & -0.2478& -0.2478       & -0.6391& -0.5670       & -0.1822& -0.1508  &   0.7692  &1.6049 & 1.1347& 2.3002         \\ 
        1.08   & -0.2380& -0.2379       & -0.6461& -0.5792       & -0.1650& -0.1402    &  0.7988  & 1.6537& 1.2220& 2.3771         \\ 
        1.12   & -0.2290& -0.2289       & -0.6541& -0.5933       & -0.1492& -0.1319      & 0.8284 & 1.6992& 1.3093& 2.4488         \\ 
%        1.16   & -0.2206& -0.2205       & -0.6631& -0.6050       & -0.1347& -0.1226        &0.8580 %&1.7504 & 1.3966& 2.5233         \\
\hline
\end{tabular}
\end{subtable}
\centering
\begin{subtable}
\centering
\begin{tabular}{|c|c|c|c|c|c|c|c|c|c|c|}\hline
\multirow{2}{*}{$\ell$} &  \multicolumn{2}{c|}{ $I^{[3](3)}_{\rm vac}$ } & \multicolumn{2}{c|}{$I^{[3](3)}_{\rm min}$} & \multicolumn{2}{c|}{$I^{[3](3)}_{\rm sat.}$} &  \multicolumn{2}{c|}{$t^{[3](3)}_{\rm min.}$}&  \multicolumn{2}{c|}{$t^{[3](3)}_{\rm sat.}$}\\
\hhline{~----------}
        &            Ana. & Num. & Ana. & Num. & Ana. & Num. & Ana. & Num. & Ana. & Num.\\\hline
%        0.42   & -0.6734& -0.6733       & -0.7701& -0.7421       & -0.5493& -0.6199   & 0.6952&  0.7553   & 1.0620& 1.1244         \\ 
        0.44   & -0.6386& -0.6384       & -0.7435& -0.7133       & -0.5227& -0.5825     & 0.7248&   0.7854& 1.1056& 1.1645         \\ 
   %     0.46   & -0.6072& -0.6070       & -0.7206& -0.6878       & -0.4986& -0.5488      &0.7544 &  0.8151& 1.1493& 1.2087         \\ 
       0.48   & -0.5787& -0.5786       & -0.7009& -0.6654       & -0.4769& -0.5177      & 0.7840&  0.8446& 1.1929& 1.2497         \\ 
 0.52   & -0.5289& -0.5287       & -0.6696& -0.6290       & -0.4396& -0.4637      & 0.8432&  0.9027& 1.2802& 1.3439         \\ 
\hline
\end{tabular}
\end{subtable}
\centering
\caption{Comparing specific values of 3-partite information:  $I^{[3](2)}$ with $h=0.1$ (first table) and 
$I^{[3](3)}$ with $h=0.1$ (second table). }
\label{tabletripartite1}
\end{table}
\begin{table}[h!]
\centering
\begin{subtable}
\centering
\begin{tabular}{|c|c|c|c|c|c|c|c|c|c|c|}\hline
\multirow{2}{*}{$\ell$} &  \multicolumn{2}{c|}{ $I^{[4](1)}_{\rm vac}$ } & \multicolumn{2}{c|}{$I^{[4](1)}_{\rm max}$} & \multicolumn{2}{c|}{$I^{[4](1)}_{\rm sat.}$}& \multicolumn{2}{c|}{$t^{[4](1)}_{\rm max}$}& \multicolumn{2}{c|}{$t^{[4](1)}_{\rm sat.}$}\\
\hhline{~----------}
        &            Ana. & Num. & Ana. & Num. & Ana. & Num. & Ana. & Num. & Ana. & Num.\\\hline
     %  2.5    & 0.0240& 0.0239         & 1.3268& 1.3086         & 0& $\mathcal{O}(10^{-4})$ & 5.7465& 5.3859         & 7.7106& 7.3228          \\ 
        2.6    & 0.0231& 0.0229         & 1.3777& 1.3253         & 0& $\mathcal{O}(10^{-4})$ & 5.9648& 5.6481         & 8.0015& 7.5837          \\ 
        2.7    & 0.0222& 0.0220         & 1.4285& 1.4027         & 0& $\mathcal{O}(10^{-4})$ & 6.1830& 5.8213         & 8.2925& 7.8826          \\ 
        2.8    & 0.0214& 0.0214         & 1.4792& 1.4471         & 0& $\mathcal{O}(10^{-4})$ & 6.4012& 6.0445         & 8.5835& 8.1135          \\
\hline
\end{tabular}
\end{subtable}
\begin{subtable}
\centering
\begin{tabular}{|c|c|c|c|c|c|c|c|c|c|c|}\hline
\multirow{2}{*}{$\ell$} &  \multicolumn{2}{c|}{ $I^{[5](1)}_{\rm vac}$ } & \multicolumn{2}{c|}{$I^{[5](1)}_{\rm min}$} & \multicolumn{2}{c|}{$I^{[5](1)}_{\rm sat.}$}& \multicolumn{2}{c|}{$t^{[5](1)}_{\rm min}$}& \multicolumn{2}{c|}{$t^{[5](1)}_{\rm sat.}$}\\
\hhline{~----------}
        &            Ana. & Num. & Ana. & Num. & Ana. & Num. & Ana. & Num. & Ana. & Num.\\\hline
%                $1.7$  & -0.0137& -0.0139    & -0.9028& -0.9128       & 0& $\mathcal{O}(10^{-4})$ & 5.3828& 5.0413         & 6.7649& 6.3957 \\
$1.8$  & -0.0129& -0.0130    & -0.9553& -0.9595       & 0& $\mathcal{O}(10^{-4})$ & 5.6738& 5.3041         & 7.1286& 6.7109\\
$1.9$  & -0.0123& -0.0123    & -1.0076& -0.9804       & 0& $\mathcal{O}(10^{-4})$ & 5.9648& 5.6481         &  7.4923& 7.0870\\
2      & -0.0117& -0.0118    & -1.0596& -1.0510       & 0& $\mathcal{O}(10^{-4})$ & 6.2557& 5.9041         &  7.8560& 7.5382\\
\hline
\end{tabular}
\end{subtable}
\caption{Comparing specific values of $I^{[4](1)}$ with $h=0.2$ (up) and $I^{[5](1)}$ with $h=0.2$
(down).}
\label{tabletripartite4}
\end{table}

\begin{figure}[h!]
\centering
\begin{subfigure}
\centering
\includegraphics[scale=.7]{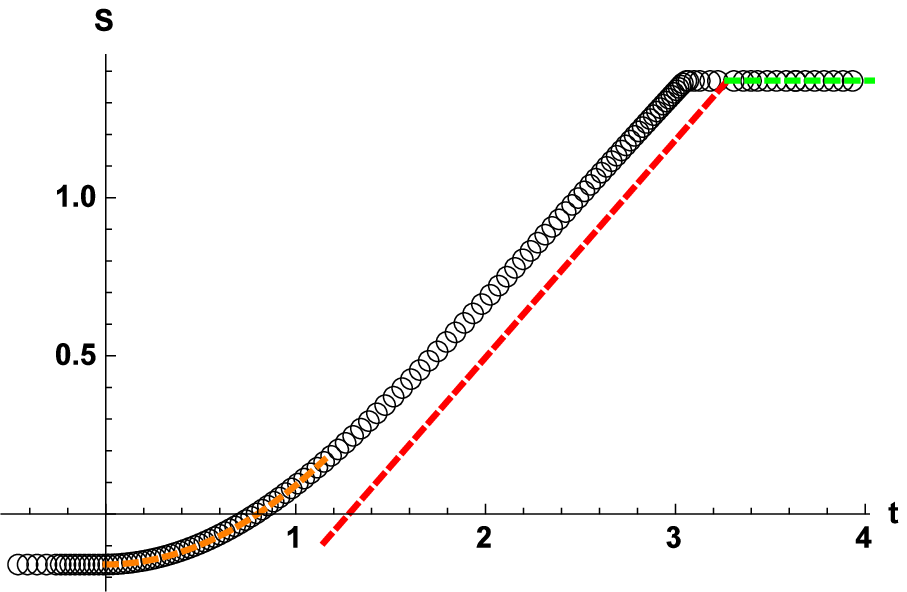}
\end{subfigure}
\begin{subfigure}
\centering
\includegraphics[scale=.7]{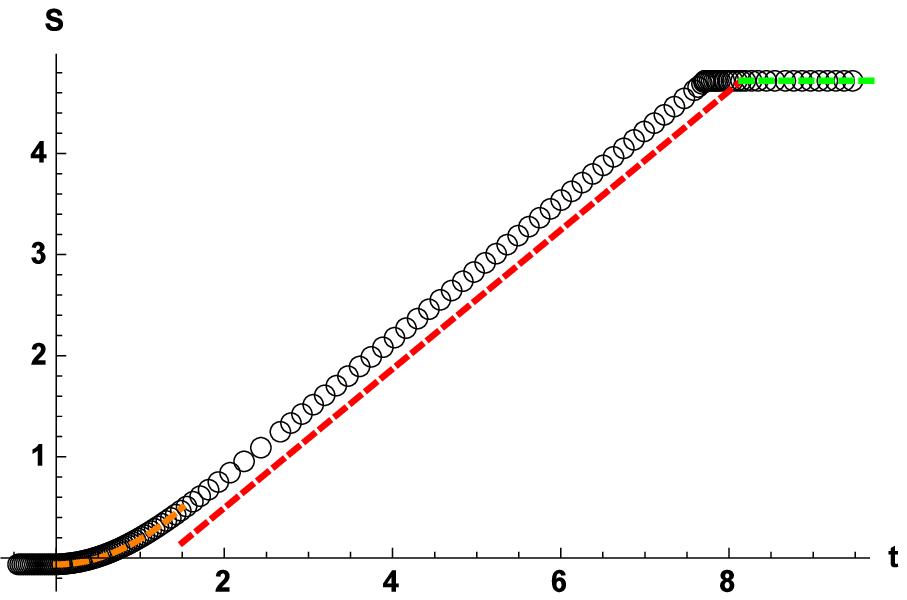}
\end{subfigure}
\caption{Evolution of HEE for $\ell=4.5, 11.2$ from left to right. In each plot the black circles show the numerical result and the dashed curves represent the analytic expansion. The dashed orange curve corresponds to the quadratic growth at early time, the dashed red curve corresponds to the intermediate linear growth and the dashed green line shows the saturation regime. In both cases the saturation time that is obtained by the semi-analytic expansion is larger than the numerical ones, but increasing $\ell$ they converge together.
}
\label{fig:Fit1}
\end{figure}

Although  the results given in tables 1-3 are enough to explore the level of agreement and range 
of validity of two different approaches which have been considered in this paper (numerical and semi-analytical), it is useful to visually present the results in different plots. To proceed it helps if one first
studies the behavior of holographic entanglement entropy.  Actually Fig.  \ref{fig:Fit1} shows 
the evolution of HEE for different values of $\ell$, {\it e.i.} $\ell=4.5$ and $11.2$. In this figure 
the  black circles show the numerical results while the dashed curves represent the results
of our  semi-analytic study. As we have already mentioned  the semi-analytic method gives just 
a piece wise plot which should be compared with certain  regions of the numerical computations.
From this figure one also observes that by increasing the width of the entangling region, the analytic expansions become more precise and smooth,  as expected. 

According to our semi-analytical expansions we would expect  that the transition between quadratic 
growth and linear growth occurs  at $t_{\rm trans.}\sim \mathcal{O}(\rho_H)=1$, though  our 
numerical results give actual value for the saturation times. For example for  $\ell=4.5$ one has 
$t_{\rm trans.}\sim 1.15$  while for  $\ell=11.2$ one gets $t_{\rm trans.}\sim 1.5$. 
It is worth noting  that in both cases the saturation time obtained by the semi-analytic expansion is
generally  larger than the numerical ones, though by  increasing $\ell$ they would  converge 
at the same time.   Since the $n$-partite information may be given in terms of entanglement entropy,
such a difference would also affect the saturation times of the $n$-partite information.

\begin{figure}[h!]
\centering
\begin{subfigure}
\centering
\includegraphics[scale=.75]{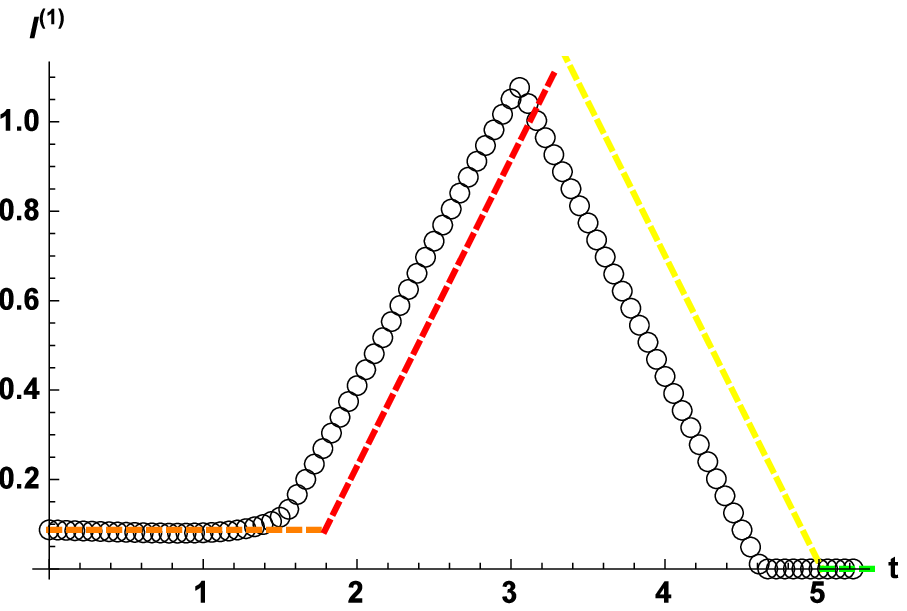}
\end{subfigure}
\begin{subfigure}
\centering
\includegraphics[scale=.75]{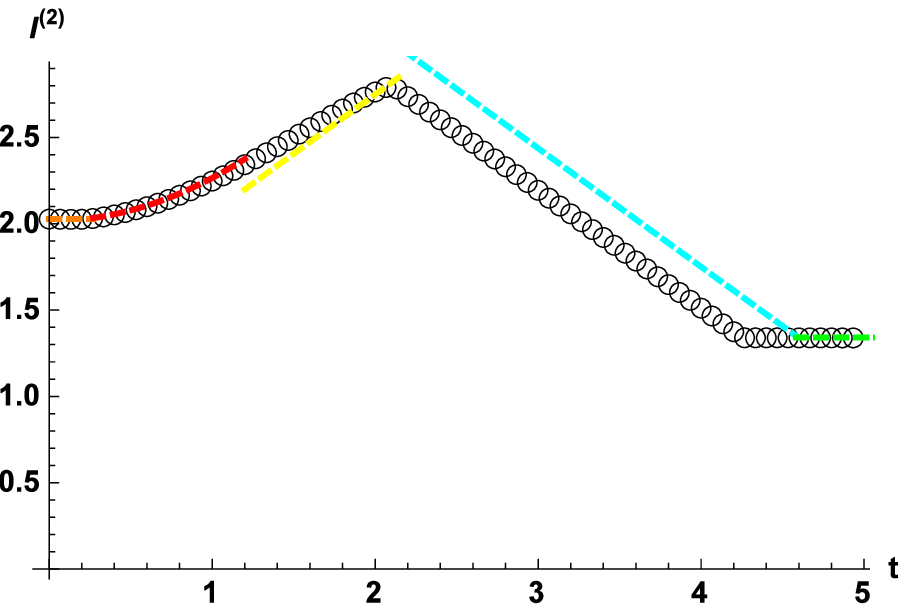}
\end{subfigure}
\begin{subfigure}
\centering
\includegraphics[scale=.75]{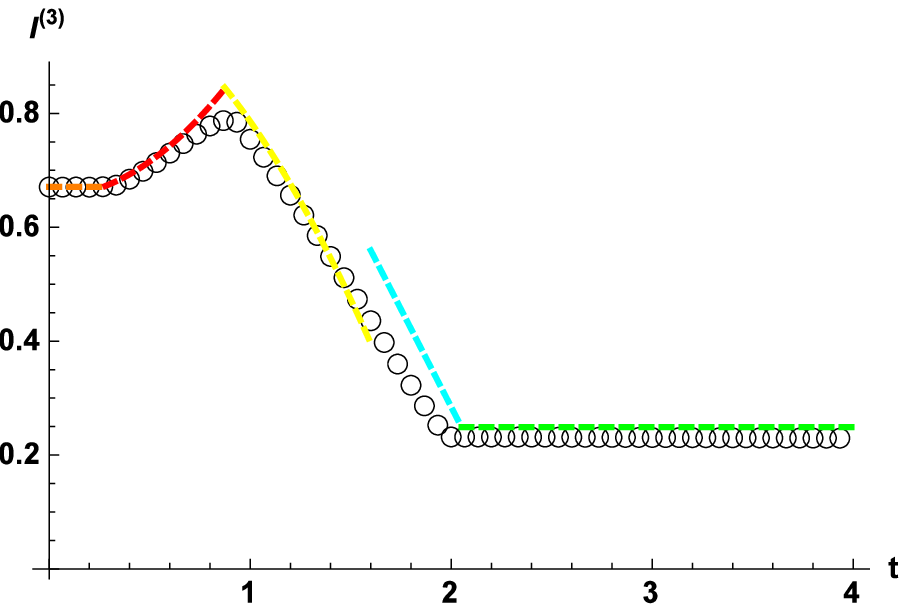}
\end{subfigure}
\begin{subfigure}
\centering
\includegraphics[scale=.75]{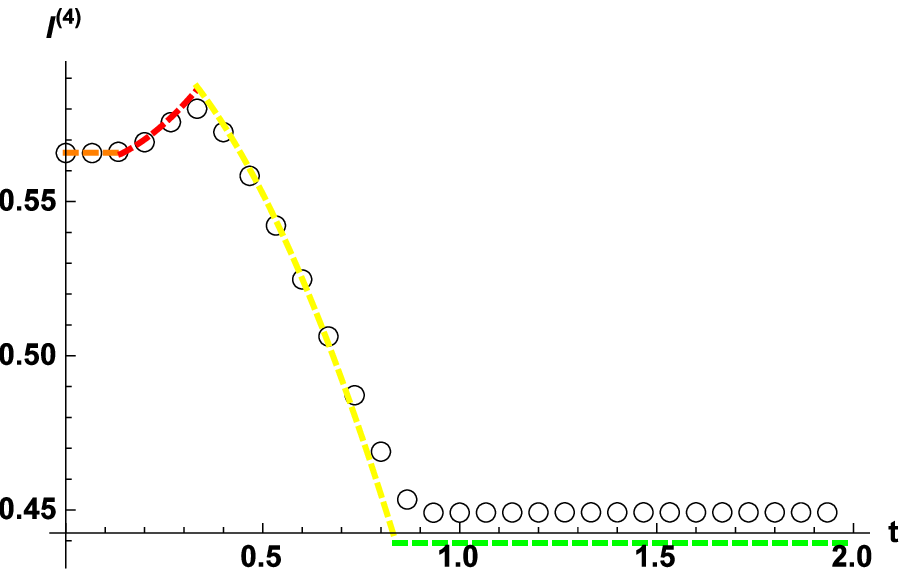}
\end{subfigure}
\caption{Comparing analytical and numerical results for $I^{(1)}$ for $\ell=4.5$ and $h=2.1$
(left up), $I^{(2)}$ for $\ell=3$ and $h=0.3$ (right up),  $I^{(3)}$ for $\ell=1.18$ and $h=0.44$
(left down) and  $I^{(4)}$ for $\ell=0.45$ and $h=0.23$ (right down). In these plots the black circles show the numerical results, the dashed colored  curves corresponds to different scaling 
behaviors we have found in our analytical studies.}
\label{fig:Fit2}
\end{figure}

\begin{figure}[h!]
\centering
\begin{subfigure}
\centering
\includegraphics[scale=.75]{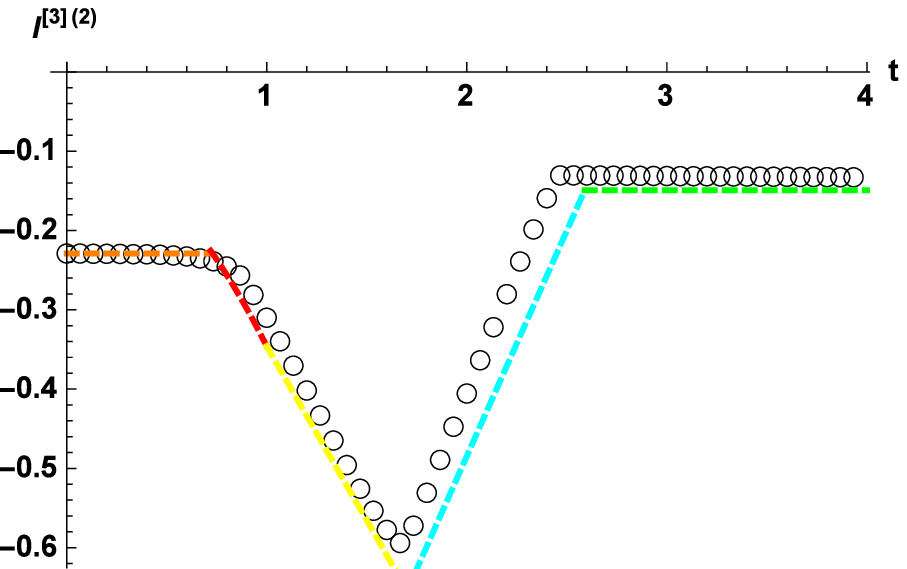}
\end{subfigure}
\begin{subfigure}
\centering
\includegraphics[scale=.75]{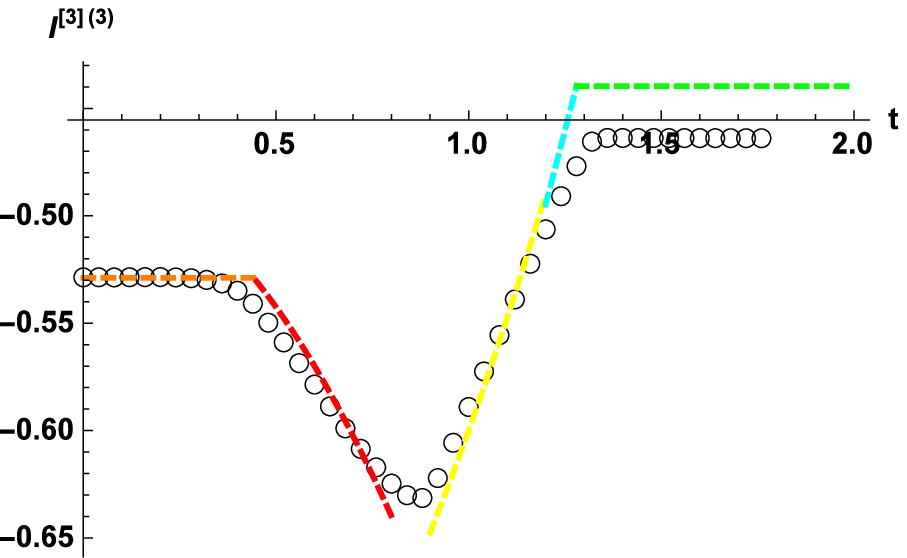}
\end{subfigure}
\centering
\begin{subfigure}
\centering
\includegraphics[scale=.75]{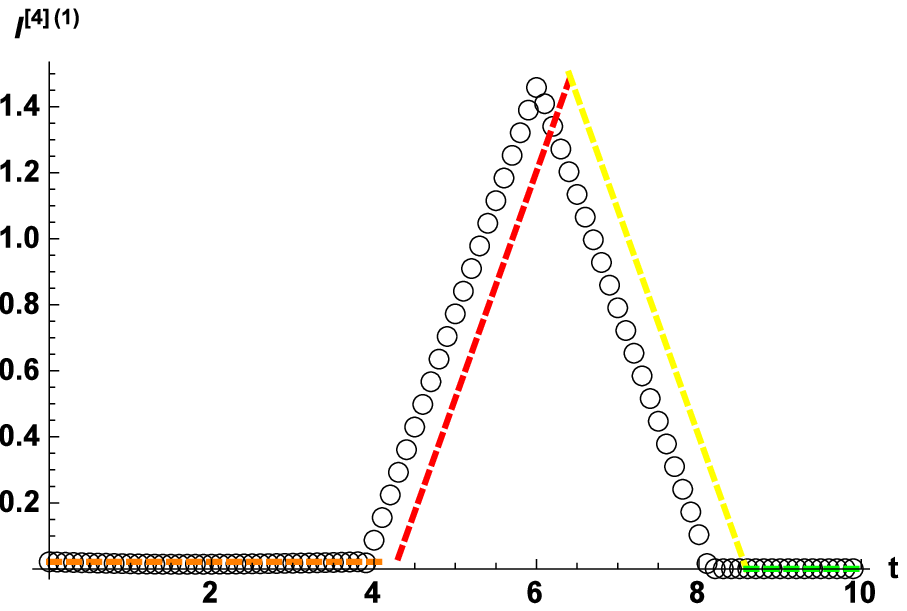}
\end{subfigure}
\begin{subfigure}
\centering
\includegraphics[scale=.75]{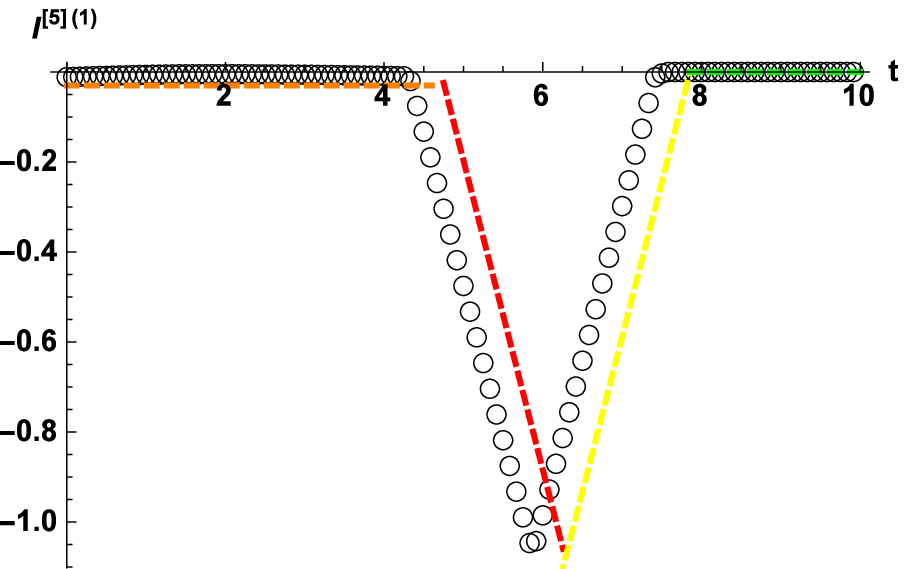}
\end{subfigure}
\caption{Comparing analytical and numerical results for $I^{[3](2)}$ for $h=0.1$ and $\ell=1.12$
(left up), $I^{[3](3)}$ for $h=0.1$ and $\ell=0.42$ (right up), $I^{[4](1)}$ for $h=0.2$ and $\ell=2.8$
(left down) and $I^{[5](1)}$ for $h=0.2$ and $\ell=2$ (right down).
In these plots the black circles show the numerical results, the dashed colored  curves corresponds to different scaling 
behaviors we have found in our analytical studies.}
\label{fig:Fit6}
\end{figure}

Let us now consider mutual and $n$-partite information. To be specific in Fig.\ref{fig:Fit2} we 
have depicted the  results for mutual information in 
different cases for particular values of parameters. In these figures the numerical results 
are compared with the semi-analytical results. Note that
for the latter approach the corresponding curves are plotted in different colors indicating 
different scaling behaviors we have found for the mutual information. In other words 
these curves have been plotted patch wise and the resulting function may not be a smooth function for whole time during the process of  thermalization. Similarly one could graphically compare other cases. For example in the case of higher $n$, we have 
depicted the results for $I^{[3](2)},\;I^{[3](3)},\;I^{[4](1)}$ and $I^{[5](1)}$ in Fig.\ref{fig:Fit6}.
%Actually these figures should be enough  to compare the two approaches. Indeed other cases 
%$I^{[n](i)}$ with different $n$'s
%and $i$'s would lead to the similar conclusion.

From these tables and figures one observes that there is a reasonable agreement between two 
approaches for values of 
$n$-partite information at different points, thought there are mismatch on the values of the times
on which different events occur, such as the saturation time or the time where 
the  linear behavior starts or terminates.  Actually the agreement and mismatch
of the values are related to the validity of our approximations in both approaches. This may be
understood as follows.

\begin{figure}[h!]
\centering
\begin{subfigure}
\centering
\includegraphics[scale=.75]{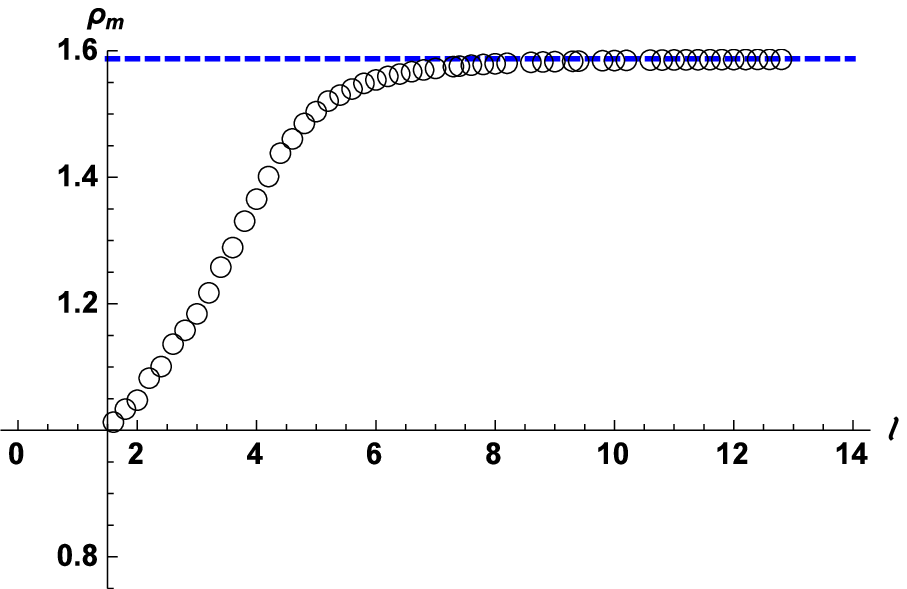}
\end{subfigure}
\begin{subfigure}
\centering
\includegraphics[scale=.75]{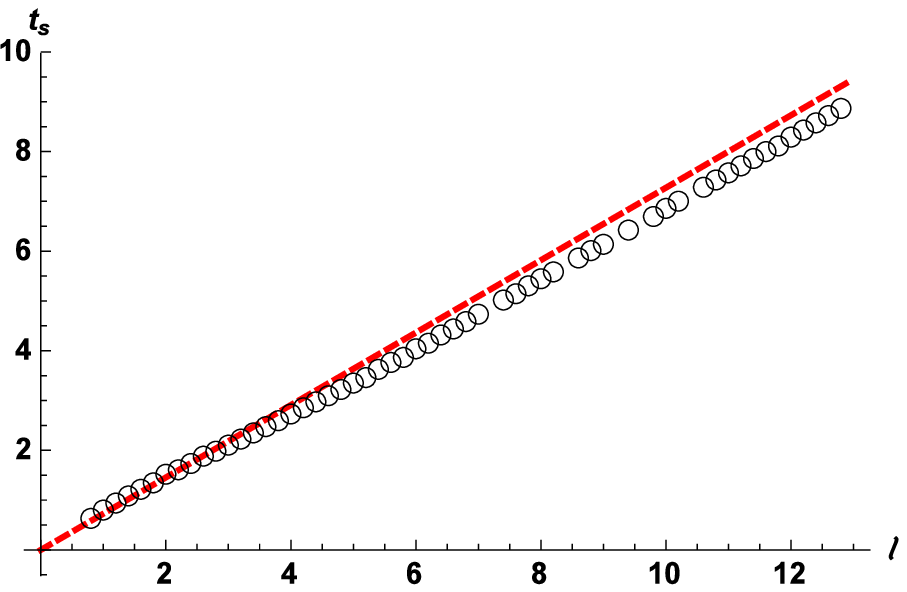}
\end{subfigure}
\caption{$\rho_m\;{\rm and}\; t_s$ as  a function of entangling width $\ell$. In each plot the solid circles are the numerical results. The dashed blue line in the left plot shows the value of $\rho_m$ approximated by \eqref{largesimp1} and the dashed red in the right plot show the value of $t_s$ approximated by \eqref{tsfinal}.}
\label{fig:rmts}
\end{figure}

As far as the values of $n$-partite information at different distinguished points are concerned 
the small mismatch is  related to our assumptions on  the relative size of the entangling widths, their
separations and the  horizon radius.  Indeed to work out our analytical results we have assumed 
a stricken inequality such as $\rho_H\ll \frac{h}{2}\ll \frac{\ell}{2}$, though in our numerical computations our parameters satisfy  $\rho_H< \frac{h}{2}< \frac{\ell}{2}$. As a result although in the
analytical considerations we could drop all higher order corrections, in the numerical one, their
effects are  also taken into account. As an explicit example, in Fig.\ref{fig:rmts} we plot two parameters that appear in the analytic calculation for linear growth regime, i.e. $\rho_m$ and $t_s$ and compare them with the numerical data. In this figure the solid circles show the numerical results. The dashed blue line in the left plot shows the value of $\rho_m$ approximated by \eqref{largesimp1}. Note that in the analytical approximations that we have used, we always assume that in the large entangling region, $\rho_m$ is constant and according to \eqref{largesimp1} does not depend on $\ell$. This plot shows that this assumption is more concrete when one considers $\ell>6$ (Note that we always consider $\rho_H=1$). The dashed red line in the right plot show the value of $t_s$ approximated by \eqref{tsfinal}, where $v_E$ is given by \eqref{VEsimple}. Note that in this plot the slope of the curve is given by the entanglement velocity. In both plots in the large entangling region limit analytic expressions converge to numerical results.
Actually as one increases the width of entangling regions our code becomes more unstable. Therefore we have some restrictions when we are  considering 
the large entangling region limit. 

On the other hand in order to compute the time scale of the distinguished  points we have assumed 
a particular behavior for  holographic entanglement entropy as it  approaches its  saturation point. 
To be precise let us recall that in order to compute the $n$-partite information, with the assumption 
we made,  one has to compute three entanglement entropies associated with the entangling regions 
$\ell_1=n\ell+(n-1)h, \ell_2=(n-1)\ell+(n-2)h$ and $\ell_3=(n-2)\ell+(n-3)h$. We note that although 
the distinguished points occur when one of these entanglement entropies saturates to its equilibrium 
value, there is a subtlety to compute the corresponding saturation time when the entangling 
region is in a shape of strip\cite{Liu:2013qca}.

\begin{figure}[h!]
\centering
\begin{subfigure}
\centering
\includegraphics[scale=.55]{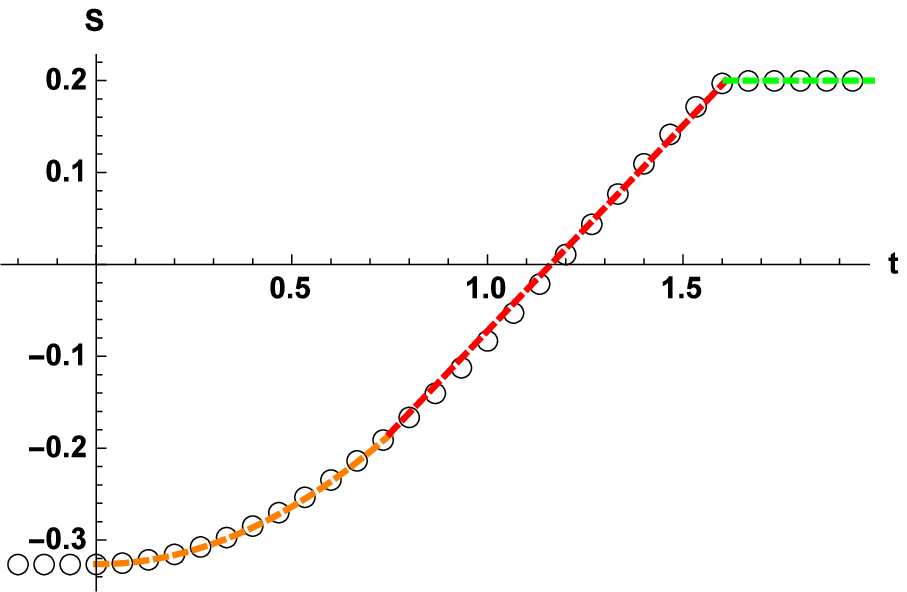}
\end{subfigure}
\begin{subfigure}
\centering
\includegraphics[scale=.46]{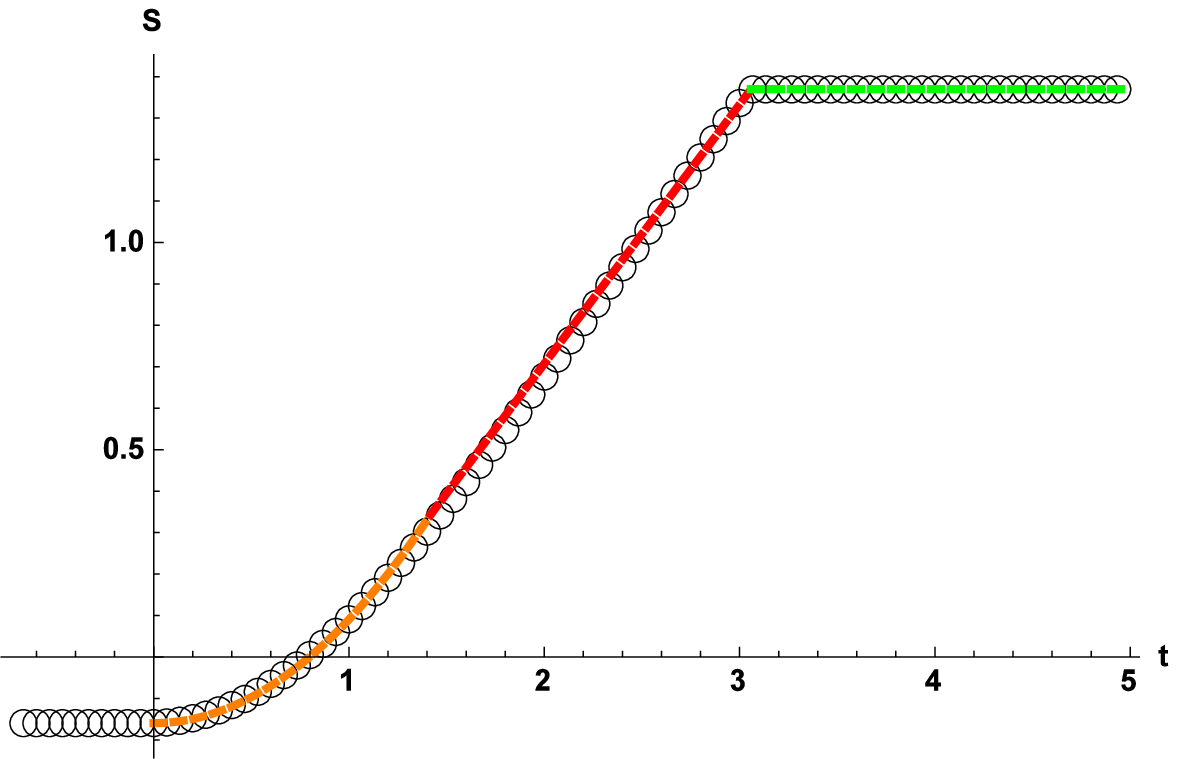}
\end{subfigure}
\begin{subfigure}
\centering
\includegraphics[scale=.46]{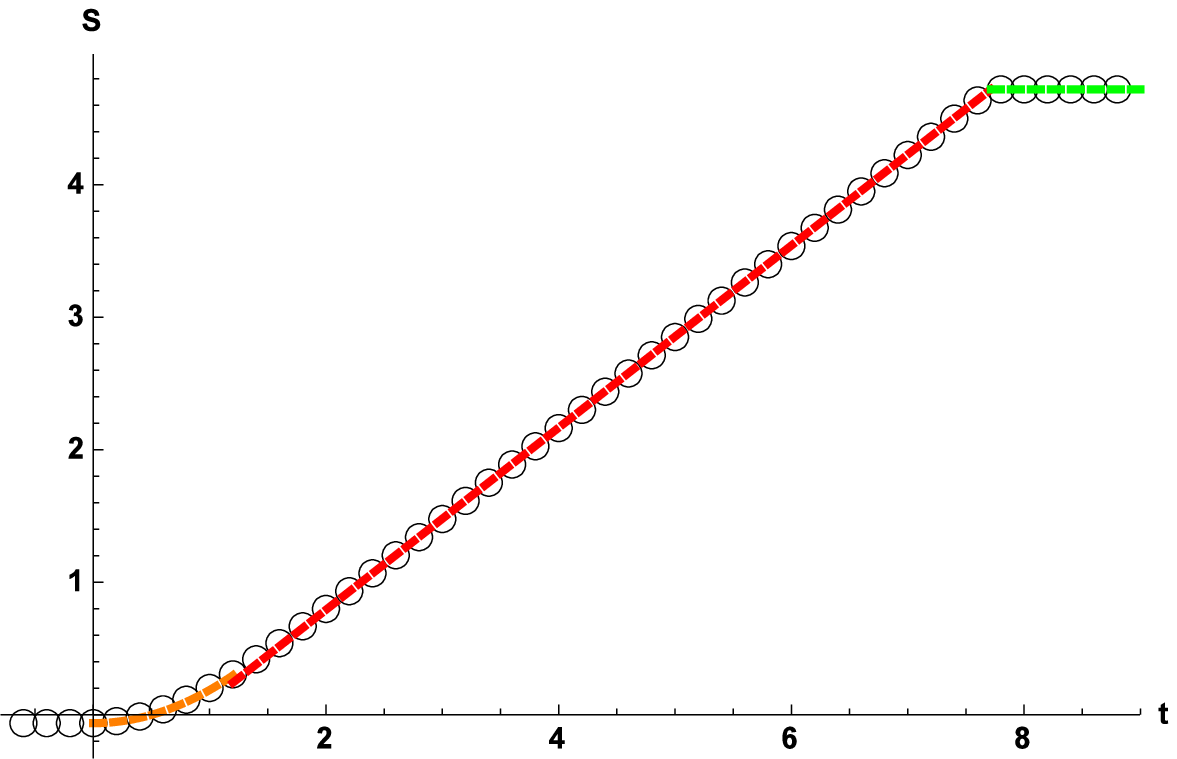}
\end{subfigure}
\caption{Evolution of HEE for $\ell=2.2, 4.5 \;{\rm and}\; 11.2$ from left to right. In each plot the solid circles are the numerical result, the dashed orange curve corresponds to the quadratic growth at early time, the dashed red curve corresponds to the intermediate linear growth and the dashed green line shows the saturation regime.}
\label{fig:Fit9}
\end{figure}

\begin{figure}[h!]
\centering
\begin{subfigure}
\centering
\includegraphics[scale=.75]{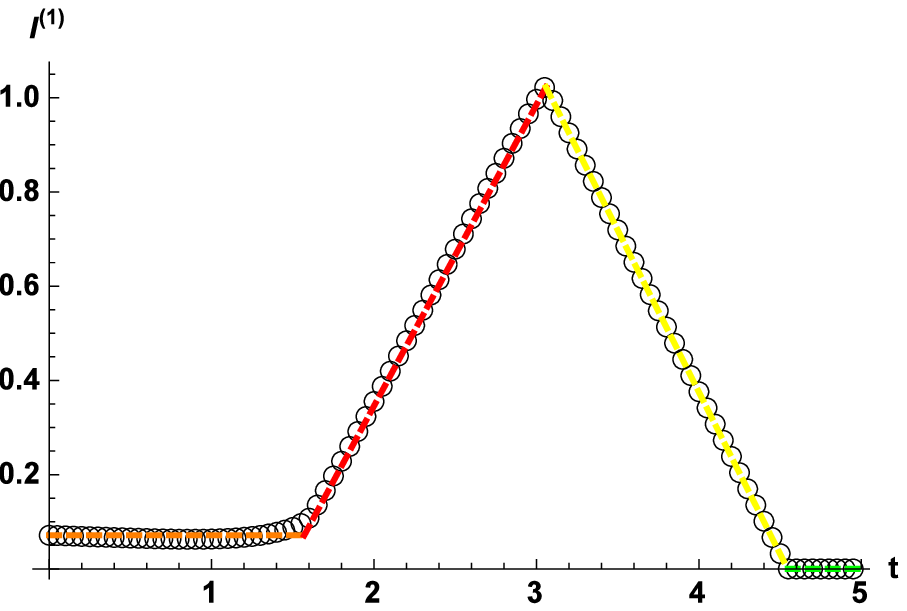}
\end{subfigure}
\begin{subfigure}
\centering
\includegraphics[scale=.75]{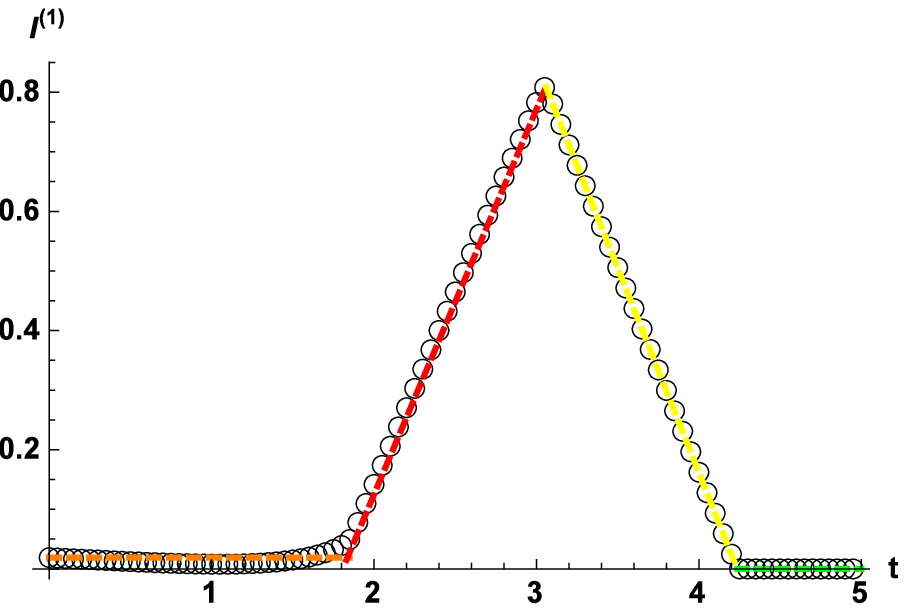}
\end{subfigure}
\caption{Evolution of $I^{(1)}$ for $\ell=4.5$ and $h=2.2, 2.6$ from left to right. In each plot the solid circles is the numerical result, the dashed orange curve corresponds to the steady behavior at early time, the dashed red curve corresponds to the linear growth, the dashed yellow curve corresponds to the linear decreasing and the dashed green line shows the saturation regime.}
\label{fig:Fit10}
\end{figure}

In order to further compare our numerical results with that of semi-analytic, in what follow we will
consider another method for the comparison to explore the regime of validity of our analytic 
expansions. Indeed using ``FindFit'' command in {\it Mathematica} and assuming certain fit 
functions, we compare  two approaches in specific examples. In Fig. \ref{fig:Fit9} we present  the 
evolution of HEE for $\ell=2.2, 4.5 \;{\rm and}\; 11.2$ numerically together with certain  piecewise 
fit functions given by 
\bea
S_{\rm reg.}^{\rm quad.}=s_1+s_2\; t^2,\;\;\;\;S_{\rm reg.}^{\rm lin.}=s_3+s_4\; t.
\eea
One can find the parameters $\{s_1,s_2,s_3,s_4\}$ and compare them with analytic expansions (e.g. eqs \eqref{SBHET} and \eqref{SBHL}), the results are summarized in table \ref{tablefit1}.
Also Fig. \ref{fig:Fit10} shows the evolution of $I^{(1)}$ for $\ell=4.5$ and $h=2.2, 2.6$. 

\begin{table}
\centering
\begin{tabular}{|c|c|c|c|c|c|c|c|c|c|}\hline
\multirow{2}{*}{$\ell$} &  \multicolumn{2}{c|}{ $s_1$ } & \multicolumn{2}{c|}{$s_2$} & \multicolumn{2}{c|}{$s_3$}& \multicolumn{2}{c|}{$s_4$}\\
\hhline{~---------}
        &            Ana. & Num. & Ana. & Num. & Ana. & Num. & Ana. & Num.\\\hline
             2.2  & -0.3262& -0.3246    & 0.25& 0.2454       & -0.5634& -0.3262 & 0.6873& 0.4792         \\
4.5  & -0.1595& -0.1534    & 0.25& 0.2391       & -0.1595& -0.6235 & 0.6873& 0.6522         \\
11.2      & -0.0640& -0.0534    & 0.25& 0.2486       & -0.0640& -0.5727 & 0.6873& 0.6848         \\\hline
\end{tabular}
\caption{Comparing specific values of the parameters $\{s_1,s_2,s_3,s_4\}$ for $\ell=2.2, 4.5 \;{\rm and}\; 11.2$.}
\label{tablefit1}
\end{table}

As we have already  mentioned  there is an excellent agreement  between analytic expansions and numerical results of vacuum and saturation values for mutual information given in 
 \ref{tablemutual1}. Nevertheless in order  to compare these two approaches in the intermediate 
 time we will consider the following fit functions
\bea
I^{(1)}_{\rm lin. grow.}=a_1+b_1\; t,\;\;\;\;I^{(1)}_{\rm lin. dec.}=a_2-b_2\; t.
\eea
Table \ref{tablefit2} shows the parameters $\{a_1,b_1,a_2,b_2\}$ for both approaches (note that for analytic expansions we use eqs. \eqref{mutuallingro} and \eqref{SD}).

\begin{figure}[h!]
\centering
\begin{subfigure}
\centering
\includegraphics[scale=.75]{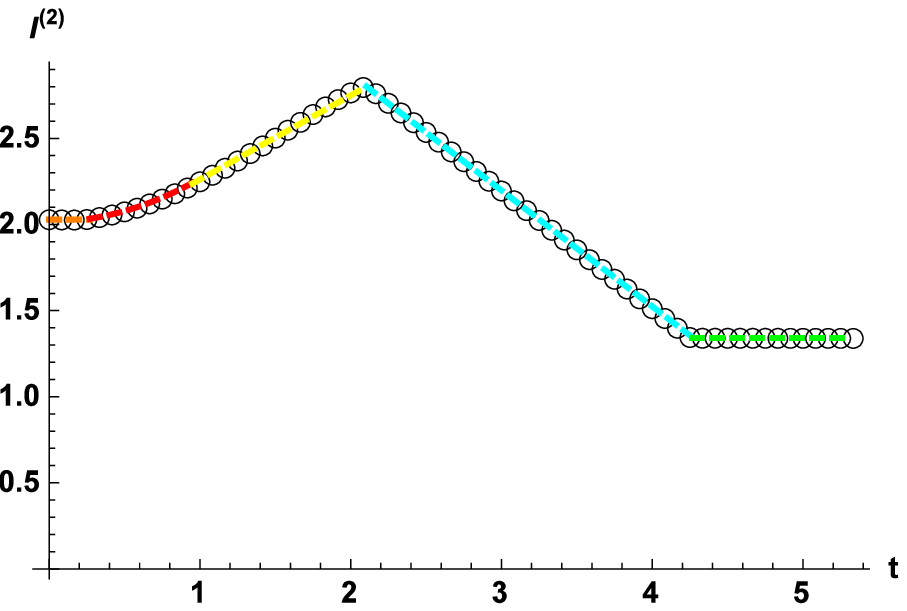}
\end{subfigure}
\begin{subfigure}
\centering
\includegraphics[scale=.75]{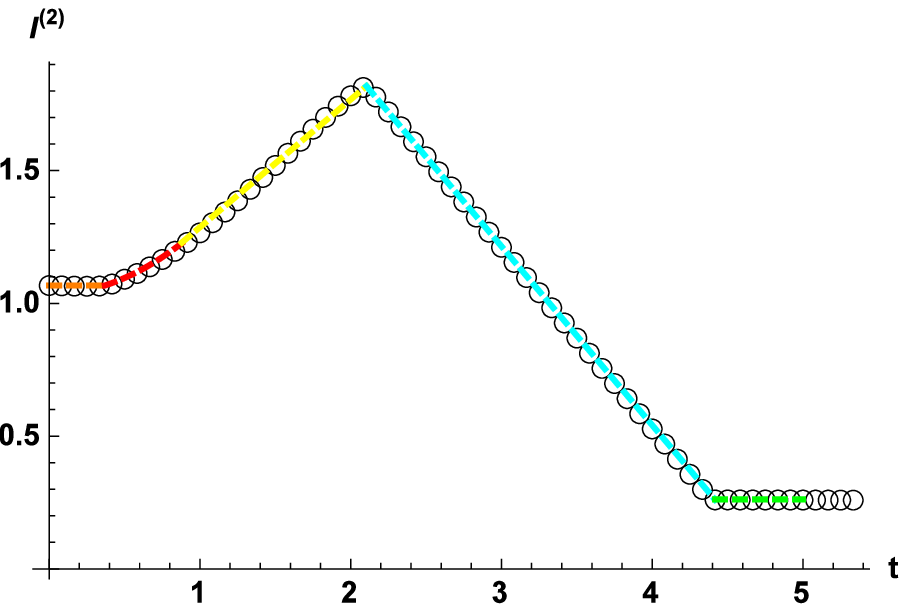}
\end{subfigure}
\caption{Evolution of $I^{(2)}$ for $\ell=3$ and $h=0.3, 0.5$ from left to right. In each plot the solid circles show the numerical result, the dashed orange curve corresponds to the steady behavior at early time, the dashed red curve corresponds to the quadratic growth, the dashed yellow curve corresponds to the linear growth, the dashed cyan curve corresponds to the linear decreasing and the dashed green line shows the saturation regime.}
\label{fig:Fit11}
\end{figure}

\begin{table}
\centering
\begin{tabular}{|c|c|c|c|c|c|c|c|c|c|}\hline
\multirow{2}{*}{$h$} &  \multicolumn{2}{c|}{ $a_1$ } & \multicolumn{2}{c|}{$b_1$} & \multicolumn{2}{c|}{$a_2$}& \multicolumn{2}{c|}{$b_2$}\\
\hhline{~---------}
        &            Ana. & Num. & Ana. & Num. & Ana. & Num. & Ana. & Num.\\\hline
             2.2  & -0.4749& -0.9221    & 0.6873& 0.6399       & 2.5841& 3.1230 & 0.6873& 0.6867         \\
2.6      & -0.6771& -1.1411    & 0.6873& 0.6419       & 2.3818& 2.9091 & 0.6873& 0.6867         \\\hline
\end{tabular}
\caption{Comparing specific values of the parameters $\{a_1,b_1,a_2,b_2\}$ for $\ell=4.5$.}
\label{tablefit2}
\end{table}
A similar analysis also works for $I^{(2)}$ (see Fig.\ref{fig:Fit11}). In this case 
using the following fit functions 
% According to table \ref{tablemutual1}, there is an excellent agreement  between analytic 
%expansions and numerical results of vacuum and saturation values for this quantity. So in this case 
%we only consider three different scaling regions, a quadratic growth, a linear growth and a linear 
%decreasing. Considering the fit functions as 
\bea
I^{(2)}_{\rm quad.}=c_1+d_1\; t^2,\;\;\;\;I^{(2)}_{\rm lin. grow.}=a_3+b_3\; t,\;\;\;\;I^{(2)}_{\rm lin. dec.}=a_4-b_4\; t,
\eea
and utilizing  eqs. \eqref{quadraticgrowth},\eqref{i2lingrowth} and \eqref{YY}, one arrives at the results presented in  table \ref{tablefit3}. One may go head to study  $I^{[n](i)}$, though 
the conclusions would be the same.

\begin{table}
\centering
\begin{tabular}{|c|c|c|c|c|c|c|c|c|c|c|c|c|c|}\hline
\multirow{2}{*}{$h$} &  \multicolumn{2}{c|}{ $c_1$ } & \multicolumn{2}{c|}{$d_1$} & \multicolumn{2}{c|}{$a_3$}& \multicolumn{2}{c|}{$b_3$}& \multicolumn{2}{c|}{$a_4$}& \multicolumn{2}{c|}{$b_4$}\\
\hhline{~------------}
        &            Ana. & Num. & Ana. & Num. & Ana. & Num. & Ana. & Num.& Ana. & Num.& Ana. & Num.\\\hline
             0.3  & 2.0156& 2.0144    & 0.25& 0.2355       & 2.0156& 1.7200 & 0.6873& 0.5221        &3.7341 &4.2385 & 0.6873&0.6816 \\
0.5      & 1.0332& 1.0325    & 0.25& 0.2354       & 1.0332& 0.7444 & 0.6873& 0.5181        &2.7517 &3.2571 & 0.6873&0.6822 \\\hline
\end{tabular}
\caption{Comparing specific values of the parameters $\{c_1,d_1,a_3,b_3,a_4,b_4\}$ for $\ell=3$.}
\label{tablefit3}
\end{table}

To conclude this section we observe that there is rather a good agreement between
numerical and semi-analytical results. We note however that due to the limitation of the numerical computation as well as the semi-analytical approximations, the actual values of distinctive points may not be precisely the same. 

\section{Discussions}

In this paper using the covariant prescription for computing the holographic entanglement entropy
we have studied mutual information and $n$-partite information (defined by equation \eqref{JAB})  for a strongly coupled field theory whose gravitational 
description is provided by an  AdS-Vaidya metric. 
We have computed 
the $n$-partite information for a system 
consisting of $n$ parallel strips (two for mutual information) with the same width $\ell$ separated
by distances $h$ with the condition $h\ll \ell$. With this assumption the expression of $n$-partite 
information is simplified so that in order to study its behavior, one  essentially needs to study entanglement entropy of three
strips with different widths.  Therefore it is possible to explore the evolution of the 
$n$-partite information during the process of thermalization after a global quantum quench, by making 
use of the results for the entanglement entropy\cite{{Liu:2013iza},{Liu:2013qca}}.

Of course  time evolution of the $n$-partite information is sensitive to 
the size of three entangling regions appearing in the computation of  $n$-partite information. 
Moreover the model has a distinctive time scale given by the horizon $\rho_H$ in which the theory
reaches a local equilibrium. Then the behavior depends on relative size of the  corresponding entangling regions and the radius of horizon: they could be  larger or smaller than $\rho_H$.  Therefore in the intermediate region  the $n$-pratite information
could increase (decrease) linearly or quadratically with time.

An interesting observation we have made is that the holographic $n$-partite information
has definite sign: it is positive for even $n$ and negative for odd $n$, though  for a generic field 
theory it could have either signs. Therefore following \cite{Hayden:2011ag} one may suspect that having definite sign for the 
$n$-partite information is, indeed, an intrinsic property of a field theory which has gravity dual.

We also examined  our analytical study by numerical computations. 
Actually for mutual information and 3-partite information some numerical computations have been performed in, {\it e.g.}  \cite{{Balasubramanian:2011at},{Allais:2011ys}}. Here, the corresponding numerical computations were studied in more details in order to explore different scaling behaviors of mutual information and 3-partite information. We have also considered 4 and 5-partite information, the generalization
to higher $n$ is straightforward. It should be mentioned that  in the range of the parameters which we are interested in, the numerical computations confirm our results. We have also numerically checked our assumptions under which the 
expression of $n$-partite information simplified drastically. We also compare these two approaches for computing time evolution of $n$-partite information. The numerical results, patch wise, are best fitted
with, quadratic, liner and constant curves, confirming the overall picture of the time dependent
behavior of n-partite information. 

Moreover we have studied the $n$-partite information \eqref{JAB} for a system consisting 
of  $n$ parallel strips with the same width separated by distances $h$. It is however,
instructive to explore the results for the case where the system is not symmetric. 
In other words, one may consider $n$ strips $A_i$ with width $\ell_i,\; i=1,\cdots, n$ separated by $h_j,\; j=1,\cdots, n-1$. Therefore the strips could have any size and are separated by 
arbitrary distances. Nevertheless, inspired by holographic mutual information, if
for arbitrary numbers $i, m, k$ and $j>1$  one assumes 
\bea
S(A_i\cup A_{i+1}\cdots \cup A_{i+k}\cup A_{i+k+j}\cup A_{i+k+j+1} \cdots \cup
A_{i+k+j+m})&=&
S(A_i\cup A_{i+1}\cdots \cup  A_{i+k})\\ &+&S( A_{i+k+j}\cup A_{i+k+j+1} \cdots \cup A_{i+k+j+m})
\nonumber
\eea
then the $n$-partite information \eqref{JAB} may be recast into the 
following form\footnote{It is important to note that the above assumption and  therefore 
such a simplification occurs due to the fact that the field theory we are considering  has a
holographic dual description in which the  holographic entanglement entropy
is given by the area of a minimal surface in the bulk.}
\be\label{genJ}
I^{[n]}=(-1)^{n-1}\bigg[S(A_2 \cdots \cup  A_{n-1})-S(A_1 \cdots \cup  A_{n-1})-S(A_2 \cdots \cup  A_{n})+S(A_1 \cdots \cup  A_{n})\bigg]
\ee
On the other hand by making use of the holographic description of entanglement entropy
and setting ${\cal L}=\ell_2+\cdots +\ell_{n-1}+h_2+\cdots +h_{n-2}$ one finds
\be\label{JABn}
I^{[n]}=(-1)^{n-1}\bigg[S({\cal L})-S({\cal L}+\ell_1+h_1)-S({\cal L}+\ell_n+h_{n-1})
+S({\cal L}+\ell_1+\ell_2+h_1+h_{n-1})\bigg],
\ee 
where, as before, $S(l)$ is the holographic entanglement entropy of a strip with the width
$l$. It is then easy to follow our discussions to find the behavior  of $n$-partite information
during a process of thermalization. Note that in this case we have five different cases depending on whether the radius of 
horizon $\rho_H$ is larger or smaller than the width of strips appearing in \eqref{JABn}.
Indeed  the general  rule is as follows. If the width of the entangling region appearing in the expression of $n$-partite information is smaller than the radius of horizon, the corresponding 
entanglement entropy grows quadratically with time and saturates before the system
reaches a local equilibrium
 \bea
\text{Early times} && S\sim S_{\rm vac}+V_{d-1}\;{\cal E} \;t^2,\cr
\text{Saturation}&& S\sim S_{\rm vac}+V_{d-1} \;{\cal E} \;\frac{\ell^2}{4},
\eea
where ${\cal E}$ is  the energy density. Note that since the system has not reached a local equilibrium, 
the energy density  is a proper  quantity one may define. On the other hand 
if the width of the entangling region is larger than the radius of horizon, the corresponding 
entanglement entropy grows quadratically with time before the system
reaches a local equilibrium, while it has linear growth after local equilibrium and then 
it saturates
 \bea
\text{Early times} && S\sim S_{\rm vac}+V_{d-1}\; {\cal E} \ t^2,\cr
\text{After local equ.} && S \sim S_{\rm vac}+V_{d-1}\;  S_{th} \; t,\cr 
\text{Saturation}&& S\sim S_{\rm vac}+V_{d-1} \;  S_{th} \;\frac{\ell}{2}-
 \frac{V_{d-2}}{\rho_H^{d-2}}.\nonumber
\eea
Note that when the system is locally equilibrated the entanglement entropy may be given in terms of the thermal entropy\footnote{Note that in the above schematic expressions  we have dropped 
the numerical factors.}. 

\begin{figure}[h]
\centering
\includegraphics[scale=0.9]{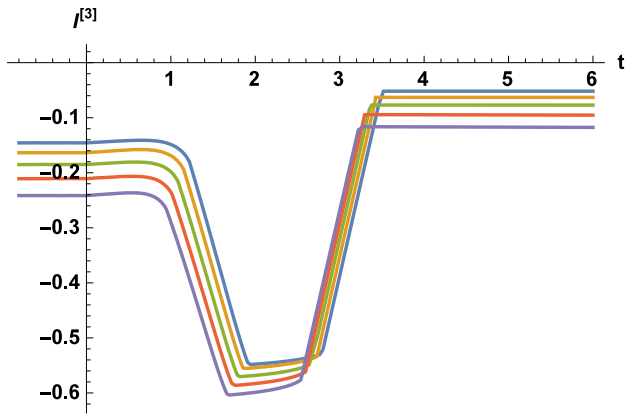}
\caption{Evolution of tripartite information for $h_1=0.1, h_2=0.5, \ell_1=1, \ell_2=2$ and $\ell_3=1.2, 1.3,..., 1.6$ increasing from bottom to top.}
\label{fig:3partitenonequalwidth}
\end{figure}

To conclude we note that for a system of $n$ parallel strips with different widths and distances between them,  one would still get the same behavior though the corresponding behavior is less symmetric around the maximum or minimum points. As an explicit example we have 
numerically computed 3-partite information for $h_1=0.1, h_2=0.5, \ell_1=1,\ell_2=2$ and 
several values for  $\ell_3$. The results are depicted in Fig.\ref{fig:3partitenonequalwidth}.

As we have already mentioned the holographic mutual information undergoes a first order
phase transition as one increases the distance between  two \cite{Headrick:2010zt}.
It is then natural to see whether such a transition would also occur for the 
$n$-partite information \eqref{JAB}. Actually in this case  it can be seen that the situation
 is very similar to that of mutual information. In other words the $n$-partite information
 vanishes as on increases the distance between the strips.  More precisely, 
if one changes the distance between given  two
 consecutive strips $A_i$ and $A_{i+1}$ in the system 
such that  $S(A_k,A_{k+1})=S(A_k)+S(A_{k+1})$, the  $n$-partite information \eqref{JAB} 
vanishes. More precisely using this identity and for $ 1\leq k<n$,  the equation \eqref{genJ} reads
\bea
I^{[n]}&=&(-1)^{n-1}\bigg[S(A_2 \cdots \cup A_k)+S(A_{k+1} \cdots \cup A_{n-1})
-S(A_1 \cdots \cup  A_{k})-S(A_{k+1} \cdots \cup  A_{n-1})\cr
&&\;\;\;\;\;\;\;\;\;\;\;\;\;\;\;\;
-S(A_2 \cdots \cup  A_{k})-S(A_{k+1} \cdots \cup  A_{n})+S(A_1 \cdots \cup  A_{k})+S(A_{k+1} \cdots \cup  A_{n})
\bigg],
\eea
which is zero, identically.

To conclude we have seen that if the mutual information of two consecutive strips of a
system consisting of $n$ parallel strip vanishes the $n$-partite information defined by
\eqref{JAB} vanishes too. Therefore there could be a phase  transition in $n$-partite
information if one increases the distance $h_i$. Since the $n$-partite information vanishes
when the mutual information vanishes, the critical  distance should be the same for both of them. More precisely for a given subsystem of $n$ parallel strips specified by $\ell_i, h_i$ and
$\ell_{i+1}$, there is a critical $h_i^{\rm c}$ over which both mutual and $n$-partite information 
vanish.  It is important to emphasis that this behavior is due to the facts that we are working with a field theory which has a holographic description and moreover  the $n$-partite information is defined by the
 equation \eqref{JAB}. In general field theory it might not be true and moreover, as we will see, this is also not true if one uses another  definition for $n$-partite such as that defined 
in equation  \eqref{IA}. Actually our numerical results concerning the contributions of different 
hypersurfaces in evaluating 3-partite information, indeed, supports the existence of the above phase transition. Definitely the phase transition of the $n$-partite information deserved more investigations. We hope to further study this phase transition in near future.

In this paper we have only considered $n$-partite information based on the definition \eqref{JAB}, though 
one could also study the behavior of multi-partite entanglement defined by equation \eqref{IA}.
Indeed in this case for the system we have been considering ( $n$ strips with width $\ell$ separated by 
$h$ with the condition $\ell\gg h$), equation \eqref{IA} reduces to the following expression
\bea
{J}^{[n]}(\ell)&=&nS(\ell)-S(n\ell +(n-1)h)-(n-1)S(h).
\eea
It is then easy to  show that  
\bea
\frac{\Delta  J^{[n]}}{\Delta E}=-\frac{8 n (n-1)\pi c_1}{d-1}\; \ell \left(1+\frac{h}{\ell}\right)^2,
\eea
where $\Delta J^{[n]}= J^{[n]}_{\text{BH}}- J^{[n]}_{\text{vac}}$. It is worth nothing that the above expression
is the same as that of mutual information \eqref{FirstM} up to a factor of $n(n-1)$. Actually the behavior 
of the above quantity in a process of thermalization is very similar to that of mutual information.
In particular one can show that it is always positive during the process of thermalization. 

It is also easy to see that unlike the previous case, in the present case when 
the mutual information of two consecutive strips becomes zero, the multi-partite 
information does not vanish and instead it breaks into two multi-partite
informations. More precisely suppose $I^{[2]}(A_k,A_{k+1})=0$, then 
from the definition of multi-partite \eqref{IA} one gets
\bea
J^{[n]}&=&\sum_{i}^kS(A_i)+\sum_{i}^{n} S(A_i)-S(A_1\cdots\cup A_{k})+S(A_{k+1}
\cdots\cup A_{n})\cr &=&
J^{[k]}+J^{[n-k]}.
\eea

\section*{Acknowledgements}
We would like to thank Ali Mollabahi and Amin Faraji Astaneh for useful discussions. M. A. would like 
to thank ICTP for very warm hospitality. M. R. M. M. would like to thank E. Tonni and P. Fonda for useful discussions about the numerical results and also J. F. Pedraza for useful comments about numerical method and also sharing a Mathematica code. M. R. M. M. would like 
to thank ICTP and SISSA for very warm hospitality during the last stage of this project. We would also like to thank the 
referee for his/her comments. In particular a comment about different sizes of the widths which leads us to 
explore a possible phase transition for the n-partite information. This work is supported by Iran
National Science Foundation (INSF).

%%%%%%%%%%%%%%%%%%%%%%%%%%%%%%%%%%%%%%%%%%%%%%%%%%%%%%%%%%%%%%%%%%%%%%%%%

\section*{Appendix}
\appendix
\section{Entanglement entropy } \label{App:AppendixA}

In this appendix  we will review the holographic computation of the entanglement entropy of a strip
 for the cases where  the dual gravitational descriptions are given by an AdS solution, AdS black brane
and AdS-Vaidya metric. From AdS/CFT corresponding point of view these will give us the entanglement 
entropy of the ground state of a CFT, a thermal state of a CFT and in the global quantum quench, respectively. It is worth noting that although in the first two cases the system is static, for the last 
one we will have to deal with a time dependent process. In what follow we will review both cases 
separately.

To proceed let us consider a strip with the width $\ell$ in a $d$-dimensional space time as follows
\be\label{strip}
-\frac{\ell}{2}\leq x_1\leq \frac{\ell}{2},\;\;\;\;\;\;\;\;t={\rm fixed},\;\;\;\;\;\;\;0\leq x_a\leq L,\;\;\;\;\;
{\rm for}\;\;a=2,d-1,
\ee
where $(t, \vec{x})$ are the space time coordinates.

\subsection{Static background}

Let us first compute the entanglement entropy for  a $d$-dimensional static system whose gravitational
dual is provided by the metric of \eqref{BB}.
Following the holographic description of the entanglement entropy \cite{{RT:2006PRL},{RT:2006}} one needs
to minimize the area of a co-dimension two hypersurface whose boundary coincides with the boundary
of the above strip. The profile of  corresponding hypersurface in the bulk may be 
parametrized by  $x_1=x(\rho)$, and therefore the area functional is
\be
A_{\text{vac}}=\frac{L^{d-2}}{2}\int{d\rho \frac{\sqrt{f^{-1}+x'^2}}{\rho^{d-1}}}.
\ee
where ``prime''  represents derivative with respect to $\rho$. It is then straightforward to minimize the above area to arrive at
\be\label{G0}
\frac{\ell}{2}=\int_0^{\rho_t}d\rho \frac{\left(\frac{\rho}{\rho_t}\right)^{d-1}}{
\sqrt{f(\rho)\left(1-\left(\frac{\rho}{\rho_t}\right)^{2(d-1)}\right)}},\;\;\;\;\;\;\;\;\;
S=\frac{L^{d-2}}{4G_N}\int_\epsilon^{\rho_t}d\rho \frac{1}{\rho^{d-1}
\sqrt{f(\rho) \left(1-\left(\frac{\rho}{\rho_t}\right)^{2(d-1)}\right)}}
\ee
where $\rho_t$ is the extremal hypersurface
turning point in the bulk and $\epsilon$ is a UV cut-off.

 For $f=1$ which corresponds to a vacuum solution one finds \cite{RT:2006}
\be\label{SV}
S_{\text{vac}}=\left\{\ba{ll} 
\frac{L^{d-2}}{4G_N}\left(\frac{1}{(d-2)\epsilon^{d-2}}-\frac{c_0}{\ell^{d-2}}\right)
&{\rm for}\;\;d> 2,\cr &\cr
\frac{1}{4G_N}\ln \frac{\ell}{\epsilon},&  {\rm for}\;\;d= 2,
\ea \right.
\ee
where $c_0=\frac{2^{d-2}}{d-2}\left(\sqrt{\pi}\Gamma(\frac{d}{2(d-1)})/\Gamma(\frac{1}{2(d-1)})\right)^{d-1}$.

For an excited state whose gravitational dual is provided by the black brane solution 
\eqref{BB} the corresponding entanglement entropy may be found by minimizing the area 
when $f\neq 1$. In this case, in general, it is not possible to find an explicit expression for 
the  entanglement entropy, though in certain limits one may extract the general behavior of 
the entanglement entropy.   In particular in the limit of  $ml^d \ll 1$, one finds
\be
\Delta {A}=\frac{L^{d-2}}{2}\int d\rho\;\; \delta_f\left(\frac{\sqrt{{f}^{-1}+x'^2}}{\rho^{d-1}}\right)\bigg|_{f=1}\Delta f,
\ee
 which leads to the following expression for the entanglement entropy 
\be\label{SBH0}
 S_{\text{BH}}=S_{\text{vac}}+\frac{L^{d-2}}{4G_N} c_1m\ell^2,
\ee
where $S_{\rm vac}$ is the entanglement entropy of the vacuum solution given in \eqref{SV}, and
\be
c_1=\frac{1}{16(d+1)\sqrt{\pi}}\;\frac{\Gamma(\frac{1}{2(d-1)})^2\Gamma(\frac{1}{d-1})}{\Gamma(\frac{d}{2(d-1)})^2\Gamma(\frac{1}{2}+\frac{1}{d-1})}.
\ee

On the other hand for $m\ell^d\gg 1$ 
 the main contributions to the entanglement entropy comes from the limit where the minimal surface
is extended all the way to the  horizon so that $\rho_t\sim \rho_H$. In this limit equation
\eqref{G0} for $d>2$ reads
\be
\frac{\ell}{2}\approx\rho_H \int_0^{1} \frac{\xi^{d-1}d\xi}{
\sqrt{(1-\xi^d)\left(1-\xi^{2(d-1)}\right)}},\;\;\;\;\;\;\;\;\;
S_{\rm BH}\approx\frac{L^{d-2}}{4G_N\rho_H^{d-2}}\int_{\frac{\epsilon}{\rho_H}}^{1} \frac{d\xi}{\xi^{d-1}
\sqrt{(1-\xi^d) \left(1-\xi^{2(d-1)}\right)}}.
\ee
Note that apart from the UV divergent term in $S_{\rm BH}$, due to the double zero in the square roots, the main contributions in the above integrals come from $\xi=1$ point. Indeed around $\xi=1$ 
the entanglement entropy $S_{\rm BH}$ may be recast to the following form 
\be
S_{\rm BH}\approx\frac{L^{d-2}}{4G_N\rho_H^{d-2} }\left(
 \int_0^{1} \frac{\xi^{d-1}d\xi}{
\sqrt{(1-\xi^d)\left(1-\xi^{2(d-1)}\right)}}+
\int_{\frac{\epsilon}{\rho_H}}^{1} d\xi\; \frac{\sqrt{1-\xi^{2(d-1)}}}{\xi^{d-1}
\sqrt{1-\xi^d }}\right).
\ee
It is now clear that the first term in the above equation is divergent at $\xi=1$ while the second
one is finite, though the second term is UV divergent. Indeed the first term is exactly the one
appears for $\ell$. Therefore one has
\be
S_{\rm BH}\approx\frac{L^{d-2}}{4G_N\rho_H^{d-1} }\;\frac{\ell}{2}
 +\frac{L^{d-2}}{4G_N\rho_H^{d-2} }
\int_{\frac{\epsilon}{\rho_H}}^{1} d\xi\; \frac{\sqrt{1-\xi^{2(d-1)}}}{\xi^{d-1}
\sqrt{1-\xi^d }}.
\ee
Now the aim is to compute the  the second integral. Of course it can not been performed analytically, though one may solve it numerically to find its finite part. Indeed using ``NIntegrate'' command in the 
{\it Mathematica} one finds
\be
\int_{\epsilon}^{1} d\xi\; \frac{\sqrt{1-\xi^{2(d-1)}}}{\xi^{d-1}
\sqrt{1-\xi^d }}=\frac{1}{(d-2)\epsilon^{d-2}}-c_2,
\ee
where $c_2$ is a positive number. For example for $d=3,4$  one gets
$c_2=0.88, 0.33$, respectively. Therefore one arrives at \cite{Fischler:2012ca}
\be\label{SBH1}
S_{\rm BH}\approx\frac{L^{d-2}}{4G_N }\left(\frac{1}{(d-2)\epsilon^{d-2}}+\frac{\ell}{2\rho_H^{d-1}}
-\;\frac{c_2}{\rho_H^{d-2}}\right).
\ee
Note that the first finite term in the above expression is proportional to the volume which is
indeed the thermal entropy, while the second finite term is proportional to the area of
the entangling region. Indeed this term plays a crucial  role in our study.

\subsection{Time dependent background}

In this subsection we will review computations of the holographic entanglement entropy in the AdS-Vaidya
background  \eqref{Vaidya} for the case where the size of the entangling region
is larger than the radius of horizon\cite{{Liu:2013iza},{Liu:2013qca}}. More  precisely the entangling region 
is given by a strip given in equation \eqref{strip} with $\ell\gg \rho_H$. As mentioned before for this time dependent background the covariant proposal
for the holographic entanglement entropy is needed and the $v(x)$ and $\rho(x)$ may be used to parametrize the corresponding co-dimension two hypersurface in the bulk. 
 Then the induced metric on the hypersurface is
\be
ds_{\rm ind}^2=\frac{1}{\rho^2}\bigg[\bigg(1-f(\rho,v){v'}^2-2  v'\rho'\bigg) dx^2+d\vec{x}^2\bigg],
\ee
where ``prime'' represents derivative with respect to $x$. Therefore, the hypersurface's area can be obtained as
\be\label{area00}
A=\frac{L^{d-2}}{2}\int_{-\ell/2}^{\ell/2} dx\; \frac{\sqrt{1-2v'\rho'- v'^2 f}}{\rho^{d-1}}\equiv\frac{L^{d-2}}{2}\int_{-\ell/2}^{\ell/2} dx\; \frac{{\cal L}}{\rho^{d-1}},
\ee
the corresponding entanglement entropy can then be found after evaluating (\ref{area00})
at the extremal surface as follows
\be\label{Qun}
S(t)=\frac{A(t)}{4G_N}.%\;\;\;\;\;\;\;\;\;\;\;\;\;\;\;\;\;\;\langle W(t)\rangle \sim e^{-\frac{\mathcal{A}(t)}{2\pi\alpha'}},
%\;\;\;\;\;\;\;\;\;\;\;\;\;\;\;\;\;\;G(\ell, t)\sim e^{-M {\mathcal{A}(\ell/2, t)}},
\ee
Note that (\ref{area00}) may be thought of as a one dimensional action for a dynamical system for the fields $v(x)$ and $\rho(x)$. Since the action is independent of $x$ its corresponding Hamiltonian is a constant of motion
\bea\label{J}
\rho^{d-1}{\cal L}=H={\rm constant}.
\eea    
This conservation law helps one to write equations of motion for $v$ and $\rho$ which read as 
\be
\partial_x P_v
=\frac{P_\rho^2}{2}\;\frac{\partial f}{\partial v},\;\;\;\;\;\;\;\;
\partial_x P_\rho=
\frac{P_\rho^2}{2}\;\frac{\partial f}{\partial \rho}+\frac{d-1}{\rho^{2d-1}}H^2
P_\rho P_v,
\ee
where $P$'s are the momenta conjugate to $v$ and $\rho$ up to a factor of $H^{-1}$ and are defined by
\be
P_v=\rho'+ v'f,\;\;\;\;\;\;\;\;\;\;\;\;P_\rho=v'.
\ee
These equations have to be solved by the following boundary conditions 
\begin{eqnarray}
\label{bdycondition}
\rho(\frac{\ell}{2})&=&0,\;\;\;\;\;\;\;\;v(\frac{\ell}{2})=t,\;\;\;\;\;\;\;\;\rho'(0)=0,\nonumber\\
v'(0)&=&0,\;\;\;\;\;\;\;\;\rho(0)=\rho_t,\;\;\;\;\;\;\;v(0)=v_t,
\end{eqnarray} 
Note that with this boundary condition one obtains $H=\rho_t^{d-1}$, where $(\rho_t,v_t)$ stands for the turning point coordinate of the extremal hypersurface in the bulk.

One should solve equations to find the extremal surface and the numerical method is actually needed, however, analytic solutions can still be found for some particular forms of $m(v)$. It is known that in a quench there is a rapid change in the theory so that, one may assume that   
$f(\rho,v)=1-\theta(v) g(\rho)$, where $\theta(v)$ is the step function. 
This implies that $f$ does not depend on $v$ in most of time and hence $\frac{\partial f(\rho,v)}{\partial v}=0$, consequently, the momentum conjugate of $v$ becomes a constant of motion
\bea\label{Pvconst}
P_v=\rho'+ v'{\tilde f}(\rho)={\rm constant}, \;\;\;\;\;{\rm with}\;\;
{\tilde f}(\rho)=1-g(\rho).
\eea
In the present case one has $g(\rho)=(\frac{\rho}{\rho_H})^{d}$ where the horizon locates at $\rho_H$ with $m=\frac{1}{\rho_H^{d}}$.

For $v<0$ one has $f=1$ and therefore the geometry is actually  an AdS geometry which corresponds to the vacuum. In this case one gets
\be
P_{(i) v}=\rho'+v'=0,
\ee
which together with the conservation law \eqref{J} yields to the following profile of the extremal surface
\be\label{eq5}
v(\rho)=v_t+\rho_t-\rho,\hspace*{1cm}x(\rho)=\int_\rho^{\rho_t}\frac{dz \;z^{d-1}}{\sqrt{\rho_t^{2(d-1)}-z^{2(d-1)}}}.
\ee
Note that since the  position of the null shell is $v=0$,  from equation \eqref{eq5}, one gets
\be
\rho_c=\rho_t+ v_t
\ee
which, indeed, gives the point where the hypersurface intersects the null shell. 
Moreover, by making use of equation \eqref{J} at $v<0$, one finds
\be
\rho_{(i)}'=-v_{(i)}'=-\sqrt{\left(\frac{\rho_t}{\rho_c}\right)^{2(d-1)}-1}
\ee

On the other hand for $v>0$ one has $f={\tilde f}(\rho)=1-g(\rho)$ and therefore the 
corresponding geometry is an AdS black brane. By  making use of equations \eqref{Pvconst} and \eqref{J}, at the back brane side, one obtains
\be\label{oneD}
\rho'^2=P_{(f) v}^2+\left(\left(\frac{\rho_t}{\rho}\right)^{2(d-1)}-1\right){\tilde f}(\rho)\equiv V_{eff}(\rho),
\ee
which can also be used to find
\be
\frac{dv}{d\rho}=-\frac{1}{{\tilde f}(\rho)}\left(1+\frac{P_{(f)v}}{\sqrt{V_{eff}(\rho)}}\right),
\ee
where $(\rho_t, v_t)$ is the extremal hypersurface
turning point in the bulk and the crossing point where 
the hypersurface intersects the null shell is given by $(\rho_c,v_c)$. 
Since one is  injecting the matter in $v$ direction, one would expect that its corresponding 
momentum conjugate jumps once one moves from the initial phase to the final phase. While
the momentum conjugate of $\rho$ must be continuous. Therefore one gets $v'_{(f)}=v'_{(i)}$.
On the other hand  by integrating equations of motion across the null shell one  arrives at 
\be
\rho_{(f)}' =\left(1-\frac{1}{2}g(\rho_c)\right) \rho_{(i)}',\hspace*{1cm}{\cal L}_{(f)}={\cal L}_{(i)}.
\ee
It is, then, easy to read the momentum conjugate of $v$ in the final phase
\be
P_{(f) v}=\frac{1}{2}g(\rho_c)\rho_{(i)}'=-\frac{1}{2}g(\rho_c)\sqrt{\left(\frac{\rho_t}{\rho_c}\right)^{2(d-1)}-1}.
\ee

Now we have all ingredients to find the area of the corresponding extremal  hypersurface in the bulk.
In general the hypersurface could extend in both $v<0$ and  $v>0$ regions 
of space-time. Therefore the width $\ell$ and the boundary time are found as follows
\be\label{general1}
\frac{\ell}{2}=\int_{\rho_c}^{\rho_t}\frac{d\rho\;\rho^{d-1}}{\sqrt{\rho_t^{2(d-1)}-\rho^{2(d-1)}}}+\int_0^{\rho_c}\frac{d\rho}{\sqrt{V_{eff}(\rho)}},\;\;\;\;\;
t=\int_0^{\rho_c}\frac{d\rho}{\tilde f(\rho)}\left(1+\frac{E}
{\sqrt{V_{eff}(\rho)}}\right),
\ee
where $E=P_{(f)v}$ which in the large entangling region limit becomes
\bea\label{largesimp2}
E=-\left(\frac{\rho_t}{\rho_m}\right)^{d-1}\sqrt{-\tilde{f}(\rho_m)}.
\eea
Finally, the area reads
\bea\label{general2}
\frac{{ A}}{L^{d-2}}=\int_{\rho_c}^{\rho_t}{\frac{\rho_t^{d-1}d\rho}{\rho^{d-1}\sqrt{\rho_t^{2(d-1)}-\rho^{2(d-1)}}}}+\rho_t^{d-1} \int_0^{\rho_c}{\frac{d\rho}{\rho^{2(d-1)}\sqrt{V_{eff}(\rho)}}}.
\eea
Using the above expressions for $t, \ell$ and $A$ one may find the scaling behavior of the 
entanglement entropy during the process of thermalization.
Here, we will only present the final results which have been obtained in \cite{{Liu:2013iza},{Liu:2013qca}}.

At the early time where  $t \ll \rho_H$ the crossing point of the hypersurfaces is very close to the boundary,
$ \frac{\rho_{c}}{\rho_H}\ll 1$. Therefore one may expand $t, \ell$, and $A$  leading to 
\bea
t&\approx& 
%= \int_0^{\rho_{c}}{d\rho \left(1+\left(\frac{\rho}{\rho_H}\right)^d+\left(\frac{\rho}{\rho_H}\right)^{2d}+...%\right)}
\rho_{c}\left(1+\frac{1}{d+1}\left(\frac{\rho_{c}}{\rho_H}\right)^d+\frac{1}{2d+1}\left(\frac{\rho_{c}}{\rho_H}\right)^{2d}+...\right),\;\;\;\;\;\;
\frac{\ell}{2}\approx\rho_{t} \left(c+\frac{m}{4d}\frac{\rho_c^{2d}}{\rho_{t}^d}+...\right),\nonumber\\
A%&\approx&\frac{L^{d-2}}{(d-2)}\left(\frac{1}{\epsilon^{d-2}}-c\frac{1}{\rho_{t}^{d-2}}\right) %+\frac{L^{d-2}}{2\rho_H^d}\int_0^{\rho_{c}}\frac{\rho d\rho}{\sqrt{1-\left(\frac{\rho}
%{\rho_{t}}\right)^{2(d-1)}}}
&\approx &\frac{L^{d-2}}{(d-2)}\left(\frac{1}{\epsilon^{d-2}}-c\frac{1}{\rho_{t}^{d-2}}\right) +\frac{L^{d-2}m}{4}\rho_{c}^{2}\left(1+\frac{1}{2d}\left(\frac{\rho_{c}}{\rho_{t}}\right)^{2(d-1)}+...\right),
\eea
where $c=\sqrt{\pi}\frac{\Gamma(\frac{d}{2(d-1)})}{\Gamma(\frac{1}{2(d-1)})}$.
So that at leading order one finds
\be\label{SBHET}
S \approx \frac{L^{d-2}}{4G_N}\left[\frac{1}{(d-2)\epsilon^{d-2}}-\frac{c_0}{\ell^{d-2}}+\frac{t^2}{4 \rho_H^d}+{\cal O}(t^{d+2})\right].
\ee

On the other hand in the intermediate time interval where $\rho_H\ll t\ll \frac{\ell}{2}$, the entanglement 
entropy growth linearly with time. Indeed it was shown \cite{{Liu:2013iza},{Liu:2013qca}}
that there is a critical extremal surface which is responsible for the linear growth in this time interval.
More precisely, $V_{eff}(\rho)$ defined in equation \eqref{oneD} might be thought of as an effective potential for a one dimensional 
dynamical system whose dynamical variable is $\rho$. Actually for a fixed extremal hypersurface turning point in the bulk, $\rho_t$, there is a free parameter in the effective potential
given by $\rho_c$ which 
may be tuned to a particular value $\rho_c=\rho^*_c$ such that the minimum of the 
effective potential becomes zero. In other words, one has
\bea\label{criticalhyper}
\frac{\partial V_{eff}(\rho)}{\partial\rho}\bigg|_{\rho_m,\rho^*_c}=0,\;\;\;\;\;\;\;\;\;\;\;\;
V_{eff}(\rho)|_{\rho_m,\rho^*_c}=0.
\eea
If the hypersurface intersects the null shell at the critical point it remains fixed at 
$\rho_m$. Therefore in the intermediate time interval the main contributions to $\ell, t$
and $A$ come from a hypersurface which is closed to the critical extremal hypersurface.
In this case assuming  $ \rho_{c}= \rho^*_{c}(1-\delta)$ for $\delta \ll1$ in the limit of  $\rho \rightarrow \rho_{m}$ and
with the conditions
$\frac{\rho_{c}^*}{\rho_{t}},\frac{\rho_{m}}{\rho_{t}}\ll 1$ equations \eqref{general1} and
\eqref{general2} may be approximated as follows\cite{{Liu:2013iza},{Liu:2013qca}}
\bea
t&\approx& -\frac{E^*}{\tilde{f}(\rho_{m})\sqrt{\frac{1}{2}V''_{eff}}}\log \delta,\;\;\;\;\;\;\;\;\;
\frac{\ell}{2}\approx c \rho_{t}+\frac{\tilde{f}(\rho_{m})}{E^*}t\nonumber\\
A&\approx&\frac{L^{d-2}}{(d-2)}\left(\frac{1}{\epsilon^{d-2}}-c\frac{1}{\rho_{t}^{d-2}}\right) -\frac{L^{d-2}\rho_{t}^{d-1}}{\rho_{m}^{2(d-1)}\sqrt{\frac{1}{2}V''_{eff}}}\log \delta
\eea
where $E^*\equiv E(\rho_{c}^*)$. Therefore using \eqref{largesimp2} the entanglement entropy reads
\be\label{SBHL}
S\approx\frac{L^{d-2}}{4G_N}\left[\frac{1}{(d-2)\epsilon^{d-2}}-\frac{c_0}{\ell^{d-2}}+ \frac{\sqrt{-\tilde{f}(\rho_{m})}}{\rho_{m}^{d-1}}t+\cdots\right].
\ee
Now using \eqref{SV} for $d>2$ and above equation one finds
\bea
S-S_{\rm vac}=L^{d-2}\mathcal{S}v_E t+\cdots,
\eea
where $\mathcal{S}=\frac{1}{4G_N\rho_H^{d-1}}$ is thermal entropy density and $v_E$ is entanglement velocity which is given by
\bea\label{VE}
v_E=\frac{\rho_{H}^{d-1}}{\rho_{m}^{d-1}}\sqrt{-\tilde{f}(\rho_{m})}.
\eea

Note that $\rho_m$ and $\rho^*_c$ can also be obtained in terms of the radius of horizon using equation \eqref{criticalhyper}. In particular for large entangling region (or large $\rho_t$) 
assuming that both $\rho_m$ and $\rho_c^*$ remain finite (which is the case in the system we are
considering) one gets
\bea\label{largesimp1}
\frac{\rho_m}{\rho_H}=\left(\frac{2(d-1)}{d-2}\right)^{1/d},\;\;\;\;\;\;\;\;\;\;\;\;
\frac{\rho_c^*}{\rho_H}=2\sqrt{\frac{d}{d-2}}\left(\frac{d-2}{2(d-1)}\right)^{1-1/d}.
\eea
In this limit the expression for the entanglement velocity simplified as follows
\bea\label{VEsimple}
v_E=\frac{\sqrt{d/(d-2)}}{(\frac{2(d-1)}{(d-2)})^{(d-1)/d}}.
\eea
Finally if one waits enough the entanglement entropy will be saturated to its thermal value which 
is essentially given by \eqref{SBH1} and the saturation time approximated by
\bea\label{tsfinal}
t_s \sim \frac{\ell}{2v_E}.
\eea


\begin{thebibliography}{99}




%\cite{Srednicki:1993im}
\bibitem{Srednicki:1993im} 
  M.~Srednicki,
  ``Entropy and area,''
  Phys.\ Rev.\ Lett.\  {\bf 71}, 666 (1993)
  [hep-th/9303048].
  %%CITATION = HEP-TH/9303048;%%
  %534 citations counted in INSPIRE as of 24 Mar 2014

%\cite{Holzhey:1994we}
\bibitem{Holzhey:1994we} 
  C.~Holzhey, F.~Larsen and F.~Wilczek,
  ``Geometric and renormalized entropy in conformal field theory,''
  Nucl.\ Phys.\ B {\bf 424}, 443 (1994)
  [hep-th/9403108].
  %%CITATION = HEP-TH/9403108;%%
  %296 citations counted in INSPIRE as of 24 Mar 2014


%\cite{Calabrese:2004eu}
\bibitem{Calabrese:2004eu} 
  P.~Calabrese and J.~L.~Cardy,
  ``Entanglement entropy and quantum field theory,''
  J.\ Stat.\ Mech.\  {\bf 0406}, P06002 (2004)
  [hep-th/0405152].
  %%CITATION = HEP-TH/0405152;%%
  %244 citations counted in INSPIRE as of 24 Mar 2014



%\cite{Caraglio:2008pk}{Furukawa:2008uk}{Calabrese:2009ez}{Calabrese:2010he}
\bibitem{Caraglio:2008pk}
  M.~Caraglio and F.~Gliozzi,
  ``Entanglement Entropy and Twist Fields,''
  JHEP {\bf 0811} (2008) 076
  [arXiv:0808.4094 [hep-th]].
  %%CITATION = ARXIV:0808.4094;%%
  %30 citations counted in INSPIRE as of 24 Mar 2014


%\cite{Furukawa:2008uk}
\bibitem{Furukawa:2008uk} 
  S.~Furukawa, V.~Pasquier and J.~'i.~Shiraishi,
  ``Mutual Information and Compactification Radius in a c=1 Critical Phase in One Dimension,''
  Phys.\ Rev.\ Lett.\  {\bf 102}, 170602 (2009)
  [arXiv:0809.5113 [cond-mat.stat-mech]].
  %%CITATION = ARXIV:0809.5113;%%
  %35 citations counted in INSPIRE as of 24 Mar 2014



%\cite{Calabrese:2009ez}{Calabrese:2010he}
\bibitem{Calabrese:2009ez} 
  P.~Calabrese, J.~Cardy and E.~Tonni,
  ``Entanglement entropy of two disjoint intervals in conformal field theory,''
  J.\ Stat.\ Mech.\  {\bf 0911}, P11001 (2009)
  [arXiv:0905.2069 [hep-th]].
  %%CITATION = ARXIV:0905.2069;%%
  %40 citations counted in INSPIRE as of 24 Mar 2014


%\cite{Calabrese:2010he}
\bibitem{Calabrese:2010he} 
  P.~Calabrese, J.~Cardy and E.~Tonni,
  ``Entanglement entropy of two disjoint intervals in conformal field theory II,''
  J.\ Stat.\ Mech.\  {\bf 1101}, P01021 (2011)
  [arXiv:1011.5482 [hep-th]].
  %%CITATION = ARXIV:1011.5482;%%
  %26 citations counted in INSPIRE as of 24 Mar 2014

%\cite{Hayden:2011ag}
\bibitem{Hayden:2011ag} 
  P.~Hayden, M.~Headrick and A.~Maloney,
  ``Holographic Mutual Information is Monogamous,''
  Phys.\ Rev.\ D {\bf 87}, no. 4, 046003 (2013)
  [arXiv:1107.2940 [hep-th]].
  %%CITATION = ARXIV:1107.2940;%%
  %27 citations counted in INSPIRE as of 23 May 2014


%\cite{Horodecki:2009zz}
\bibitem{Horodecki:2009zz} 
  R.~Horodecki, P.~Horodecki, M.~Horodecki and K.~Horodecki,
  ``Quantum entanglement,''
  Rev.\ Mod.\ Phys.\  {\bf 81}, 865 (2009)
  [quant-ph/0702225].
  %%CITATION = QUANT-PH/0702225;%%
  %101 citations counted in INSPIRE as of 26 Jun 2014


\bibitem{CC}
 P.~Calabrese and J.~L.~Cardy,  ``Evolution of Entanglement Entropy in One-Dimensional Systems,''
 J.\ Stat.\ Mech.\ {\bf 0504}, P04010 (2005), 
arXiv:cond-mat/0503393 [cond-mat].


\bibitem{Maldacena:1997}
J. M. Maldacena,
"The large N limit of superconformal field theories and supergravity,''
Adv.\ Theor.\ Math.\ Phys.\  {\bf 2}, 231 (1998)
[Int.\ J.\ Theor.\ Phys.\  {\bf 38}, 1113 (1999)]
[hep-th/9711200].
%%CITATION = IJTPB,38,1113;%%



\bibitem{RT:2006PRL}
S.~Ryu and T.~Takayanagi,
"Holographic Derivation of Entanglement Entropy from AdS/CFT,''
Phys. Rev. Lett. {\bf 96} (2006) 181602
[hep-th/0603001].
%%CITATION = PHRVA,L96,181602;%%

\bibitem{RT:2006}
S.~Ryu and T.~Takayanagi,
"Aspects of Holographic Entanglement Entropy,''
JHEP {\bf 0608} (2006) 045
[hep-th/0605073].
%%CITATION = JHEPA,0608,045;%%


%\cite{Hubeny:2007xt}
\bibitem{Hubeny:2007xt} 
  V.~E.~Hubeny, M.~Rangamani and T.~Takayanagi,
  ``A Covariant holographic entanglement entropy proposal,''
  JHEP {\bf 0707}, 062 (2007)
  [arXiv:0705.0016 [hep-th]].
  %%CITATION = ARXIV:0705.0016;%%
  %148 citations counted in INSPIRE as of 19 Dec 2013



%\cite{AbajoArrastia:2010yt}
\bibitem{AbajoArrastia:2010yt} 
  J.~Abajo-Arrastia, J.~Aparicio and E.~Lopez,
  ``Holographic Evolution of Entanglement Entropy,''
  JHEP {\bf 1011}, 149 (2010)
  [arXiv:1006.4090 [hep-th]].
  %%CITATION = ARXIV:1006.4090;%%
  %72 citations counted in INSPIRE as of 03 Jan 2014


%\cite{Albash:2010mv}
\bibitem{Albash:2010mv} 
  T.~Albash and C.~V.~Johnson,
  ``Evolution of Holographic Entanglement Entropy after Thermal and Electromagnetic Quenches,''
  New J.\ Phys.\  {\bf 13}, 045017 (2011)
  [arXiv:1008.3027 [hep-th]].
  %%CITATION = ARXIV:1008.3027;%%
  %62 citations counted in INSPIRE as of 03 Jan 2014

%\cite{Balasubramanian:2010ce}
\bibitem{Balasubramanian:2010ce} 
  V.~Balasubramanian, A.~Bernamonti, J.~de Boer, N.~Copland, B.~Craps, E.~Keski-Vakkuri, B.~Muller and A.~Schafer {\it et al.},
  ``Thermalization of Strongly Coupled Field Theories,''
  Phys.\ Rev.\ Lett.\  {\bf 106}, 191601 (2011)
  [arXiv:1012.4753 [hep-th]].
  %%CITATION = ARXIV:1012.4753;%%
  %91 citations counted in INSPIRE as of 03 Jan 2014



%\cite{Aparicio:2011zy}{Galante:2012pv}{Caceres:2012em}{Baron:2012fv}
\bibitem{Aparicio:2011zy} 
  J.~Aparicio and E.~Lopez,
  ``Evolution of Two-Point Functions from Holography,''
  JHEP {\bf 1112}, 082 (2011)
  [arXiv:1109.3571 [hep-th]].
  %%CITATION = ARXIV:1109.3571;%%
  %28 citations counted in INSPIRE as of 23 Aug 2013

%\cite{Galante:2012pv}
\bibitem{Galante:2012pv} 
  D.~Galante and M.~Schvellinger,
  ``Thermalization with a chemical potential from AdS spaces,''
  JHEP {\bf 1207}, 096 (2012)
  [arXiv:1205.1548 [hep-th]].
  %%CITATION = ARXIV:1205.1548;%%
  %38 citations counted in INSPIRE as of 03 Jan 2014

%\cite{Caceres:2012em}
\bibitem{Caceres:2012em} 
  E.~Caceres and A.~Kundu,
  ``Holographic Thermalization with Chemical Potential,''
  JHEP {\bf 1209}, 055 (2012)
  [arXiv:1205.2354 [hep-th]].
  %%CITATION = ARXIV:1205.2354;%%
  %31 citations counted in INSPIRE as of 03 Jan 2014

%\cite{Baron:2012fv}
\bibitem{Baron:2012fv} 
  W.~Baron, D.~Galante and M.~Schvellinger,
  ``Dynamics of holographic thermalization,''
  JHEP {\bf 1303}, 070 (2013)
  [arXiv:1212.5234 [hep-th]].
  %%CITATION = ARXIV:1212.5234;%%
  %17 citations counted in INSPIRE as of 03 Jan 2014

%\cite{Fischler:2012ca}
\bibitem{Fischler:2012ca} 
  W.~Fischler and S.~Kundu,
``Strongly Coupled Gauge Theories: High and Low Temperature Behavior of Non-local Observables,''
  JHEP {\bf 1305}, 098 (2013)
  [arXiv:1212.2643 [hep-th]].
  %%CITATION = ARXIV:1212.2643;%%
  %12 citations counted in INSPIRE as of 15 Jan 2014


%\cite{Fischler:2012uv}
\bibitem{Fischler:2012uv} 
  W.~Fischler, A.~Kundu and S.~Kundu,
``Holographic Mutual Information at Finite Temperature,''
  Phys.\ Rev.\ D {\bf 87}, 126012 (2013)
  [arXiv:1212.4764 [hep-th]].
  %%CITATION = ARXIV:1212.4764;%%
  %13 citations counted in INSPIRE as of 15 Jan 2014

  %\cite{Caputa:2013eka}
\bibitem{Caputa:2013eka} 
  P.~Caputa, G.~Mandal and R.~Sinha,
``Dynamical entanglement entropy with angular momentum and U(1) charge,''
  JHEP {\bf 1311}, 052 (2013)
  [arXiv:1306.4974 [hep-th]].
  %%CITATION = ARXIV:1306.4974;%%
  %8 citations counted in INSPIRE as of 15 Jan 2014

  %\cite{Fischler:2013fba}
\bibitem{Fischler:2013fba} 
  W.~Fischler, S.~Kundu and J.~F.~Pedraza,
``Entanglement and out-of-equilibrium dynamics in holographic models of de Sitter QFTs,''
  arXiv:1311.5519 [hep-th].
  %%CITATION = ARXIV:1311.5519;%%
  %2 citations counted in INSPIRE as of 15 Jan 2014



%\cite{Pedraza:2014moa}
\bibitem{Pedraza:2014moa} 
  J.~F.~Pedraza,
``Evolution of non-local observables in an expanding boost-invariant plasma,''
  arXiv:1405.1724 [hep-th].
  %%CITATION = ARXIV:1405.1724;%%



%\cite{Mukherjee:2014gia}
\bibitem{Mukherjee:2014gia} 
  D.~Mukherjee and K.~Narayan,
  ``AdS plane waves, entanglement and mutual information,''
  arXiv:1405.3553 [hep-th].
  %%CITATION = ARXIV:1405.3553;%%
  %1 citations counted in INSPIRE as of 29 Jun 2014


%\cite{Alishahiha:2014cwa}{Fonda:2014ula}
\bibitem{Alishahiha:2014cwa} 
  M.~Alishahiha, A.~F.~Astaneh and M.~R.~M.~Mozaffar,
  ``Thermalization in Backgrounds with Hyperscaling Violating Factor,''
  arXiv:1401.2807 [hep-th].
  %%CITATION = ARXIV:1401.2807;%%
  %1 citations counted in INSPIRE as of 24 Mar 2014


%\cite{Fonda:2014ula}
\bibitem{Fonda:2014ula} 
  P.~Fonda, L.~Franti, V.~Keranen, E.~Keski-Vakkuri, L.~Thorlacius and E.~Tonni,
  ``Holographic thermalization with Lifshitz scaling and hyperscaling violation,''
  arXiv:1401.6088 [hep-th].
  %%CITATION = ARXIV:1401.6088;%%


%\cite{Keranen:2014zoa}
\bibitem{Keranen:2014zoa} 
  V.~Keranen, H.~Nishimura, S.~Stricker, O.~Taanila and A.~Vuorinen,
  ``Universality in holographic entropy production,''
  arXiv:1405.7015 [hep-th].
  %%CITATION = ARXIV:1405.7015;%%












%\cite{Headrick:2010zt}{Hubeny:2007re}{Tonni:2010pv}
\bibitem{Headrick:2010zt} 
  M.~Headrick,
  ``Entanglement Renyi entropies in holographic theories,''
  Phys.\ Rev.\ D {\bf 82}, 126010 (2010)
  [arXiv:1006.0047 [hep-th]].
  %%CITATION = ARXIV:1006.0047;%%
  %112 citations counted in INSPIRE as of 24 Mar 2014

%\cite{Hubeny:2007re}{Tonni:2010pv}
\bibitem{Hubeny:2007re} 
  V.~E.~Hubeny and M.~Rangamani,
  ``Holographic entanglement entropy for disconnected regions,''
  JHEP {\bf 0803}, 006 (2008)
  [arXiv:0711.4118 [hep-th]].
  %%CITATION = ARXIV:0711.4118;%%
  %30 citations counted in INSPIRE as of 24 Mar 2014


%\cite{Tonni:2010pv}
\bibitem{Tonni:2010pv} 
  E.~Tonni,
  ``Holographic entanglement entropy: near horizon geometry and disconnected regions,''
  JHEP {\bf 1105}, 004 (2011)
  [arXiv:1011.0166 [hep-th]].
  %%CITATION = ARXIV:1011.0166;%%
  %17 citations counted in INSPIRE as of 24 Mar 2014



%\cite{Balasubramanian:2011at}{Allais:2011ys}{Callan:2012ip}{Li:2013sia}
\bibitem{Balasubramanian:2011at} 
  V.~Balasubramanian, A.~Bernamonti, N.~Copland, B.~Craps and F.~Galli,
  ``Thermalization of mutual and tripartite information in strongly coupled two dimensional conformal field theories,''
  Phys.\ Rev.\ D {\bf 84}, 105017 (2011)
  [arXiv:1110.0488 [hep-th]].
  %%CITATION = ARXIV:1110.0488;%%
  %29 citations counted in INSPIRE as of 24 Mar 2014

%\cite{Allais:2011ys}{Callan:2012ip}{Li:2013sia}
\bibitem{Allais:2011ys} 
  A.~Allais and E.~Tonni,
  ``Holographic evolution of the mutual information,''
  JHEP {\bf 1201}, 102 (2012)
  [arXiv:1110.1607 [hep-th]].
  %%CITATION = ARXIV:1110.1607;%%
  %33 citations counted in INSPIRE as of 24 Mar 2014










%\cite{Callan:2012ip}{Li:2013sia}
\bibitem{Callan:2012ip} 
  R.~Callan, J.~-Y.~He and M.~Headrick,
  ``Strong subadditivity and the covariant holographic entanglement entropy formula,''
  JHEP {\bf 1206}, 081 (2012)
  [arXiv:1204.2309 [hep-th]].
  %%CITATION = ARXIV:1204.2309;%%
  %25 citations counted in INSPIRE as of 24 Mar 2014

%\cite{Li:2013sia}
\bibitem{Li:2013sia} 
  Y.~-Z.~Li, S.~-F.~Wu, Y.~-Q.~Wang and G.~-H.~Yang,
  ``Linear growth of entanglement entropy in holographic thermalization captured by horizon interiors and mutual information,''
  JHEP {\bf 1309}, 057 (2013)
  [arXiv:1306.0210 [hep-th]].
  %%CITATION = ARXIV:1306.0210;%%
  %5 citations counted in INSPIRE as of 24 Mar 2014






%\cite{Liu:2013iza}{Liu:2013qca}
\bibitem{Liu:2013iza} 
  H.~Liu and S.~J.~Suh,
  ``Entanglement Tsunami: Universal Scaling in Holographic Thermalization,''
  arXiv:1305.7244 [hep-th].
  %%CITATION = ARXIV:1305.7244;%%
  %16 citations counted in INSPIRE as of 20 Dec 2013



%\cite{Liu:2013qca}
\bibitem{Liu:2013qca} 
  H.~Liu and S.~J.~Suh,
  ``Entanglement growth during thermalization in holographic systems,''
  arXiv:1311.1200 [hep-th].
  %%CITATION = ARXIV:1311.1200;%%
  %3 citations counted in INSPIRE as of 06 Dec 2013

%\cite{Fischler:2012uv}
%\bibitem{Fischler:2012uv} 
%  W.~Fischler, A.~Kundu and S.~Kundu,
%  ``Holographic Mutual Information at Finite Temperature,''
%  Phys.\ Rev.\ D {\bf 87}, no. 12, 126012 (2013)
%  [arXiv:1212.4764 [hep-th]].
  %%CITATION = ARXIV:1212.4764;%%
  %14 citations counted in INSPIRE as of 24 Mar 2014




%\cite{Bhattacharya:2012mi}{Allahbakhshi:2013rda}{Blanco:2013joa}{Wong:2013gua}
\bibitem{Bhattacharya:2012mi} 
  J.~Bhattacharya, M.~Nozaki, T.~Takayanagi and T.~Ugajin,
  ``Thermodynamical Property of Entanglement Entropy for Excited States,''
  Phys.\ Rev.\ Lett.\  {\bf 110}, no. 9, 091602 (2013)
  [arXiv:1212.1164].
  %%CITATION = ARXIV:1212.1164;%%
  %24 citations counted in INSPIRE as of 18 Dec 2013

%\cite{Allahbakhshi:2013rda}{Blanco:2013joa}{Wong:2013gua}
\bibitem{Allahbakhshi:2013rda} 
  D.~Allahbakhshi, M.~Alishahiha and A.~Naseh,
  ``Entanglement Thermodynamics,''
  JHEP {\bf 1308}, 102 (2013)
  [arXiv:1305.2728 [hep-th]].
  %%CITATION = ARXIV:1305.2728;%%
  %16 citations counted in INSPIRE as of 18 Dec 2013


%\cite{Blanco:2013joa}{Wong:2013gua}
\bibitem{Blanco:2013joa} 
  D.~D.~Blanco, H.~Casini, L.~-Y.~Hung and R.~C.~Myers,
  ``Relative Entropy and Holography,''
  JHEP {\bf 1308}, 060 (2013)
  [arXiv:1305.3182 [hep-th]].
  %%CITATION = ARXIV:1305.3182;%%
  %16 citations counted in INSPIRE as of 18 Dec 2013

%\cite{Wong:2013gua}
\bibitem{Wong:2013gua} 
  G.~Wong, I.~Klich, L.~A.~Pando Zayas and D.~Vaman,
  ``Entanglement Temperature and Entanglement Entropy of Excited States,''
  arXiv:1305.3291 [hep-th].
  %%CITATION = ARXIV:1305.3291;%%
  %13 citations counted in INSPIRE as of 18 Dec 2013




%\cite{Nishioka:2009un}
\bibitem{Nishioka:2009un} 
  T.~Nishioka, S.~Ryu and T.~Takayanagi,
  ``Holographic Entanglement Entropy: An Overview,''
  J.\ Phys.\ A {\bf 42}, 504008 (2009)
  [arXiv:0905.0932 [hep-th]].
  %%CITATION = ARXIV:0905.0932;%%
  
 
\end{thebibliography}
\end{document}